\documentclass[11pt]{article}

\usepackage[usenames,dvipsnames,svgnames,table]{xcolor} 
\usepackage[obeyspaces,hyphens,spaces]{url}
\usepackage{jcapmod}

\usepackage{booktabs}
\usepackage[english]{babel}
\usepackage{amsmath, amssymb, amsbsy, amstext, amsthm, simplewick}
\usepackage{hyperref}
\usepackage{graphicx}
\usepackage{amsfonts}
\usepackage{enumitem}
\usepackage{upgreek}
\usepackage{framed}
\usepackage{tensor}
\usepackage{pifont}
\usepackage{latexsym, mathrsfs}
\usepackage{array}
\usepackage{hyperref}
\usepackage{xspace}
\usepackage{longtable}
\usepackage{multirow}
\usepackage{cases}
\usepackage{empheq}
\usepackage{bm}
\usepackage{bbm}

\usepackage{braket}

\usepackage[load-configurations=astronomy, range-units=brackets, range-phrase=-, per-mode=reciprocal, mode=math]{siunitx}
\usepackage{threeparttable}
\usepackage{subfig}

\usepackage{ragged2e}

\usepackage{hhline}
\usepackage{array}
\newcolumntype{P}[1]{>{\centering\arraybackslash}p{#1}}
\usepackage{booktabs}
\usepackage{tikz}
\usetikzlibrary{calc,shadings,patterns,tikzmark,fadings}

\definecolor{Blue}{rgb}{0.25, 0.41, 0.88}
\definecolor{Red}{rgb}{0.92,0.,0.}
\definecolor{darkorange}{rgb}{1.0,0.549,0.}
\definecolor{cobalt}{RGB}{44, 98, 120}
\definecolor{Mathematica1}{rgb}{0.368417, 0.506779, 0.709798}
\definecolor{Mathematica2}{rgb}{0.880722, 0.611041, 0.142051}
\definecolor{Mathematica3}{rgb}{0.560181, 0.691569, 0.194885}
\definecolor{Mathematica4}{rgb}{0.922526, 0.385626, 0.209179}
\definecolor{Mathematica5}{rgb}{0.528488, 0.470624, 0.701351}
\definecolor{Mathematica6}{rgb}{0.772079, 0.431554, 0.102387}
\definecolor{Mathematica7}{rgb}{0.363898, 0.618501, 0.782349}
\definecolor{Mathematica8}{rgb}{1, 0.75, 0}
\definecolor{Mathematica9}{rgb}{0.647624, 0.37816, 0.614037}
\definecolor{plotBlue}{RGB}{94, 130, 181}
\definecolor{plotRed}{RGB}{233, 85, 54}
\definecolor{plotGreen}{RGB}{142, 176, 50}
\definecolor{plotPurple}{RGB}{135, 120, 178}

\definecolor{cornellRed}{HTML}{B31B1B}
\definecolor{cornellBlue}{HTML}{0068AC}
\definecolor{cornellGreen}{HTML}{6EB43F}

\newcolumntype{C}[1]{>{\centering\let\newline\\\arraybackslash\hspace{0pt}}m{#1}}

\def\d{{\rm d}}

\newcommand{\es}{\hspace{0.5pt}}

\makeatletter
\newlength{\apb@width}
\newcommand{\autoparbox}[2][c]{\settowidth{\apb@width}{#2}\parbox[#1]{\apb@width}{#2}}

\makeatother

\numberwithin{equation}{section}

\def\beq{\begin{equation}}
\def\eeq{\end{equation}}

\def\bea{\begin{eqnarray}}
\def\eea{\end{eqnarray}}

\def\d{{\rm d}}

\def\beq{\begin{equation}}
\def\eeq{\end{equation}}
\def\bea{\begin{eqnarray}}
\def\eea{\end{eqnarray}}

\def\d{{\rm d}}

\def\d{{\rm d}}

\newcommand{\ud}{\mathrm{d}}
\newcommand{\lab}[1]{{\mathrm{#1}}}
\newcommand{\mb}[1]{{\mathbf{#1}}}
\newcommand{\minus}{{\scalebox{0.75}[1.0]{$-$}}}

\DeclareRobustCommand{\SkipTocEntry}[4]{}

\usepackage{colortbl}
\definecolor{blue2}{cmyk}{1, 0.1, 0.1, 0}

\definecolor{pyBlue}{RGB}{31, 119, 180}
\definecolor{pyRed}{RGB}{214, 39, 40}
\definecolor{pyGreen}{RGB}{44, 160, 44}
\definecolor{pyBlue2}{RGB}{0, 111, 237}
\definecolor{pyRed2}{RGB}{224, 52, 36}

\makeatletter
\def\Ddots{\mathinner{\mkern1mu\raise\p@
\vbox{\kern7\p@\hbox{.}}\mkern2mu
\raise4\p@\hbox{.}\mkern2mu\raise7\p@\hbox{.}\mkern1mu}}
\makeatother

\newcommand{\res}{\Omega} 
\newcommand{\Or}{\Omega_r} 

\begin{document}

\pagenumbering{roman}
\begin{titlepage}
\baselineskip=15.5pt \thispagestyle{empty}
\begin{flushright}
DESY 19-221
\end{flushright}
\vspace{1cm}
\begin{center}
{\fontsize{22}{24}\selectfont  \bfseries Gravitational Collider Physics}
\end{center}
\vspace{0.1cm}
\begin{center}
{\fontsize{12}{18}\selectfont Daniel Baumann,$^{1}$ Horng Sheng Chia,$^{1}$ Rafael A.~Porto,$^{2,3}$ and John Stout$^{1}$} 
\end{center}

\begin{center}
\vskip8pt
\textsl{$^1$ Institute for Theoretical Physics, University of Amsterdam,\\Science Park 904, Amsterdam, 1098 XH, The Netherlands}

\vskip8pt
\textsl{$^2$ Deutsches Elektronen-Synchrotron DESY,\\
Notkestra$\beta$e 85, D-22607 Hamburg, Germany} \vskip8pt
\textsl{$^3$ The Abdus Salam International Center for Theoretical Physics,\\ Strada Costiera, 11, Trieste 34151, Italy}

\end{center}

\vspace{1.2cm}
\hrule \vspace{0.3cm}
\noindent {\bf Abstract}\\[0.1cm]
We study the imprints of new ultralight particles on the gravitational-wave signals emitted by binary black holes. Superradiant instabilities may 
create large clouds of scalar or vector fields around rotating black holes. The presence of a binary companion then induces transitions between different states of the cloud, which become resonantly enhanced when the orbital frequency matches the energy gap between the states. We~find that the time dependence of the orbit 
significantly impacts the cloud's dynamics during a transition. Following an analogy with particle colliders, we introduce an S-matrix formalism to describe the evolution through multiple resonances. We show that the state of the cloud, as~it approaches the merger, carries vital information about its spectrum via time-dependent finite-size effects. Moreover, due to the transfer of energy and angular momentum between the cloud and the orbit, a dephasing of the gravitational-wave signal can occur which is correlated with the positions of the resonances. Notably, for intermediate and extreme mass ratio inspirals, long-lived floating orbits are possible, as well as kicks that yield large eccentricities. Observing these effects, through the precise reconstruction of waveforms, 
has the potential to unravel the internal structure of the boson clouds, ultimately probing the masses and spins of~new~particles. 

\vskip10pt
\hrule
\vskip10pt

\end{titlepage}

\thispagestyle{empty}
\setcounter{page}{2}
\tableofcontents

\newpage
\pagenumbering{arabic}
\setcounter{page}{1}

\clearpage
\section{Introduction}
 \label{sec:introduction}

\begin{figure}[t]
  \centering
  \includegraphics{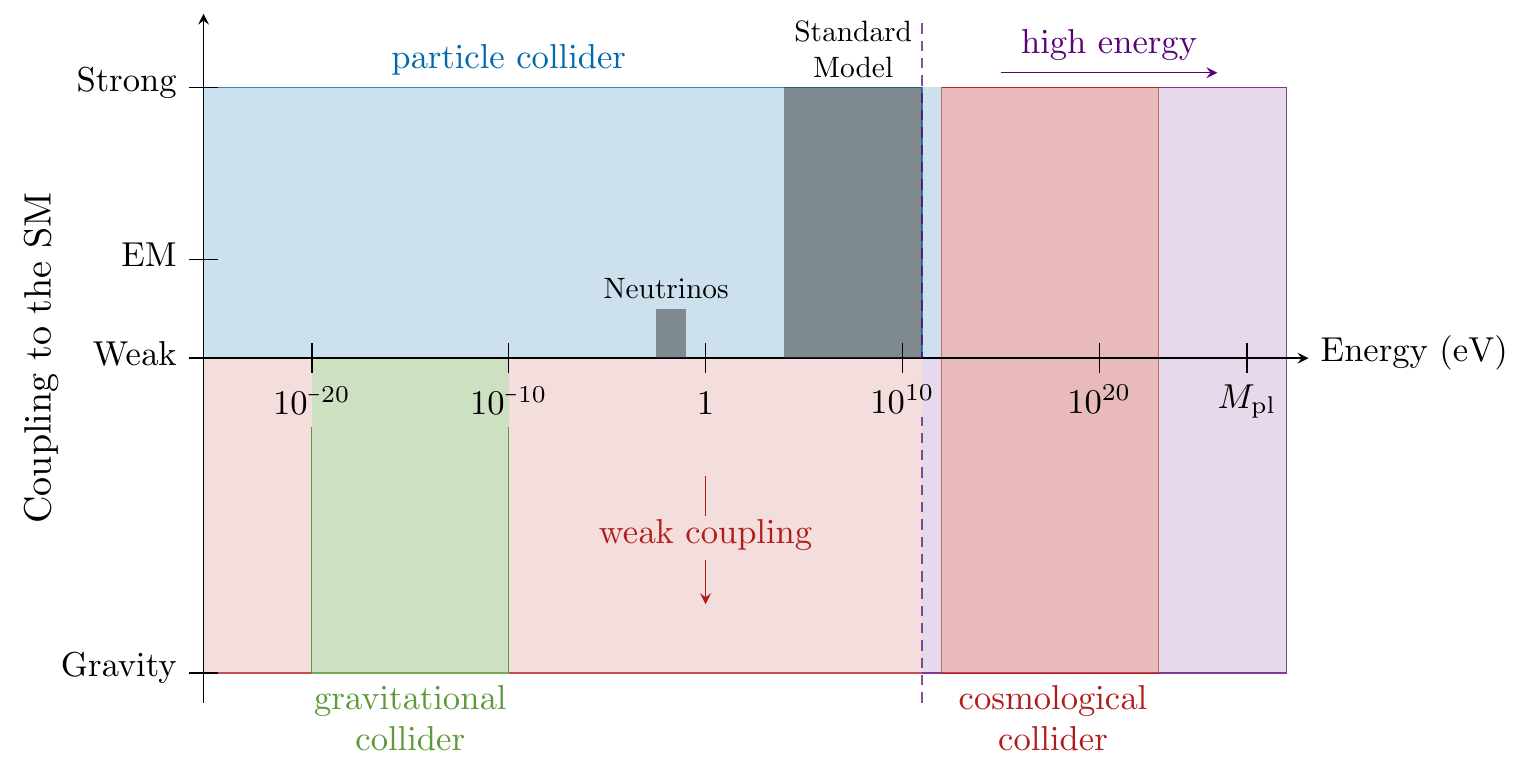}
  \caption{Particle colliders are blind to new particles that are either too heavy or too weakly-coupled to be produced in sufficient numbers. Very massive particles (up to $10^{14}$\,GeV) can be created in the ``cosmological collider" that was active during inflation~\cite{Chen:2009zp, Baumann:2011nk, Noumi:2012vr, Flauger:2013hra,Arkani-Hamed:2015bza, Arkani-Hamed:2018kmz, Baumann:2019oyu}.  
  In this paper, we explore instead the weak-coupling frontier, through searches of ultralight particles with precision gravitational-wave observations. These bosons, with masses in the range $[10^{-20}, 10^{-10}]$\,eV, can be produced around astrophysical 
   black holes,
 leaving distinct imprints in the gravitational waves emitted by black hole binaries.  Like in ordinary colliders, the signals in ``gravitational colliders" are sensitive to both the masses and spins of the bosonic fields. } 
  \label{fig:col1}
\end{figure}

The search for fundamental particles has 
  played a key role in the history of particle physics. In~high-energy scattering processes, new particles produced during the collision may either be long-lived and observed directly, or detected indirectly as resonances. The position of the resonance determines the mass of the new particle, while the angular dependence of the decay products encodes its spin.  
While particle colliders have helped to uncover the fundamental building blocks of nature~\cite{Aad:2012tfa,Chatrchyan:2012xdj}, 
 this way of probing new physics relies on having appreciable interactions with the Standard Model particles involved in the collision. Traditional collider experiments are therefore blind to ``dark sectors" which couple very weakly to ordinary matter, even if the associated new particles are very light~\cite{Essig:2013lka}  (see Fig.~\ref{fig:col1}).  In that
  case, we must find creative new ways to probe these particles. 
  As we will show,
  the detection of gravitational waves~\cite{Abbott:2016blz, Abbott:2016nmj, Abbott:2017vtc, Abbott:2017oio, TheLIGOScientific:2017qsa} 
  has not only initiated a new era for multi-messenger astronomy~\cite{GBM:2017lvd, Sathyaprakash:2019rom}, but also provides a new 
 opportunity to explore this
 weak-coupling frontier~\cite{review,Porto:2016zng,Porto:2017lrn, Barack:2018yly, Sathyaprakash:2019yqt,Bertone:2019irm}. In particular, the gravitational waves emitted by 
 binary black holes may carry the fingerprints of the masses and spins of hypothetical new particles, making these systems effectively  ``gravitational colliders." 

\vskip 4pt 
Although the new particles may couple to the Standard Model only through gravity, 
they can still be copiously produced in astrophysical environments, 
such as around spinning black holes. 
Specifically, if the Compton wavelength of an ultralight bosonic field is larger than the black hole's gravitational radius, a large condensate, or ``boson cloud,'' can form through a superradiant instability~\cite{Zeldovich:1971a, Zeldovich:1972spj, Brito:2015oca}. 
 The energy eigenstates of this cloud are similar to those of the hydrogen atom, which is why the system is often  
   called a ``gravitational atom''~\cite{Arvanitaki:2009fg, Arvanitaki:2010sy}.
In~isolation, these boson clouds are difficult to observe, as we must rely
on their feeble continuous gravitational-wave emission; see e.g.~\cite{Arvanitaki:2010sy, Yoshino:2014}. 
However, as it was demonstrated in~\cite{Baumann:2018vus}, when 
 gravitational atoms are part of binary systems, they leave characteristic imprints in the gravitational waves emitted during the inspiral. 
The presence of the binary companion can induce transitions between the different states of the boson~cloud, which are resonantly enhanced when the orbital frequency matches the energy gap between the states. 
These signals are much louder than those from the cloud itself~\cite{Arvanitaki:2010sy, Yoshino:2014}, making them interesting new targets for gravitational-wave~searches.

\vskip 4pt
In this paper, we extend our previous work \cite{Baumann:2018vus} in a number of important directions. 
First, we take into account more precisely the time dependence of the orbital frequency during the binary inspiral. 
We find that this time dependence, however small,  
has an important effect on the dynamics of the cloud near a resonance, leading to an analog of the 
Landau-Zener transition in quantum mechanics~\cite{Landau, Zener}.
Second, we study the backreaction of the cloud on the binary's orbital motion and discuss the associated gravitational-wave signatures, which includes the {\it dephasings} produced by ``floating" or ``sinking" orbits. 
Finally, we study the dynamics of both ultralight scalar and vector clouds,  
thus allowing for the new particles to carry spin.\footnote{String compactifications typically generate hundreds of four-dimensional $p$-form fields, some of which are plausibly ultralight \cite{Demirtas:2018akl}. 
Since massive two- and three-forms can be dualized into massive one-forms and scalars~\cite{Buchbinder:2008jf}, we cover a wide range of fields, with large theoretical prior, by considering scalars and vectors.} 
For vector fields, the intrinsic spin leads to additional, nearly degenerate states in the spectrum~\cite{Baumann:2019eav}, 
which will play a key role in distinguishing the phenomenology of scalar and vector~clouds.

\vskip 4pt
\noindent
As first highlighted in \cite{Baumann:2018vus}, resonant transitions are a key feature of the evolution of boson clouds in binary systems. For scalar clouds and quasi-circular orbits, the relevant mixings are predominantly between two states. In this case, the dynamics near the resonance are rather simple and can be studied in detail.  However, qualitatively new features appear when multiple, nearly degenerate states are involved in the transition, which is the case for vector clouds.  As we will see, when the binary moves slowly through the resonance band, a complete transfer of population from the initial state to a linear combination of allowed states may occur, a process known as a Landau-Zener transition (or ``adiabatic following")~\cite{Landau, Zener}. While the transition populates only a single state for scalars, a vector cloud can evolve into a superposition of states. 
This can result in neutrino-like oscillations between different shapes, with characteristic frequencies set by the energy splitting between the states.  These oscillations persist after the 
 resonance, and are therefore a key characteristic of clouds carrying intrinsic spin. Because the transitions are localized in time (or frequency), they can also be treated as ``scattering events" and described individually by an S-matrix. This approach will be particularly useful when we consider multiple, sequential resonances (see below).

\begin{figure}[t]
        \centering
        \includegraphics[scale=0.9]{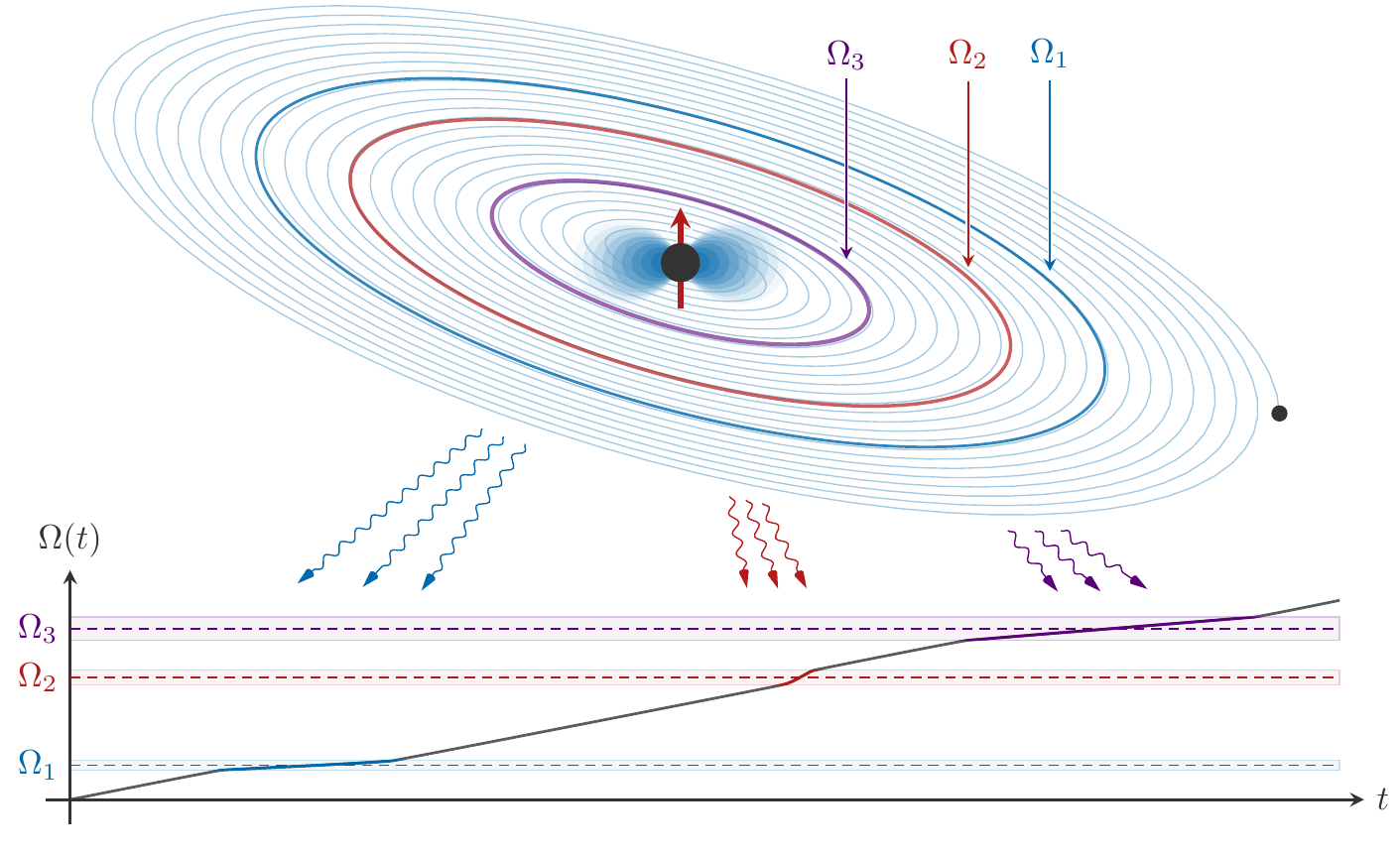}
    \caption{Illustration of the gravitational atom in a binary system. During the inspiral, the companion induces level mixings between different states of the cloud. This mixing is strongest when the orbital frequency $\Omega(t)$, which slowly increases due to gravitational-wave emission, matches the energy gap between the coupled states. During these resonant transitions, the change in the state of the cloud produces a backreaction on the binary's orbital dynamics. These effects either induce floating or sinking orbits across these resonance bands, thus leaving distinct imprints in the gravitational waves emitted by the binary. Moreover, minute time-dependent finite-size effects can further distinguish the nature of the boson cloud, including its  intrinsic spin.} \label{fig:AtomInBinary}
\end{figure}

\vskip 4pt
As we elaborate in this paper, the dynamics of the boson clouds can have a significant impact on the binary's orbital motion and  
associated gravitational-wave signals. 
For instance, any change in the cloud's angular momentum (which is most prominent during a resonance) must be balanced by a change in the orbital angular momentum of the binary, so that the total angular momentum is conserved. Depending on the orientation of the orbit and the nature of the transition, the binary can either absorb angular momentum from the cloud or release it.
 The induced changes in the orbital motion can produce a significant dephasing in the gravitational waves emitted by the binary, relative to the evolution without a cloud (see Fig.~\ref{fig:AtomInBinary}).\footnote{For intermediate and extreme mass ratio inspirals, with the cloud carried by the larger black hole, the backreaction on the orbit can be a dramatic effect.  If angular momentum is transfered from the cloud to the binary, long-lived floating orbits are naturally induced, creating a new source of continuous monochromatic gravitational waves.  In contrast, when the binary transfers angular momentum to the  cloud, it can experience a sudden kick instead. This greatly enriches the dynamics of the binary system, for instance by producing eccentric orbits. While these kicks are difficult to model quantitatively, they are distinctive qualitative signatures of the existence of a gravitational atom in a binary system.} 
This dephasing is correlated with the positions of the resonances and therefore probes the spectral properties of the cloud. 

\vskip 4pt
Even though the backreaction on the orbit is a smoking gun for the existence of boson clouds, many degeneracies between the scalar and vector case still remain, since the transfer of angular momentum is not very sensitive to the 
internal structure of the clouds. To break these degeneracies, 
we need to probe 
the clouds during the later stages
of the inspiral and towards merger. 
This is where the S-matrix formalism we develop in this paper becomes a useful tool, as it allows us to follow the state of the cloud through a sequence of resonances. As we shall see, given an initial state $|i\rangle$, the final state after $N$ resonances can be written as 
\begin{equation}
|f \rangle  = \prod_{n=1}^N S_n\, |i \rangle \, , \nonumber
\end{equation} 
where $S_n$ is a unitary operator that evolves the state across the $n$-th resonance. Given the differences in the interaction Hamiltonian and eigenstates, the state of the cloud after multiple resonances will be different for scalar and vector clouds. In particular, as was anticipated in~\cite{Baumann:2018vus}, time-dependent finite-size effects develop which, due to the extended nature of the cloud, may be observed in the early phase of the binary inspiral. Moreover, strong mixing with decaying modes may also rapidly deplete the cloud as it approaches merger. The evolving shape of the cloud thus creates a distinct fingerprint in the binary's gravitational-wave signal. 

\vskip 4pt
As in ordinary collider physics, signals from the gravitational collider are sensitive to the masses and spins of the new particles. In particular, the positions of the resonant transitions depend on the mass of the bosonic field, while its spin can be inferred by linking the dephasing during these
transitions with the oscillations in the shape of the cloud, as measured through 
time-dependent finite-size effects in the waveform. Although these effects are difficult to model in detail, they are correlated with the position of the resonances, and therefore represent robust qualitative signatures of the existence of gravitational atoms in binary systems. 
Our findings motivate revisiting current waveform models in order to target the unique signatures of ultralight particles in binary gravitational-wave searches.

\paragraph{Outline}  The plan of the paper is as follows: In Section~\ref{sec:Binary}, we review the spectra of bound states of scalar and vector fields around  
rotating black holes. We also describe the mixing between different states in the presence of a binary companion. In~Section~\ref{sec:gcollider}, we study 
the dynamics of this level mixing. We show that
resonant transitions can be described as scattering events and derive the transition ``probabilities" between coupled states. We explain that this S-matrix approach is particularly well suited to chain together a sequence of multiple resonances.
In Section~\ref{sec:Backreaction}, we study the backreaction of 
 time-dependent boson clouds on the orbital motion. 
 In Section~\ref{sec:unravel}, we explore the observational consequences of boson clouds in binary systems, and study the effects that can help us ultimately unravel 
 their atomic structure. Our conclusions and outlook are presented in Section~\ref{sec:Conclusions}. 

\vskip 4pt
A number of appendices contain additional details: In Appendix~\ref{app:GP}, we elaborate on the gravitational perturbations induced by the binary companion.   In Appendix~\ref{app:LZ}, we review the analytic solution for the two-level Landau-Zener transition. We also introduce Floquet theory 
 to deal with more general orbital configurations. Finally, in Appendix~\ref{app:AMTransfer}, we provide further details on the angular momentum transfer between the cloud and the orbit.

\paragraph{Notation and conventions} Our metric signature will be $(-,+, +, +)$ and, unless stated otherwise, we will work in natural units with $G = \hbar = c = 1$. Greek letters will stand for spacetime indices.  The gravitational radius of a black hole is $r_g \equiv G M / c^2$.  Quantities associated to the boson clouds will be denoted by the subscript~$c$.
For example, the mass and angular momentum of the clouds are $M_c$ and $S_c$, respectively.
The gravitational fine-structure constant is $\alpha \equiv r_g/\lambda_c$, where $\lambda_c \equiv \hbar / (\mu c)$ is the (reduced) Compton wavelength of a boson field with mass $\mu$.

\vskip 4pt
The Kerr metric in Boyer-Lindquist coordinates is 
\beq
\d s^2 = - \frac{\Delta}{\rho^2}\left(\d t - a \sin^2 \theta\, \d \phi \right)^2 + \frac{\rho^2}{\Delta} \d r^2 + \rho^2\, \d \theta^2 + \frac{\sin^2 \theta}{\rho^2}  \left(a\, \d t - (r^2 + a^2) \, \d \phi \right)^2\, , \label{equ:Kerr}
\eeq
where $\Delta \equiv r^2 -  2 Mr +a^2$ and $ \rho^2 \equiv r^2 + a^2 \cos^2 \theta$. 
The roots of $\Delta$ determine the inner and outer horizons, located at $r_\pm = M \pm \sqrt{M^2 -a^2}$, and the angular velocity of the black hole at the outer horizon is $\Omega_\lab{H} \equiv a/ 2 M r_+$. Dimensionless quantities, defined with respect to the black hole mass~$M$, are labeled by tildes. For example, the dimensionless spin of the black hole is~$\tilde{a} \equiv a/M$.

\vskip 4pt
The scalar and vector eigenstates are denoted by $|\es n \es \ell \es m\rangle$ and $| \es n \es \ell \es j \es m\rangle$, with the integers $\{n, \ell, j, m \}$ corresponding to the principal, orbital angular momentum, total angular momentum, and azimuthal angular momentum numbers, respectively.

\vspace{0.5cm}
\section{Gravitational Atoms in Binaries} 
\label{sec:Binary}

The key parameter underlying the superradiance phenomenon is the ratio of the gravitational
radius of the rotating black hole, $r_g$, to the (reduced) Compton wavelength of the field, $\lambda_c$. This is often called the gravitational fine-structure constant, whose typical values are of order
\begin{equation}
  \alpha \simeq 0.04 \left(\frac{M}{60 M_\odot}\right) \left(\frac{\mu}{10^{-13}\, \lab{eV}}\right).
\end{equation}
Superradiant growth can only occur when the parameter $\alpha$ is smaller than order unity. 
At the same time, $\alpha$ also determines the energy spectrum of the quasi-bound states of a bosonic field around the black hole, playing a similar role to the fine-structure constant in quantum electrodynamics.

\vskip 4pt 
We start, in \S\ref{sec:ScalarVector}, by reviewing the bound-state spectra of gravitational atoms for both scalar and vector fields. We then describe, in \S\ref{sec:gravlevelmix}, how gravitational perturbations associated with the binary companion can induce transitions between different energy eigenstates. The~time dependence of these perturbations and the conditions for exciting resonant transitions are discussed in~\S\ref{sec:TimeDependentTidalMoments}.

\subsection{Scalar and Vector Clouds} 
\label{sec:ScalarVector}

There is an extensive literature on superradiance with scalar~\cite{Detweiler:1980uk, Dolan:2007mj, Arvanitaki:2010sy, Yoshino:2012kn, Dolan:2012yt, Okawa:2014nda, Yoshino:2014, Baumann:2018vus, Brito:2014wla, Arvanitaki:2014wva, Arvanitaki:2016qwi, Brito:2017wnc, Brito:2017zvb} and vector fields~\cite{Witek:2012tr, Pani:2012vp, Pani:2012bp, Endlich:2016jgc, East:2017ovw, East:2017mrj, East:2018glu, Baryakhtar:2017ngi, Cardoso:2018tly, Dolan:2018dqv, Baumann:2019eav, Siemonsen:2019cr}. We will briefly review the key features that will be relevant for the analysis in this paper. 

\subsubsection*{Scalar clouds}

The Klein-Gordon equation for a scalar field of mass $\mu$ in a curved spacetime is
\beq
\left(g^{\alpha \beta} \nabla_\alpha \nabla_\beta - \mu^2 \right) \Phi (t, \textbf{r}) = 0 \, . \label{equ:KG} 
\eeq
Remarkably, the bound-state solutions of (\ref{equ:KG}) in the Kerr background \eqref{equ:Kerr} are similar to the states of the hydrogen atom in quantum mechanics. To make this manifest, it is useful to consider the ansatz 
\begin{equation}
  \Phi(t, \bm{r}) = \frac{1}{\sqrt{2 \mu}} \left[\psi(t, \mb{r})\, e^{-i \mu t} + \psi^*(t, \mb{r}) e^{+i \mu t}\right] ,
  \label{equ:NRansatz}
\end{equation}
where $\psi$ is a complex scalar field which varies on a timescale that is longer than $\mu^{-1}$. Substituting this ansatz into (\ref{equ:KG}) and expanding in powers of $\alpha$, we find
\begin{equation}
  i \frac{\partial}{\partial t} \psi (t, \textbf{r}) =  \left( -\frac{1}{2\mu} \nabla^2  - \frac{\alpha}{r} + \Delta V\right) \psi (t, \textbf{r}) \, , \label{eqn:NonRelScalar}
\end{equation}
where $\Delta V$ represents higher-order corrections in $\alpha$.  At~leading order, this is identical to the Schr\"{o}dinger equation for the hydrogen atom, whose eigenfunctions are labeled by the principal  
``quantum" number $n$, the orbital angular momentum number~$\ell$, and the azimuthal angular momentum number~$m$.
These quantum numbers satisfy $n \geq \ell+1$, $\ell \geq 0$, and $ \ell \geq |m|$. The bound-state solutions of the scalar cloud at leading order are thus given by
 \beq
\psi_{n \ell m}(t, \textbf{r} ) = R_{n \ell}(r) Y_{\ell m}(\theta, \phi ) \, e^{- i (\omega_{n \ell m}-\mu) t}  \, , \label{eqn:ScalarEigenstate}
\eeq
where $Y_{\ell m}$ are the scalar spherical harmonics and $R_{n \ell}$ are the hydrogenic radial functions.  For notational simplicity, we will  
denote the normalized eigenstates (\ref{eqn:ScalarEigenstate}) by $|n \es \ell \es m\rangle$, with $\langle n \es \ell \es m | n' \es \ell' \es m'\rangle = \delta_{n n'} \delta_{\ell \ell'} \delta_{m m'}$. The overall amplitude of (\ref{eqn:ScalarEigenstate}), determined by the total mass of the cloud, will be restored when necessary. Notice that for small values of $\alpha$, following the analogy with the hydrogen atom, the radial profile peaks at the ``Bohr radius'' $r_\lab{c} \equiv (\mu \alpha)^{-1}$. The cloud is thus concentrated at a distance which is larger than both the gravitational radius of the black hole and the Compton wavelength of the~field.

\vskip 4pt
There is, however, an important difference between the hydrogen atom and the gravitational atom. While the former has wavefunctions that are regular at $r = 0$, the latter must satisfy purely ingoing boundary conditions at the black hole's outer horizon. As a consequence, the eigenstates of a boson cloud have eigenfrequencies which are generally complex,
\beq
\omega = E + i \es \Gamma \, ,
\eeq
where $E$ and $\Gamma$ denote the energies and instability rates, respectively.  
At small values of $\alpha$, one has~\cite{Baumann:2018vus, Baumann:2019eav, Detweiler:1980uk} 
\begin{align}
E_{n \ell m}  &= \mu \left( 1 - \frac{\alpha^2}{2n^2} - \frac{\alpha^4}{8n^4}   - \frac{(  3n - 2\ell - 1) \, \alpha^4}{n^4 \left( \ell + 1/2\right)} +  \frac{2  \tilde a m\hskip 2pt \alpha^5}{n^3 \ell (\ell+1/2)(\ell+1)} + \mathcal{O}(\alpha^6) \right)   , \label{eqn:scalarspectrum} \\
\Gamma_{n \ell m} &= 2 \tilde{r}_+ C_{n \ell} \, g_{\ell m}(\tilde{a}, \alpha,\omega) \, (m \Omega_\lab{H}   - \omega_{n \ell m})\hskip 1pt \alpha^{4\ell+5}  \, , \label{eqn:ScalarRate}
\end{align}
where the numerical coefficients $C_{n \ell}$ and $g_{\ell m}$ can be found in~\cite{Baumann:2019eav}. 
Continuing the analogy with the hydrogen atom, we refer to the energy splittings as Bohr $(\Delta n \neq 0)$, fine $(\Delta \ell \neq 0)$, and hyperfine $(\Delta m \neq 0)$, respectively.  Notice that the dominant growing mode is $|n \es  \ell \es m\rangle  = |2\es 1\es 1\rangle$, with $\Gamma_{211} \propto \mu \alpha^8$. 

\begin{figure}[t]
\centering
\includegraphics[scale=1, trim = 0 0 0 0]{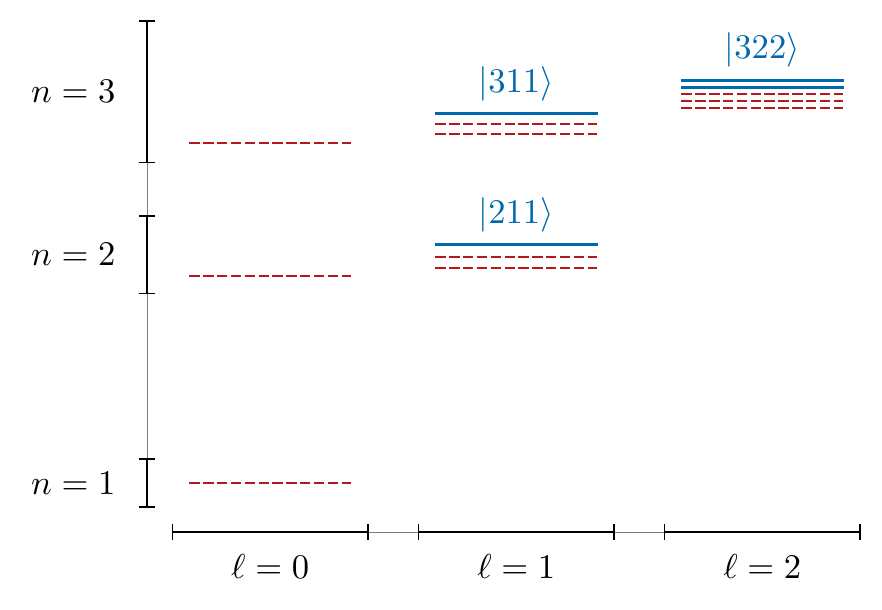}
\caption{Schematic illustration of the energy spectrum of a massive scalar field. Each state is denoted by $|\es n \es \ell \es m\rangle$. The solid blue lines are growing modes, while the dashed red lines are decaying modes. The dominant scalar mode is $|2 \es 1\es 1\rangle$. } 
\label{fig:ScalarSpectra}
\end{figure}

\subsubsection*{Vector clouds} 

A massive vector field obeys the Proca equation
\beq
\left(g^{\alpha \beta} \nabla_\alpha \nabla_\beta - \mu^2 \right) A^\mu = 0 \, , 
\label{equ:Proca}
\eeq
subject to the Lorenz condition  $ \nabla_\mu A^\mu = 0$. 
The Lorenz condition completely determines $A^0$, and so we may focus entirely on the spatial components $A^i$. Just like in the scalar case, it is convenient to introduce the analog of the ansatz (\ref{equ:NRansatz}) for $A^i$, where the slowly-varying field is now denoted by $\psi^i$. The Proca equation then leads to 
\beq
i \frac{\partial }{\partial t} \psi^i (t, \textbf{r})  =  \left[ -\frac{\delta^{il}}{2\mu} \nabla^2  - \delta^{il} \, \frac{\alpha }{r}   + \Delta V^{il}   \right] \psi^l (t, \textbf{r})  \, , \label{eqn:NonRelVector}
\eeq
where the terms in (\ref{eqn:NonRelVector}) share the same interpretation as their scalar analogs in (\ref{eqn:NonRelScalar}). Spherical symmetry is restored far from the black hole and, to leading order in $\alpha$, it is straightforward to find the eigenfunctions,
\beq
\begin{aligned}
\psi_{n\ell j m}^i (t, \textbf{r}) & = R_{n \ell}(r) Y^i_{\ell, j m} (\theta, \phi) \, e^{- i (\omega_{n \ell j m} -\mu)t} \,  , 
\label{eqn:FarLOAsol}
\end{aligned}
\eeq
where the radial function $R_{n \ell}$ is again hydrogenic and the angular functions $Y_{\ell, j m}^i$ are vector spherical harmonics~\cite{Baryakhtar:2017ngi, Thorne:1980ru,Baumann:2019eav}. While the analogy between the hydrogen and gravitational atoms still holds, the vector cloud has more elaborate angular structure.
The eigenstates of massive vector fields are now labeled by four ``quantum numbers'' $\{ n, \ell, j, m \}$, where $n$ and $\ell$ have the same interpretation as in the scalar case, and $j = \ell\pm1$, $\ell$ is the total angular momentum with $m$ its azimuthal component. Since (\ref{eqn:FarLOAsol}) acquires a factor of $(-1)^{\ell+1}$ under a parity transformation, the $j = \ell \pm 1$ states are called electric-type modes, while the $j = \ell$ states are magnetic-type.  For notational simplicity, we will denote the normalized states (\ref{eqn:FarLOAsol}) by $|n \es \ell \es j \es m\rangle$.

\begin{figure}[t]
\centering
\makebox[\textwidth][c]{\hspace{0.7cm} \includegraphics[scale=0.99, trim = 0 0 0 0]{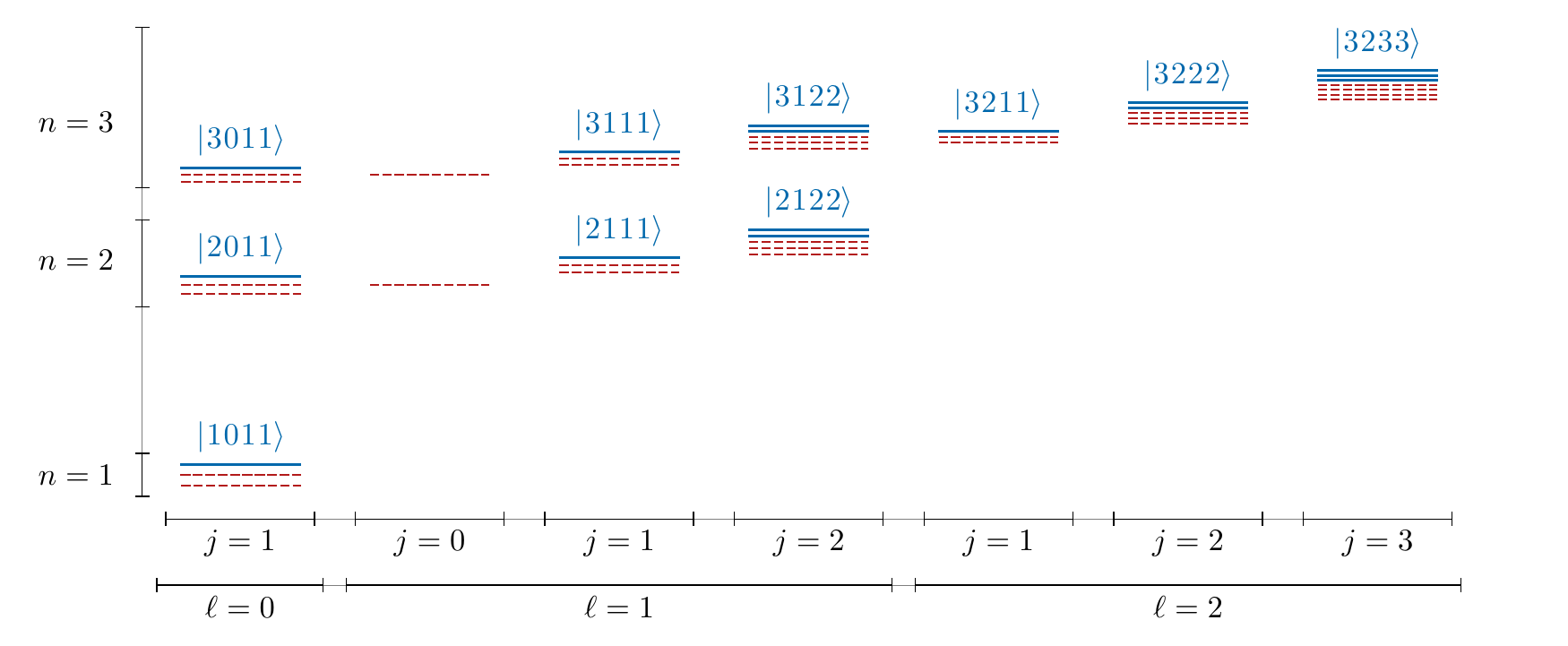}}
\caption{Schematic illustration of the energy spectrum of a massive vector field. Each state is denoted by 
 $|\es  n \es  \ell \es   j \es   m\rangle$. The dominant vector mode is $|1\es 0\es 1\es 1\rangle$. } 
\label{fig:VectorSpectra}
\end{figure}

\vskip 4pt
The energy eigenvalues and instability rates of the vector quasi-bound states are~\cite{Baumann:2019eav}
\begin{align}
E_{n \ell j m} &= \mu \left(  1 -\frac{\alpha^2}{2n^2} - \frac{\alpha^4}{8n^4} + \frac{f_{n \ell j}}{n^3} \hskip 2pt \alpha^4 + \frac{h_{n \ell j}}{n^3}\hskip 2pt \tilde a m\hskip 2pt \alpha^5 \right) , \label{eqn:vectorspectrum}\\
\Gamma_{n\ell jm} &= 2 \tilde{r}_+ C_{n \ell j } \, g_{j m}(\tilde{a}, \alpha, \omega)  \left( m \Omega_H - \omega_{n \ell j m} \right) \alpha^{2 \ell + 2 j + 5} \, ,\label{eqn:VectorRates}
\end{align}
where 
\begin{equation}
		f_{n \ell j} = \frac{2}{n} - \frac{4(2 + 3 \ell + 3 j(2 \ell +1))}{(j+\ell)(j + \ell +1 )(j + \ell + 2)}\, , \label{eq:fineStructure}
\end{equation} 
are the fine-structure splittings, which will be relevant later. The expressions for the coefficients $h_{n \ell j}$, $C_{n \ell j}$ and $g_{jm}$ can be found in~\cite{Baumann:2019eav}. Crucially, the scaling of (\ref{eqn:VectorRates}) with $\alpha$ depends on both $\ell$ and $j$. The dominant growing mode for the vector field is $|1 \es 0\es 1\es 1\rangle$, which has $\Gamma_{1011} \propto \mu \alpha^6$. This is enhanced, by two powers of $\alpha$, with respect to the dominant growing mode for the scalar field. The 
spectrum of the vector cloud is illustrated in Fig.~\ref{fig:VectorSpectra}.

\subsection{Gravitational Level Mixings}
 \label{sec:gravlevelmix}
 
Next, we describe how the presence of a binary companion modifies the evolution of the boson cloud. After a brief outline of the geometry of the problem, we describe how the gravitational perturbation enables the mixing between different states of the spectra. As we shall see, the associated selection rules are different for scalar and vector transitions.

 \begin{figure}[t]
\centering
\includegraphics[scale=1, trim = 0 10 0 0]{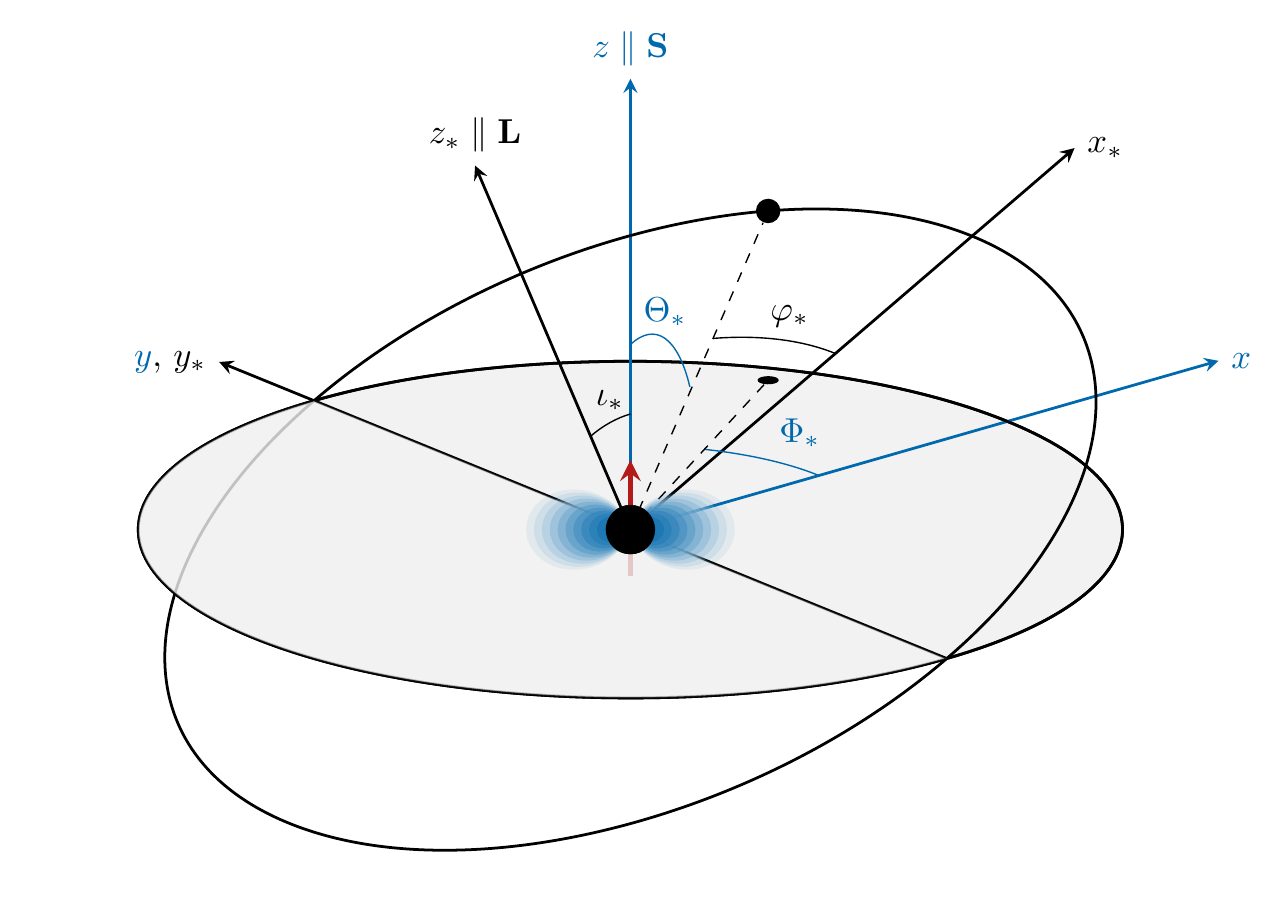}
\caption{Illustration of the coordinates used to describe the binary system. The Cartesian bases $\{x, y, z \}$ and $\{ x_*, y_*, z_* \}$ are adapted to the cloud's equatorial plane and the binary's orbital plane, respectively. The equatorial plane (gray) intersects the orbital plane at an inclination angle~$\iota_*$, and we have chosen
the pericenter to be located on the $x_*$-axis, such that $\varphi_*$ denotes the true anomaly of the binary. } 
\label{fig:BinaryPlane}
\end{figure}

\vskip 4pt
To investigate the dynamics of a boson cloud in a binary system, we must work in the cloud's Fermi co-moving frame \cite{Baumann:2018vus}. The relative motion of the companion is most conveniently described by the coordinates $\textbf{R}_* \equiv \{R_*, \iota_*, \varphi_*\}$, where $R_*$ is the 
separation between the members of the binary, $\iota_*$ is the \textit{inclination} angle between the orbital plane and the cloud's equatorial plane, and $\varphi_*$ is the \textit{true anomaly}, which represents the angle between the binary separation and the pericenter of the orbit (see  Fig.~\ref{fig:BinaryPlane}). 
 These angles are related to the usual angular spherical coordinates of the binary companion, $\{ \Theta_*, \Phi_*\}$, via 
\beq
 \begin{aligned}
 \cos \Theta_* & = \sin \iota_* \cos  \varphi_* \, , \\
  \tan \Phi_* & = \sec \iota_* \tan  \varphi_*  \, . \label{eqn:Angles}
 \end{aligned}
 \eeq
Since $0 \leq \Theta_* \leq \pi$, we can choose $-\pi/2 \leq \iota_* \leq \pi/2$, such that the spin of the cloud, $\textbf{S}_c $, projected on the $z_*$-axis of the orbital frame, satisfies $\textbf{z}_* \cdot \textbf{S}_c \geq 0$. While $\varphi_*$ increases in magnitude during the orbital evolution, the sign of $\varphi_*$ determines the orientation of the orbit. 
Orbits with $\varphi_* > 0$ are {\it co-rotating}, while those with $\varphi_* < 0$ are {\it counter-rotating}.

\subsubsection*{Metric perturbations}

The gravitational perturbation due to the binary companion can be encoded by additional potential terms in the Schr\"{o}dinger equations,~(\ref{eqn:NonRelScalar}) or~(\ref{eqn:NonRelVector}), for scalar and vector fields, respectively. At leading order, these terms are
\begin{align}
V_* &=  - \frac{1}{4} \mu \bar{h}^{00} \mathrlap{\qquad \quad (\text{scalar})\, ,} \label{eqn:ScalarPert} \\
V_*^{il}  &=  - \frac{1}{4} \mu   \bar{h}^{00} \delta^{il}   \mathrlap{\qquad (\text{vector})\, ,} \label{eqn:VectorPert} 
\end{align}
where $\bar{h}^{00}$ is the Newtonian trace-reversed metric perturbation. 
Denoting the spatial coordinates of the cloud in the Fermi frame by $\textbf{r} \equiv \{r, \theta, \phi\}$,\footnote{Since the center-of-mass of the cloud is the same as that of the isolated central black hole, the Fermi coordinates coincide with the Boyer-Lindquist coordinates $\{t, r, \theta, \phi \}$ at leading order in the post-Newtonian expansion.}  
we can write the metric perturbation as 
\beq 
\begin{aligned}
\bar{h}^{00} &=  4 M_* \sum_{\ell_*\neq 1} \sum_{|m_*| \leq \ell_*} \mathcal{E}_{\ell_* m_*} (\iota_*, \varphi_*)  Y_{\ell_* m_*}  (\theta, \phi) \Bigg( \frac{r^{\ell_*}}{R^{\ell_* + 1}_*} \Theta (R_* - r) + \frac{R_{*}^{\ell_*}}{r^{\ell_* + 1}_*} \Theta (r - R_*)  \Bigg)   \,, \label{eqn:h00Expansion}
\end{aligned}
\eeq
where $M_*$ is the mass of the binary companion, $\Theta$ is the Heaviside step-function and $\mathcal{E}_{\ell_* m_*}$ are the tidal moments, whose explicit forms depend on the geometry of the binary (see \S\ref{app:TidalMoments} for a detailed discussion). Crucially, the dipole moment $\ell_*=1$ is absent in the Fermi frame of the cloud \cite{Baumann:2018vus}. We will concentrate on the cloud's dynamics when the binary companion is located {\it outside} of the cloud ($R_* > r_c$), where the metric perturbation is dominated by the first term in \eqref{eqn:h00Expansion}. However, when the binary separation approaches the Bohr radius of the cloud, $R_* \sim r_c$, the second term in \eqref{eqn:h00Expansion} can provide non-negligible support to the perturbation.

\vskip 4pt
In this paper, we will focus only on the dynamics of the boson cloud perturbed by the leading-order potentials. In principle, we could also include couplings in (\ref{eqn:ScalarPert}) and (\ref{eqn:VectorPert}) that are higher order in~$\alpha$, such as those arising from interaction terms with spatial and temporal gradients acting on the non-relativistic fields, or from the post-Newtonian (PN) corrections to the metric perturbation.\footnote{Notice that gradients acting on the non-relativistic field are $\alpha$-suppressed, $\partial_i \psi \sim \mu \alpha \psi$ and $\partial_0 \psi  \sim \mu \alpha^2 \psi$. Similarly, the virial theorem demands that the companion's velocity is roughly $v \sim \sqrt{( M + M_*)/R_*}$, so that velocity-dependent PN corrections are also $\alpha$-suppressed whenever $R_* \gtrsim r_c$ \cite{Baumann:2018vus}.\label{footnote:alpha-scaling}}
These subleading potential terms are given explicitly in \S\ref{app:HigherOrder}, and we defer a detailed study of their impact on the evolution of the cloud to future work.

\subsubsection*{Selection rules}

The perturbation (\ref{eqn:h00Expansion}) induces level mixings between different states of the gravitational atom. However, not all levels couple to one another as certain {\it selection rules} must be obeyed. To deduce the allowed transitions for both scalar and vector clouds, we must compute the overlap \cite{Baumann:2018vus}
\beq
\langle a' | V_*(t) | a \rangle = - M_* \, \mu \sum_{\ell_* \neq 1} \sum_{m_* \leq |\ell_*|} \mathcal{E}_{\ell_* m_*} \!\times I_{r} \times I_{\Omega}  \, , \label{eqn:MatrixElement}
\eeq
where $V_*$ schematically represents both (\ref{eqn:ScalarPert}) and (\ref{eqn:VectorPert}), and the states $|a\rangle$ and $|a^\prime\es \rangle$ can  denote either the scalar $|n \es \ell \es m\rangle$ or vector $|n  \es \ell \es  j \es m\rangle$ eigenstates. The radial integral in both cases is
\begin{align}
I_{r} & \equiv \int_0^{R_*} \!\d r \, r^2 \left( \frac{r^{\ell_*}}{R_*^{\ell_* + 1}} \right) R_{n^\prime \ell^\prime } R_{n \ell }, + \int_{R_*}^{\infty} \!\d r \, r^2 \left( \frac{R_*^{\ell_*}}{r^{\ell_* + 1}} \right) R_{n^\prime \ell^\prime } R_{n \ell} \, , 
\label{eqn:I_r}
\end{align}
where $R_{n \ell}$ are the radial functions described in \S\ref{sec:ScalarVector}. However, the angular integrals $I_{\Omega}$ are different for the scalar and vector:
\begin{align}
I_{\Omega} &\equiv \int \!\d \Omega \,\, Y^*_{\ell^\prime m^\prime} (\theta, \phi)\, Y_{\ell_* m_*} (\theta, \phi) \, Y_{\ell m} (\theta, \phi) \mathrlap{\quad \quad \qquad \ \  (\text{scalar})\, ,} \label{eqn:I_Omega_scalar} \\
I_{\Omega} &\equiv \int \d \Omega \ Y_{\ell_* m_*} (\theta, \phi)\, \textbf{Y}^*_{\ell^\prime, j^\prime m^\prime }(\theta, \phi)  \cdot \textbf{Y}_{\ell, j m} (\theta, \phi)  \mathrlap{\qquad (\text{vector})\, .}  \label{eqn:I_OmegaVec}
\end{align}
The first integral (\ref{eqn:I_Omega_scalar}) is only non-vanishing when the following selection rules are satisfied~\cite{Baumann:2018vus} 
\beq
\begin{aligned}
\text{({\rm S1})} & \hskip 8pt -m^\prime + m_* + m = 0 \, , \\
\text{({\rm S2})} & \hskip 8pt  \ell + \ell_*  + \ell^\prime = 2p \, ,  \  \text{for} \ p\in \mathbb{Z} \, , \\
\text{({\rm S3})} & \hskip 8pt  |\ell - \ell^\prime | \le \ell_* \le \ell + \ell^\prime  \, .
\label{eqn:GeneralTransitionRulesScalar}
\end{aligned}
\eeq
This implies that the dominant scalar mode $|2 \es 1\es 1\rangle$ can only couple to states with $\ell^\prime =1$ and $\ell^\prime =3$, if we assume mixing through the $\ell_* = 2$ quadrupole perturbation. 
Similarly, the integral (\ref{eqn:I_OmegaVec}) implies the following selection rules for the vector cloud:
\beq
\begin{aligned}
\text{(V1)} & \hskip 8pt -m^\prime + m_{*} + m = 0 \, , \\
\text{(V2)} & \hskip 8pt  \ell + \ell_*  + \ell^\prime = 2p \, , \  \text{for} \ p\in \mathbb{Z} \, ,  \\
\text{(V3)} & \hskip 8pt  |\ell - \ell^\prime | \le \ell_* \le \ell + \ell^\prime  \, , \\
\text{(V4)} & \hskip 8pt  |j - j^\prime | \le \ell_* \le j + j^\prime  \, .
 \label{eqn:GeneralTransitionRulesVector2}
\end{aligned}
\eeq
\vskip 6pt
\noindent
While (V1) -- (V3) are similar to (S1) -- (S3), there is a new rule~(V4) for the vector, which  reduces to (S3) when $j = \ell$.\footnote{There are special cases where $I_\Omega=0$ even when the selection rules (V1) -- (V4) are naively satisfied. These occur when the inner product between the vector spherical harmonics in the integrand (\ref{eqn:I_OmegaVec}) vanishes, i.e. when $\ell = \ell^\prime, m = m^\prime= 0$, and the inner product is taken between an electric $j = \ell \pm 1 $ and a magnetic $j^\prime = \ell^\prime $ vector spherical harmonic.}  
The rule (V3) implies that the dominant state $|1 \es 0 \es 1 \es 1\rangle$ of the vector cloud  can only transition to  modes with $\ell^\prime \geq 2$. We describe the phenomenological consequences of the different selection rules for scalar and vector clouds in Section~\ref{sec:unravel}.

\subsection{Dynamical Perturbation} 
\label{sec:TimeDependentTidalMoments}

We now discuss how the binary's orbital motion forces 
the gravitational perturbation on the cloud to evolve in time, which leads to novel dynamical effects.

\subsubsection*{Shrinking orbits}

As the coordinates of the binary $\mb{R}_*(t) = \{R_*(t), \iota_*(t), \varphi_*(t)\}$ evolve, so does the metric perturbation (\ref{eqn:h00Expansion}).
For general orbit, the overlap in (\ref{eqn:MatrixElement}) is (see \S\ref{app:TidalMoments} for details) 
\beq
\bra{a} V_*(t) \ket{b} \equiv \sum_{ m_\varphi\in \mathbbm{Z}} \eta_{ab}^{(m_\varphi)} (R_*(t), \iota_*(t)) \, e^{-i m_\varphi \varphi_* (t)} \, , \label{eqn:etaDef}
\eeq
where $\eta_{ab}^{(m_\varphi)}$ characterizes the strength of the perturbation, and the oscillatory factors $e^{-i m_\varphi \varphi_* (t)}$ arise from the tidal moments $\mathcal{E}_{\ell_* m_*}$.\footnote{As we will illustrate in \S\ref{app:TidalMoments}, the sum over $m_\varphi$ can be understood as a sum over polarizations. Depending on the inclination $\iota_*$ of the orbit, different polarizations may contribute.} Crucially, the presence of these oscillatory terms means that the perturbation acts like a periodic driving force, which greatly enriches the dynamics of the boson cloud.  In principle, the couplings $\eta_{ab}^{(m_\varphi)}$ receive contributions from \emph{all} multipoles in the expansion (\ref{eqn:h00Expansion}). However, we will concentrate on the $\ell_* = 2$ quadrupole coupling, which dominates the perturbation.

\vskip 4pt 
It is therefore necessary to understand the behavior of $\varphi_*(t)$, including the effects induced by the shrinking of the orbit due to gravitational-wave emission.
 In terms of the instantaneous orbital frequency, $\Omega(t) > 0$, we have \cite{Baumann:2018vus}
\beq
\varphi_*(t) = \pm \int_{0}^t \d t^\prime \, \Omega(t^\prime) \, ,  \label{eqn:GeneralTrueAnomaly}
\eeq
where $t \equiv 0$ is an initial reference time and the upper (lower) sign denotes co-rotating (counter-rotating) orbits. In general, $\varphi_*(t)$ and $\Omega(t)$ evolve in a complicated manner. However, for quasi-circular equatorial orbits, the orbital frequency is determined by~\cite{PhysRev.131.435}
\beq
\frac{\d \Omega}{\d t} = \gamma \left(\frac{\Omega}{\Omega_0} \right)^{11/3} \, , \label{eqn:Omega-Quasi-Circular}
\eeq
where $\Omega_0$ is a reference orbital frequency and $\gamma$ is the rate of change due to the gravitational-wave emission,
\beq
\gamma \equiv \frac{96}{5} \frac{q}{(1+q)^{1/3}} (M \Omega_0)^{5/3} \Omega^2_0\, , \label{eqn:circleRate}
\eeq
with $q \equiv M_* / M$ the mass ratio of the black holes.

\vskip 4pt 
When the orbital frequency does not change appreciably near $\Omega_0$, the solution of (\ref{eqn:Omega-Quasi-Circular}) can be approximated by
\beq
\Omega(t) = \Omega_0 + \gamma t \, , \label{eqn:OmegaLinear}
\eeq
where we have dropped nonlinearities that become important on the timescale $t \sim \Omega_0/\gamma$.
The effects of a time-periodic perturbation on the gravitational atom, with fixed frequency $\Omega_0$, were studied in~\cite{Baumann:2018vus}, where \textit{Rabi oscillations} were found. The oscillations are enhanced when $\Omega_0$ matches the difference between two energy levels of the clouds, such that resonances are excited. 
However, as we discuss in~Section~\ref{sec:gcollider}, 
the presence of a new timescale associated to the shrinking of the orbit introduces additional coherent effects, that qualitatively change the behavior of the cloud during the resonant transition.

\subsubsection*{Resonances}

As we will elaborate in Section~\ref{sec:gcollider}, resonant transitions can be excited when the orbital frequency, or its overtones, matches the energy difference $\Delta E_{ab} \equiv E_a - E_b$ between two states $|a \rangle$ and $| b \rangle$ connected by the gravitational perturbation,
  \begin{equation}
    \dot{\varphi}_*(t) = \pm \Omega(t) = \frac{\Delta E_{ab}}{m_\varphi}\,. \label{eqn:resCondition}
  \end{equation}
  The weak gravitational perturbation is resonantly enhanced at these frequencies, and can force the cloud to evolve into an entirely different configuration. In most of this paper, we will focus on quasi-circular equatorial orbits for which $m_\varphi = \Delta m_{ab}$.\footnote{In \S\ref{app:TidalMoments}, we show that, for equatorial orbits, the summation over $m_\varphi$ in (\ref{eqn:etaDef}) reduces to a single term with $m_\varphi = m_*$. The selection rules (S1) and (V1) then imply that $m_\varphi = \Delta m_{ab} \equiv m_a - m_b$. \label{footnote:equatorial}} This restriction is convenient, because fewer transitions are allowed by the selection rules, and transitions between a pair of states may only occur at a single frequency. The resonance condition (\ref{eqn:resCondition}) then simply becomes $\dot{\varphi}_* = \Delta E_{ab} / \Delta m_{ab}$.
  
  \vskip 4pt  
  Since $\Delta E_{ab}/\Delta m_{ab}$ can be either positive or negative, the orientation of the orbit determines which resonances are excited. Specifically, co-rotating orbits can only excite resonances for which $\Delta E_{ab}$ and $\Delta m_{ab}$ have the same sign, while counter-rotating orbits require the signs to be opposite. Resonances do not occur if either $\Delta E_{ab} = 0$ or $\Delta m_{ab} = 0$, since energy and angular momentum must be transferred in the process. 
These constraints, in addition to the selection rules, dictate whether or not a transition is allowed. We will denote the resonance frequencies by
  \begin{equation}
    \res_{ab} = \left|\frac{\Delta E_{ab}}{\Delta m_{ab}}\right| , \label{eqn:resCondition2}
  \end{equation}
  though we stress that not all transitions between states are accessible during a particular inspiral.
  
  \vskip 4pt
In order to understand the dynamics of the cloud near the resonances, we will thus set our reference frequency to be $\Omega_0 \equiv \res_{ab}$, such that (\ref{eqn:OmegaLinear}) becomes 
  \begin{equation}
    \Omega(t) = \res_{ab} + \gamma_{ab} \hskip 1pt t\,, \label{eqn:ResonanceLinear}
  \end{equation}
where $\gamma_{ab}$ is (\ref{eqn:circleRate}) evaluated at $\Omega_0=\res_{ab}$. The linear approximation (\ref{eqn:ResonanceLinear}) is justified if the fractional change in the orbital frequency is small throughout the transition. We find that this is indeed the case for virtually all transitions, for both the scalar and vector clouds. When there is no risk of confusion, we will often label a general resonance frequency as $\Omega_r$.

\vskip 4pt
Before moving on, it will be useful to understand the order of magnitudes and scalings of the  relevant quantities involved in these transitions. For concreteness, let us consider \textit{Bohr transitions} between states with different principal quantum numbers $n_a$ and  $n_b$.  Compared to the resonance frequency~$\res_{ab}$, the typical values of $\eta_{ab}$ and $\gamma_{ab}$ are
  \begin{align}
  \frac{\gamma_{ab}}{\res_{ab}^2} & \simeq 3.2 \times 10^{-6} \, \frac{q}{(1+q)^{1/3}} \left( \frac{\alpha}{0.07}\right)^5 \left( \frac{2}{|\Delta m_{ab}|} \right)^{5/3} \left| \frac{1}{n^2_a} - \frac{1}{n^{ 2}_b } \right|^{5/3} \, , \label{eqn:gamma} \\[4pt]
  \frac{\eta_{ab}}{ \res_{ab}} & \simeq 0.3 \left( \frac{R_{ab}}{0.3} \right) \frac{q}{1+q} \, ,\label{eqn:Eta}
  \end{align}
where we have assumed that the transition is mediated by the quadrupole, $\ell_* = 2$, and introduced $R_{ab}$, a dimensionless coefficient that characterizes the overlap between different states in the spectrum. For the relevant states, we typically have $R_{ab} \lesssim 0.3$.\footnote{We assumed that the binary separation is much larger than the Bohr radius of the cloud, such that the first term in (\ref{eqn:I_r}) dominates the $q$-scaling. The general dependence on $q$ is actually more complicated, and is properly taken into account in the numeric results of Section~\ref{sec:unravel}.} 
\vskip 4pt
Because the transitions do not happen instantaneously, an initial state can, in principle, resonate with several states at the same time. This will occur if both the selection rules allow it and the couplings $\eta_{ab}$ are strong enough to excite states that are slightly off-resonance. In practice, the resonance between states $a$ and $b$ has a \emph{bandwidth} set by the coupling~$\eta_{ab}$.\footnote{This implies that the timescale of the transition is $\Delta t \sim \eta_{ab}/\gamma_{ab}$. 
The linear approximation (\ref{eqn:ResonanceLinear}) is therefore justified during the transition as long as $\eta_{ab}/\res_{ab} \ll 1$, which in light of (\ref{eqn:Eta}) is always the case.} Whenever the resonance frequency of a different state $c$ falls within this bandwidth, it may also participate in the transition. 

Remarkably, Bohr transitions for the scalar atom typically involve only two states, whereas boson clouds with vector fields can access multiple states. 
To illustrate this point, it is instructive to consider transitions mediated by the quadrupole 
in counter-rotating orbits. Since perturbations have $m_*=\pm 2$ (see \S\ref{app:TidalMoments}), we are thus restricted to transitions with $\Delta E_{ab} > 0$, which must 
satisfy $\Delta m_{ab} = -2$. Due to the selection rule  (\ref{eqn:GeneralTransitionRulesScalar}), the scalar $|2\es 1\es 1 \rangle$ state  
only couples to the $| 3\es 1\es \, \minus 1 \rangle$ state. 
In contrast, the vector $| 2\es 1\es 2 \es 2 \rangle$ state
 simultaneously resonates with all three $|3\es 1 \es j \es 0 \rangle$ modes with the same azimuthal quantum number, 
since their energy differences, of order $\mu \alpha^4$, are small compared to the size of the bandwidth $\sim \mu \alpha^2$. While we have illustrated this difference with a specific example, we find that virtually all relevant scalar and vector transitions follow this behavior.

\vskip 4pt
As we shall see momentarily, there are qualitative differences for transitions involving more than two states. We study next how these differences manifest themselves in the dynamics of the cloud during the resonance, and how we can encode it into an S-matrix formalism. We return to this point in Section \ref{sec:unravel}, where we discuss in more detail how the different evolution trees for scalar and vector clouds can help us measure the spin of the ultralight particles.

\vskip 4pt

\section{The Gravitational Collider} 
\label{sec:gcollider}

The goal of this section is to characterize the dynamics of the cloud under a perturbation whose frequency gradually increases with time, as in the binary system. We will find that a type of ``collision event'' (or resonance) occurs when the orbital frequency matches the energy difference between two or more states of the cloud.  As we will show, the dynamics of the cloud through this event can be captured by an S-matrix, which describes how a state defined far before the resonance evolves long after it has passed it. The evolution throughout the entire inspiral can then be described as a series of scattering events, with the S-matrix formulation providing a convenient and simple description of a generally complicated process.

  \vskip 4pt
 A general bound state can be written as $|\psi(t)\rangle = \sum_a c_a (t) | a \rangle$, where $a$ ranges over all states in the spectrum.  In this eigenstate basis, the Schr\"odinger equations (\ref{eqn:NonRelScalar}) and (\ref{eqn:NonRelVector}) reduce to
 \begin{equation}
      i \es \frac{\ud c_a}{\ud t} = \sum_{a} \mathcal{H}_{ab}(t) \, c_b\, ,\label{eqn:Schr}
    \end{equation}
 where the Hamiltonian $\mathcal{H}_{ab} = E_a \delta_{ab} + V_{ab}(t)$ splits into a constant, diagonal matrix of eigenstate energies, (\ref{eqn:scalarspectrum}) and (\ref{eqn:vectorspectrum}), and a time-varying off-diagonal piece encoding the gravitational mixings induced by the companion, (\ref{eqn:MatrixElement}). Conservation of the total occupation density demands that $\sum_{a} |c_a(t)|^2 = 1$. We describe the qualitative behavior of (\ref{eqn:Schr}) in this section.

 \vskip 4pt
We begin, in \S\ref{sec:twoState}, with the simple case of a two-state system. 
  We show that the dynamics is characterized by the Landau-Zener (LZ) transition, as it occurs in quantum mechanics.
 In~\S\ref{sec:MultiState}, we  generalize the formalism to a multi-state system.  
 We work out in detail the case of a three-state system, which captures all qualitative features of the generic case. Finally, in \S\ref{sec:Smatrix}, we discuss how the same evolution can also be described by an S-matrix, which provides a convenient way to describe a sequence of LZ transitions.

\subsection{Two-State Transitions} 
\label{sec:twoState}

To illustrate the main features of the cloud's dynamics, it is convenient to first study a simple two-level system. This is more than just a pedagogical device, since the transitions in  scalar atoms typically involve only two states (see \S\ref{sec:TimeDependentTidalMoments}).

\vskip 4pt
To avoid unnecessary complications, we will truncate the level mixing (\ref{eqn:etaDef}) to a single quadrupolar interaction term. This is equivalent to assuming that the companion travels on a large, equatorial, quasi-circular orbit, so that the dominant gravitational perturbation connecting the different energy eigenstates oscillates with a definite phase.  Specifically, if the states $|1\rangle$ and $|2\rangle$ have azimuthal angular momenta $m_1$ and $m_2$, with $\Delta m_{21} \equiv m_2 - m_1$, the gravitational mixing (\ref{eqn:etaDef}) takes the form 
    \begin{equation}
      V_{12}(t) \equiv \langle 1 | V_*(t) | 2 \rangle = \eta_{12}(t)\hskip 2pt e^{i \Delta m_{21} \hskip 1pt \varphi_*(t)}\, ,
    \end{equation}
    where the strength of the interaction $\eta_{12}(t)$ varies slowly in time. As discussed in \S\ref{sec:TimeDependentTidalMoments}, the true anomaly $\varphi_*(t)$ for quasi-circular orbits near the resonances can be approximated as\hskip 1pt\footnote{To avoid clutter, we have dropped the subscripts on $\res_{ab}$ and $\gamma_{ab}$.}
    \begin{equation}
      \varphi_*(t) = \pm \int_{0}^{t}\!\ud t'\, \Omega(t') = \pm \left(\res \hskip 1pt t + \frac{\gamma}{2} \hskip 1pt t^2 \right) , \label{eq:linearizedOmeg}
    \end{equation} 
    where the sign depends on the orientation of the orbit. 
 It will be instructive to first consider the simplest case, where (\ref{eqn:Schr}) describes two states separated by an energy gap $\Delta E \equiv E_2 - E_1 > 0$, with $\Delta m_{21} \equiv \Delta m$ and constant $ \eta_{12} \equiv \eta$. 
    The Hamiltonian in the ``Schr\"{o}dinger frame'' then reduces~to 
    \begin{equation}
      \mathcal{H} = \begin{pmatrix} -\Delta E/2 & \eta \hskip 1pt e^{i  \Delta m \es \varphi_*(t)} \\ \eta \hskip 1pt e^{-i \Delta m \es \varphi_*(t)} & \Delta E/2 \end{pmatrix} . \label{eqn:Two-State-Schr-H}
    \end{equation}
    While the off-diagonal terms oscillate rapidly, they do so at a slowly increasing frequency, and it will be useful to work in a \emph{dressed frame} that isolates this slow behavior.  Said differently, while the companion's motion in a fixed frame is complicated, 
     its motion simplifies drastically if we rotate along with it. We thus define a time-dependent unitary transformation
    \begin{equation}
      \mathcal{U}(t) = \begin{pmatrix} e^{i \Delta m \es \varphi_*/2} & 0 \\ 0 & e^{-i \Delta m \es \varphi_*/2} \end{pmatrix} , \label{eq:timeDependentUnitary}
    \end{equation}
    so that the Schr\"{o}dinger frame coefficients can be written as $c_{a}(t) = \mathcal{U}_{ab}(t) d_{b}(t)$. 
Assuming a co-rotating orbit, such that $\dot{\varphi}_*(t) = \Omega(t)$, the dressed frame coefficients $d_a(t)$ then evolve according to the following Hamiltonian,
    \begin{equation}
      \mathcal{H}_\lab{D}(t) = \mathcal{U}^\dagger \mathcal{H} \hskip 1pt \mathcal{U} - i\hskip 1pt  \mathcal{U}^\dagger \frac{\ud \mathcal{U}}{\ud t} = \begin{pmatrix} 
          (\Delta m \es \Omega(t)-\Delta E)/2 & \eta \\
          \eta & -(\Delta m \es \Omega(t) - \Delta E)/2
      \end{pmatrix} . \label{eq:dressedFrameHam}
    \end{equation}
Note the useful fact that the structure of the transformation (\ref{eq:timeDependentUnitary}) implies that the magnitudes of the coefficients  in the Schr\"{o}dinger and dressed frames are equal, $|c_a(t)|^2 = |d_a(t)|^2$. For this reason, we will abuse notation and use $|a \rangle$ to denote the energy eigenstates in  both the Schr\"{o}dinger and dressed frames. Since our main interest will be in the overall population of the states $|a\rangle$, we only need to work with dressed frame quantities.

\begin{figure}
      	\centering
        \includegraphics{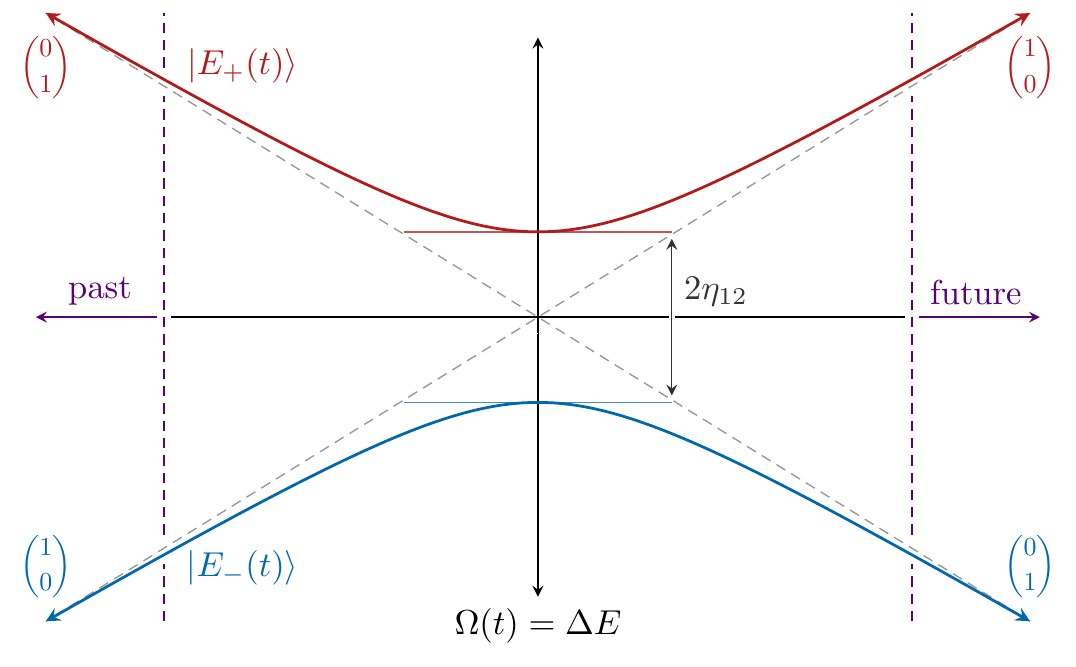}
        \caption{Instantaneous energy eigenvalues of the dressed frame Hamiltonian (\ref{eq:dressedFrameHam}) as a function of time. If $\eta = 0$, these energies cross when the frequency of the perturbation matches the energy difference between the two states, $\Omega(t) = \Delta E$. For non-vanishing $\eta$, these levels avoid each other. \label{fig:2by2eigen}}
   \end{figure}

  \vskip 4pt
  It is clear from (\ref{eq:dressedFrameHam}) that there is a ``collision'' or ``resonance'' time  $t_r \equiv 0$ when the frequency of the companion matches the energy difference between these two states,
  \begin{equation}
    \Omega(t_r) = \frac{\Delta E}{\Delta m} \,,
  \end{equation}
  and the diagonal entries of the dressed frame Hamiltonian vanish. Note that, since $\Omega(t)$ is always positive, a resonance in a co-rotating orbit can only happen if $\Delta E$ has the same sign as $\Delta m$ (see \S\ref{sec:TimeDependentTidalMoments} for more discussion). 
 To reduce clutter, we will set $\Delta m = 1$ in what follows.\footnote{Restoring factors of $\Delta m$ will simply mean replacing $\gamma \to \left| \Delta m \right| \gamma$.} With these simplifications, the dressed frame Hamiltonian is
  \begin{equation}
    \mathcal{H}_\lab{D}(t) =  \frac{\gamma t}{2} \begin{pmatrix} 
     1 & 0 \\
      0 & -1
      \end{pmatrix}  + \eta \begin{pmatrix} 
      0 & 1 \\
      1& 0
      \end{pmatrix} , \label{eq:dressedFrameHamSimp}
  \end{equation}
  whose instantaneous energy eigenvalues $E_{\pm}(t) = \pm \sqrt{(\gamma t/2)^2 + \eta^2}$ we depict in Fig.~\ref{fig:2by2eigen}. The asymptotic states shown in the figure are $\ket{1} \equiv (1, 0)$ and $\ket{2} \equiv (0, 1)$.

  \vskip 4pt

  We can use this figure to understand qualitatively how the instantaneous energy eigenstates  evolve in time. In the far past, $t \to -\infty$, the first term in (\ref{eq:dressedFrameHamSimp}) dominates and the instantaneous eigenstates are simply 
  \begin{equation}
    |E_{+}(-\infty) \rangle = \binom{0}{1}   \, , \qquad |E_-(-\infty)\rangle = -\binom{1}{0} \,.
    \label{equ:before}
  \end{equation}
  As we then move toward the collision, $t \to 0$, these states become nearly degenerate, the second term dominates, and the eigenvalues $E_\pm(t)$ are forced to repel. In the far future, $t \to + \infty$, the first term in (\ref{eq:dressedFrameHamSimp}) again dominates, and so the eigenstates take the same form. However, this repulsion event forces the states to have permuted their identities, so that in the far future
  \begin{equation}
    |E_{+}(+\infty) \rangle = -|E_{-}(-\infty)\rangle  \, ,\qquad  |E_-(+\infty) \rangle = |E_{+}(-\infty)\rangle \,.
  \label{equ:after}
  \end{equation}
  This can also be seen explicitly from the exact form of the time-dependent eigenstates,
  \begin{equation}
    \begin{aligned}
      |E_{\pm}(t) \rangle &= \mathcal{N}^{-1}_\pm \left(\gamma t/2 \pm \sqrt{(\gamma t/2)^2 + \eta^2}\,, \eta \right) ,
       \label{eq:2by2eigensystem}
        \end{aligned}
    \end{equation}
 where $ \mathcal{N}_\pm(t)$ is the appropriate normalization. 
 
  \vskip 4pt
  
As long as the evolution is adiabatic---meaning that the dressed frame Hamiltonian $\mathcal{H}_\lab{D}(t)$ evolves slow enough---the system tracks its instantaneous eigenstates.  If the system begins its life in the dressed frame's $\ket{1}$
 state, this implies that there is a \emph{complete} transfer of population into the $\ket{2}$ 
state after the resonance, cf.~Fig.~\ref{fig:2by2eigen}. This is a key characteristic of the LZ transition for a two-level system.  
  
  \begin{figure}
    \centering
      \includegraphics{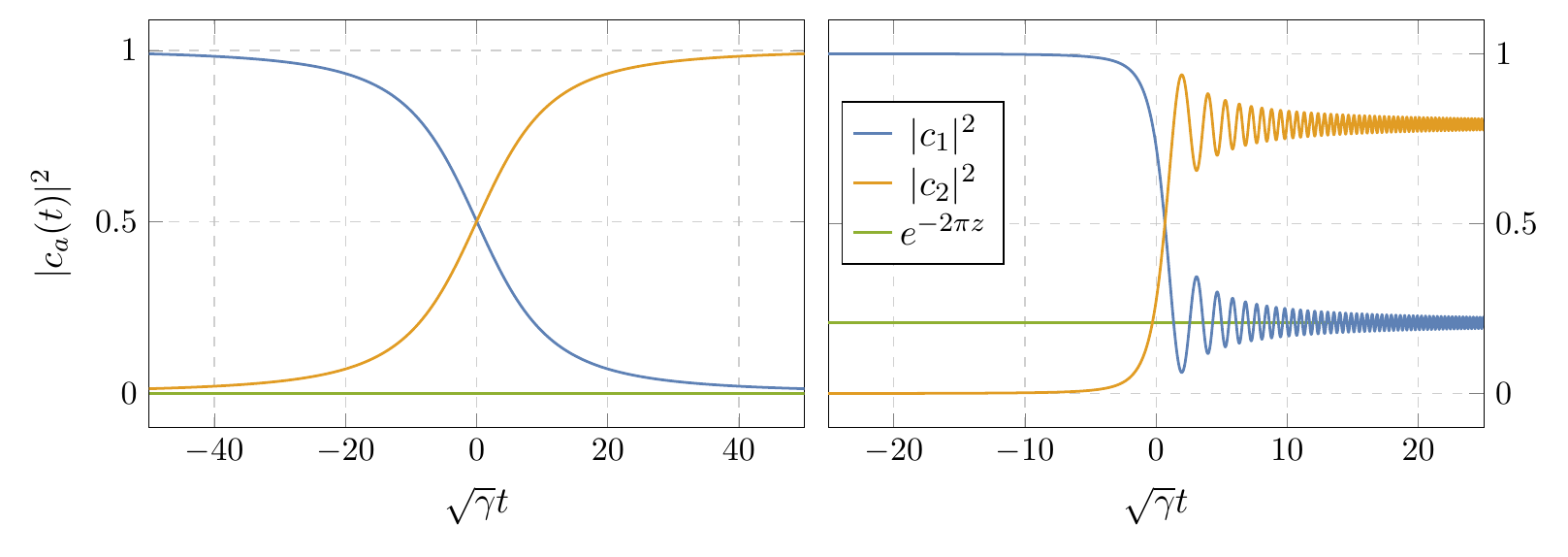}
      \caption{Adiabatic (\emph{left}) and non-adiabatic (\emph{right}) behavior of the Schr\"{o}dinger frame coefficients for transitions with Landau-Zener parameters $z = 25$ and $z = 1/2$, respectively. \label{fig:ad2by2}}
  \end{figure}

\vskip 4pt
The model described by (\ref{eq:dressedFrameHamSimp}) has the advantage that it can be solved exactly, even away from the adiabatic regime.  As detailed in \S\ref{app:twoState}, assuming that the state initially fully occupies the lowest energy eigenstate $|\psi(-\infty)\rangle \propto |E_{-}(-\infty)\rangle$, the population contained in the other eigenstate long after the transition is  
\begin{equation}
    |\langle E_+(\infty) | \psi(\infty)\rangle|^2  = \exp\left(-2 \pi z\right) , \label{eq:lzProb}
\end{equation}
where we have defined the (dimensionless) Landau-Zener parameter \cite{Landau,Zener,landau2013quantum} 
\begin{equation}
  z \equiv \frac{\eta^2}{\gamma}\, , \label{eqn:LZParam}
\end{equation}
which measures how (non-)adiabatic a given transition is.  

\vskip 4pt
In an adiabatic transition, with $z \gg 1$, the dressed frame Hamiltonian evolves slowly enough that we can ignore the other instantaneous eigenstate entirely. Intuitively, the system has enough time (as measured by $\eta$) to respond to a change in the dressed frame Hamiltonian, so that any fluctuations in the state's energy can decay.
As shown in the left panel of Fig.~\ref{fig:ad2by2}, an adiabatic transition causes the system to completely transfer its population from one state to another, a process known as \emph{adiabatic following}. Since the system tracks an instantaneous eigenstate, the magnitudes of the Schr\"{o}dinger frame coefficients change smoothly, and from the explicit expression (\ref{eq:2by2eigensystem}) we see that this transition happens on a timescale set by 
  \begin{equation}
    \Delta t \sim \frac{2 \eta}{\gamma}\,. \label{eq:lzTimeScale}
  \end{equation}
Intuitively, there is a finite ``resonance band'' of width $\Delta \Omega \sim 2 \eta$ in which this transition is active, and (\ref{eq:lzTimeScale}) is the time spent moving through it.

\vskip 4pt
A non-adiabatic transition, with $z \lesssim 1$, is qualitatively different (see the right panel of Fig.~\ref{fig:ad2by2}). Before passing through the resonance (i.e.~before the ``scattering event''), the system remains in an instantaneous eigenstate and the Schr\"{o}dinger frame coefficients change smoothly. During the transition, however, the system will partly evolve into the other instantaneous eigenstate, not dissimilar from particle production via high-energy scattering. After this event, the system exists in a linear combination of eigenstates and they oscillate among one another at a frequency set by the energy difference $E_{+} - E_{-} \sim \gamma t$.
These oscillations eventually decay away on a timescale again set by (\ref{eq:lzTimeScale}), so that the Schr\"{o}dinger (dressed) frame coefficients have a well-defined limit as $t\to \infty$. Finally, if the transition is extremely non-adiabatic, $z \ll 1$, the system has no time to respond to the changing Hamiltonian, and so its state is unaffected by the~transition.

  \subsection{Multi-State Transitions}
  \label{sec:MultiState}
  
  A  qualitatively new feature appears when multiple, nearly degenerate states are involved in a transition. To illustrate this, we study a three-state extension of (\ref{eq:dressedFrameHamSimp}), described by the following dressed frame Hamiltonian
  \begin{equation}
    \mathcal{H}_\lab{D}(t) = \frac{\gamma t}{2} \begin{pmatrix} 
    1 & 0 & 0 \\
    0 & \minus 1 & 0 \\
    0 & 0 & \minus 1 \end{pmatrix} + 
    \begin{pmatrix}
      0 & \eta_{12} & \eta_{13} \\
      \eta_{12} & 0 & \eta_{23} \\
      \eta_{13} & \eta_{23} & 0 
    \end{pmatrix}. \label{eq:3state}
  \end{equation}
As we described in \S\ref{sec:TimeDependentTidalMoments}, this type of multi-state transition will be relevant for vector clouds, as they tend to have multiple states that are nearly degenerate in energy and connected via the gravitational perturbation. Indeed, (\ref{eq:3state}) describes any transition which involves a single state interacting with two other states that have the same azimuthal angular momentum and unperturbed energy. However, our focus on the three-state system is primarily pedagogical, as vector cloud transitions tend to involve four or more states. Fortunately, we will find that the main features of these more complicated multi-state transitions can be explained using the simpler three-state model.

  \begin{figure}[t]
          \centering
            \includegraphics[width=0.7\textwidth]{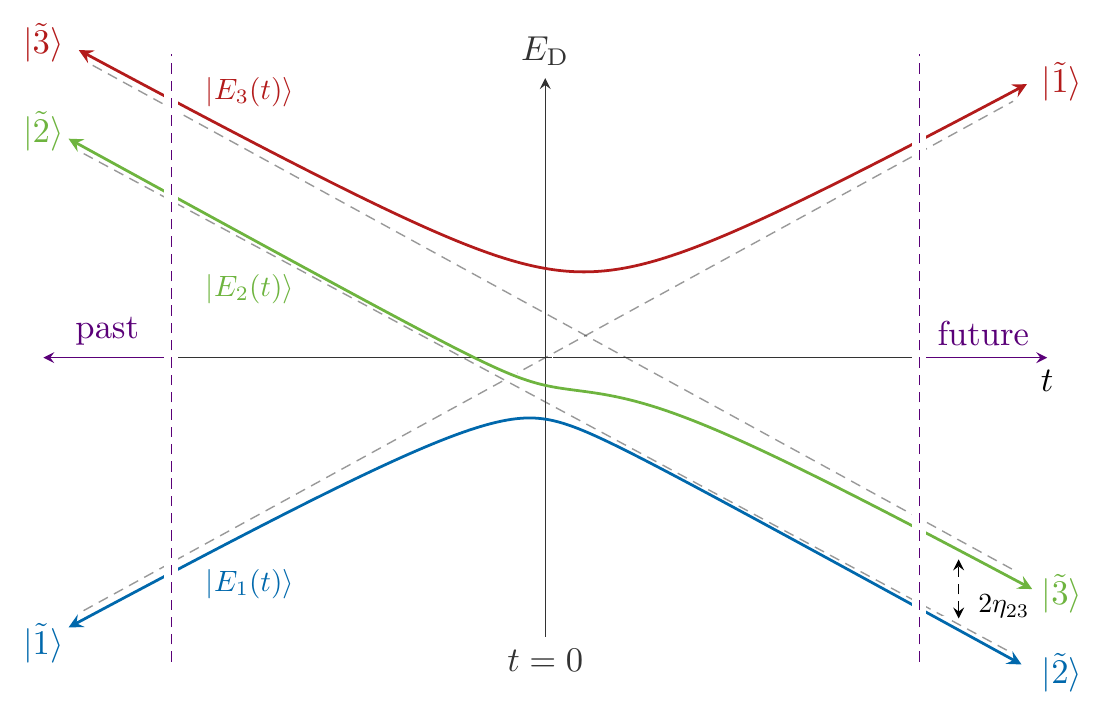}
            \caption{ Evolution of the instantaneous eigenstates of a coupled three-state system. \label{fig:threeStateEigVals}}
  \end{figure}

\begin{figure}
      \includegraphics{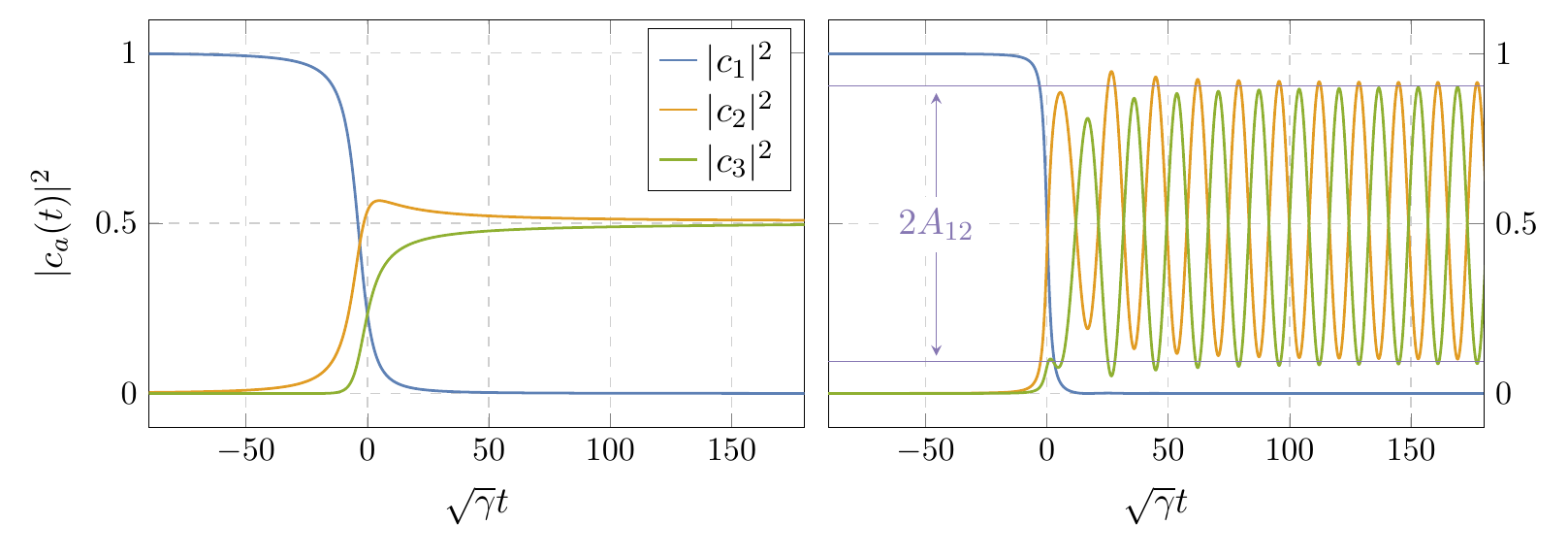}
      \caption{
      Adiabatic (\emph{left})
       and non-adiabatic (\emph{right}) evolution of the three-state system. Specifically, the adiabatic and non-adiabatic evolution are simulated with $(\eta_{12}, \eta_{13}, \eta_{23}) = (2.5, 1.5, 0.25)$ and $(\eta_{12}, \eta_{13}, \eta_{23}) = (4.75, 0.75, 3.5)$, respectively, in units of $\sqrt{\gamma}$.  The final state in the right panel oscillates with a frequency of about $2 \eta_{23}$.  } \label{fig:3state}
  \end{figure}

  \vskip 4pt
  It is again helpful to plot the instantaneous energy eigenvalues as a function of time, shown in~Fig.~\ref{fig:threeStateEigVals}. At late times, we see that there are two instantaneous energy eigenstates, $|E_1(t) \rangle$ and $|E_2(t)\rangle$, whose energy difference approaches a \emph{constant} as $t \to +\infty$. This is significant because, if the motion is non-adiabatic enough to excite the system into a combination of these two instantaneous eigenstates, there can be large, coherent oscillations with a \emph{fixed} frequency set by the interaction strength~$\eta_{23}$. Such oscillations are indeed seen in the right panel of Fig.~\ref{fig:3state}. This is impossible in a two-state transition (\ref{eq:dressedFrameHamSimp}), as the difference in energies necessarily diverges as $t \to +\infty$, and any oscillatory behavior induced by non-adiabaticity rapidly increases in frequency and decays away, as seen in the right panel of  Fig.~\ref{fig:ad2by2}. 

  \vskip 4pt
  Let us understand this transition more quantitatively. At very early times, the first term in (\ref{eq:3state}) again dominates. However, there is now a degenerate subspace that is only lifted by $\eta_{23}$ and we cannot ignore it. It will thus be convenient to explicitly diagonalize this subspace.  We therefore introduce a new basis of states $|\tilde{a}\rangle$, which is related to the dressed frame basis $|a\rangle$ by  
  \begin{equation}
    |\tilde{1} \rangle = |1\rangle \,, \quad |\tilde{2}\rangle = \frac{1}{\sqrt{2}} \left(- |2 \rangle + |3 \rangle \right) , \quad\text{and}\quad |\tilde{3} \rangle = \frac{1}{\sqrt{2}} \left(|2 \rangle + |3 \rangle \right) .
  \end{equation}
  In this basis, the dressed frame Hamiltonian becomes 
    \begin{equation}
    \tilde{\mathcal{H}}_\lab{D} = \begin{pmatrix} 
      \gamma t & \tilde{\eta}_{12} & \tilde{\eta}_{13} \\
      \tilde{\eta}_{12} & - \gamma t - \eta_{23} & 0 \\
      \tilde{\eta}_{13} & 0 & - \gamma t + \eta_{23} 
    \end{pmatrix} , \label{eq:diagDressFrameHam}
  \end{equation}
  where we have defined the effective couplings
  \begin{equation}
    \tilde{\eta}_{12} = \frac{1}{\sqrt{2}} (\eta_{12} - \eta_{13}) \qquad \text{and} \qquad \tilde{\eta}_{13} = \frac{1}{\sqrt{2}} (\eta_{12} + \eta_{13})\,. \label{eqn:EffectiveEta}
  \end{equation}
  We see that the role of the coupling $\eta_{23}$ is to break the degeneracy between $|\tilde{2}\rangle$ and $|\tilde{3}\rangle$, and force $|\tilde{1}\rangle$ to become nearly degenerate with $|\tilde{2}\rangle$ at a different time than it becomes degenerate with~$|\tilde{3}\rangle$. As we will argue in~\S\ref{sec:Smatrix}, we can thus treat this three-state transition as a combination of two-state transitions.

  \vskip 4pt
  In the asymptotic past, the instantaneous energy eigenstates are then
  \begin{equation}
    |E_{1}(\minus \infty) \rangle =  \left(\begin{array}{@{\mkern3mu} r @{\mkern3mu}} 1 \\ 0 \\ 0 \end{array}\right)\,, \,\,\quad |E_{2}(\minus \infty) \rangle = \frac{1}{\sqrt{2}}\left(\begin{array}{@{\mkern-2mu} r @{\mkern3mu}} 0 \\ -1 \\ 1 \end{array}\right)\,, \quad \text{and} \quad |E_3(\minus \infty) \rangle = \frac{1}{\sqrt{2}}\left(\begin{array}{@{\mkern3mu} r @{\mkern3mu}} 0 \\ 1 \\ 1 \end{array}\right),
  \end{equation}
  and from Fig.~\ref{fig:threeStateEigVals} we see that they permute in the asymptotic future,
  \begin{equation}
    |E_{1}(+\infty) \rangle \propto |E_{2}(\minus \infty)\rangle\,, \,\,\quad |E_{2}(+\infty) \rangle \propto |E_{3}(-\infty)\rangle\,, \quad \text{and} \quad  |E_{3}(+\infty)\rangle \propto |E_{1}(\minus \infty)\rangle\,.
  \end{equation}
  Such a cyclic permutation is present in any transition which involves one state going into many degenerate states, as is visually apparent from Fig.~\ref{fig:threeStateEigVals}.

  \vskip 4pt
Where do the large, coherent oscillations in Fig.~\ref{fig:3state} come from?
Let us focus on the transition mediated by $\tilde{\eta}_{12}$, with associated LZ parameter $\tilde{z}_{12} \equiv \tilde{\eta}_{12}^2/\gamma$. Note that, even if all of the original LZ parameters are very large, $z_{ab}  = \eta_{ab}^2/\gamma \gg 1$, the transition may still be non-adiabatic if there is a cancellation between $\eta_{12}$ and $\eta_{13}$ in (\ref{eqn:EffectiveEta}), such that $\tilde{z}_{12} \lesssim 1$. We assume that the transition is fast enough ($\tilde{z}_{12} \sim 1$) to excite $|E_{2}(t)\rangle$, but slow enough ($\tilde{z}_{13} \gg 1$) that we can ignore  $|E_3(t)\rangle$. If the initial state is $|\psi(\minus \infty)\rangle \propto |E_{1}(\minus \infty)\rangle$, then at late times it becomes 
  \begin{equation}
    |\psi(t)\rangle \xrightarrow{\ t\to \infty \ }   e^{-i \gamma t^2/2 - i \eta_{23} t}\sqrt{1 - e^{-\pi \tilde{z}_{12}}}\, |E_1(\infty)\rangle + e^{-i \gamma t^2/2 + i \eta_{23} t + i \delta} e^{-\pi \tilde{z}_{12}/2}\, |E_{2}(\infty)\rangle \,,
  \end{equation}
  where the phase difference $\delta$ depends on $\gamma$ and $\eta_{ab}$. We thus see that, in the asymptotic future, there are oscillations in the Schr\"{o}dinger frame populations with frequency $2 \eta_{23}$.
  For example, we have 
  \begin{equation}
    |c_{2}(t)|^2 = \frac{1}{2} -  A_{12} \cos(2 \eta_{23} t + \delta)\, , \label{eqn:NeutrinoOsc}
  \end{equation}
  where $A_{12} = e^{-\pi \tilde{z}_{12}/2} \sqrt{1 - e^{-\pi \tilde{z}_{12}}}$. This amplitude achieves its maximum $A_{12} = \tfrac{1}{2}$ when $\tilde{z}_{12} = \tfrac{1}{\pi} \log 2$, and we see that these oscillations disappear both when the transition is either very adiabatic, so that the  state $|E_2(t)\rangle$ is never excited, or very non-adiabatic, so that the state $|E_1(t)\rangle$ does not survive.
  
  \vskip 4pt
  These slow oscillations will persist even if the state $|E_{3}(t)\rangle$ is excited by the transition, i.e. when $\tilde{z}_{12} \sim \tilde{z}_{13} \sim 1$. In this case, there will then be transient oscillations, like those seen in the two-state system and in the right panel of Fig.~\ref{fig:ad2by2}, which will eventually decay and leave only these fixed frequency oscillations. 
  However, if the transition is extremely non-adiabatic, such that both $\tilde{z}_{12}$ and $\tilde{z}_{13} \ll 1$, the system will always almost entirely occupy the state $|E_{3}(+\infty)\rangle = |E_{1}(\minus \infty)\rangle$ and the system will evolve very little in time. 

  \vskip 4pt
  The adiabatic and (mildly) non-adiabatic evolution of this three-state system are shown in the left and right panels of Fig.~\ref{fig:3state}, respectively. The slow  oscillations created by a somewhat non-adiabatic transition are seen on the right. On the left, the completely adiabatic transition creates a final state that is a linear superposition of Schr\"{o}dinger frame states, i.e.~energy eigenstates. It is a special feature of this three-state system that the relative final state populations are independent of the coupling $\eta_{23}$. Indeed, if four or more states are involved in the transition, the final state populations do depend on the couplings $\eta_{ab}$, since the asymptotic energy eigenstates clearly depend on how the dressed frame Hamiltonian's degenerate subspace is lifted. 
 
 \vskip 4pt
 We find that multi-state transitions generically yield final states that are superpositions of the dressed basis states (and thus the energy eigenstates). Almost all of the transitions that we will analyze in Section~\ref{sec:unravel} will be adiabatic, so this superposition can be easily approximated by an eigenvector of the coupling matrix $\eta_{ab}$, restricted to the degenerate subspace. As we mentioned at the outset of this section, these multi-state transitions qualitatively distinguish scalar and vector clouds, and we will exploit this to infer the spin of the boson from gravitational-wave observations in \S\ref{sec:modulatingfinite} and \S\ref{sec:final}.

\subsection{An S-Matrix Approach} 
\label{sec:Smatrix}

   In \S\ref{sec:twoState}, we found that a two-state system initially occupying the lowest instantaneous energy eigenstate $|\psi(\minus \infty)\rangle \propto |E_{-}(\minus \infty)\rangle$ could transition into the other instantaneous eigenstate with an occupation ``probability''~(\ref{eq:lzProb}) with a well-defined asymptotic limit. Given the system's causal structure---the nature of the transition provides a well-defined notion of both asymptotic past and asymptotic future---it is tempting to interpret (\ref{eq:lzProb}) as defining the element of a scattering matrix, 
   \begin{equation}
       |\langle E_+(+\infty)| E_-(\minus \infty)\rangle|^2 = |S_{+-}|^2 = e^{-2\pi z}\,.
   \end{equation}
   However, a rigorous definition of the scattering matrix must be more subtle, as the three-state system does not necessarily have a well-defined, static final state. For instance, what is the scattering matrix describing the coherent oscillations in the right panel of Fig.~\ref{fig:3state}?

   \vskip 4pt
   The goal of this section is three-fold. We first generalize our discussion of the LZ transition beyond the simplified toy models like (\ref{eq:dressedFrameHamSimp}) and toward a description of the boson cloud throughout the entire binary inspiral. We will then introduce an ``interaction picture,'' leading to an interaction Hamiltonian that is well-localized in time. This is crucial for constructing a well-defined S-matrix.  We will leverage this localization to argue that different resonances during the inspiral decouple from one another, allowing us to ignore all but the finite number of states involved in a particular resonance. The behavior of the boson cloud during the full inspiral can thus be described as a series of localized ``scattering events,'' each of which can be analyzed individually. 
   
    \subsubsection*{Adiabatic decoupling}

      Let us first consider the simplest inspiral configurations: large, equatorial, quasi-circular orbits for which the Hamiltonian (\ref{eqn:Schr}) takes the form
      \begin{equation}
        \mathcal{H}_{ab} = E_{a} \delta_{ab} + \eta_{ab}(t) e^{-i \Delta m_{ab} \varphi_*(t)}\, ,\label{eq:multiStateDefPhase}
      \end{equation}
where $\Delta m_{ab} = m_a - m_b$ is the difference in azimuthal angular momentum between the states $|a \rangle$ and $|b\rangle$ and $\eta_{ab}(t)$ is the matrix of cross-couplings. 
The latter evolves slowly in time. Unlike in the previous sections, we will not truncate (\ref{eq:multiStateDefPhase}) to a finite set of states $\{|a\rangle\}$, but instead argue that such a truncation is possible. The indices in this equation should thus be understood to run over all states in the atomic spectrum. 

      \vskip 4pt
      As before, we may define a time-dependent unitary transformation 
      \begin{equation}
        \mathcal{U}_{ab}(t) = e^{-i m_a \varphi_*(t)} \delta_{ab} \, ,\label{eq:multiDressedUnitary}
      \end{equation}
      that moves our system into a frame that rotates along with the binary companion. This isolates the ``slow'' motion responsible for LZ transitions and discards the perturbation's distracting fast motion. In this dressed frame, the Hamiltonian is given by
      \begin{equation}
        \left(\mathcal{H}_\lab{D}\right)_{ab} = \delta_{ab}\left(E_a  - m_a \dot{\varphi}_*\right)+ \eta_{ab}\,. \label{eq:multiStateDressedHam}
      \end{equation}
      This is analogous to (\ref{eq:dressedFrameHam}), though here we have not shifted by an overall reference energy.
      \vskip 4pt

      Because the gravitational perturbations $\eta_{ab}$ are always small compared to the energy eigenvalues~$E_a$, the instantaneous eigenstates $|E_{i}(t) \rangle$ are almost always\footnote{This argument fails for states that form a nearly-degenerate subspace that is predominantly lifted by the gravitational perturbation $\eta_{ab}$, as in the three-state toy model in \S\ref{sec:MultiState}. As we saw there, the perturbation cannot be ignored and the instantaneous eigenstates far away from the transition do not asymptote to a single dressed basis state. However, the following discussion also applies to these degenerate subspaces---as long as the system is not excited into a subspace, it decouples from the dynamics.} well-approximated by the gravitational atom's energy eigenstates---that is, $|E_{i}(t) \rangle \approx |a\rangle$ for some dressed state $a$.  We will 
      denote the state of the cloud by $|\psi (t)\rangle$. If the system evolves adiabatically, its population in each instantaneous eigenstate, $|\langle E_i(t) | \psi(t)\rangle|^2$, remains unchanged. Since these instantaneous eigenstates are well-approximated by the dressed states $|a \rangle$, the cloud's population in each dressed state, $|\langle a | \psi(t)\rangle|^2$, similarly remains unchanged. This means that, unless the system already has an appreciable population in a particular dressed state, we can simply ignore it.\footnote{Strictly speaking, this is only true at leading order, and this state does have a small effect on the states we do not ignore. However, this small effect can be incorporated using standard perturbative techniques.} 

    \vskip 4pt
    This will always be the case unless there is a point in time when a subspace becomes nearly degenerate with the subspace the system occupies, in which case the approximation $|E_{i}(t)\rangle \approx |a \rangle$ fails. Because of the assumed hierarchy between the energies $E_a$ and the couplings $\eta_{ab}$, this can only happen when the instantaneous frequency $\Omega(t)$ is such that the diagonal elements for two or more states are approximately equal, 
    \begin{equation}
      \dot{\varphi}_* \equiv \pm \Omega(t) \approx \frac{E_a - E_b}{m_a - m_b}\,. \label{eq:resonanceCondition}
    \end{equation}
    To determine which states we need to keep track of in a given frequency range, we therefore simply need to find for which states this \emph{resonance condition} is met and include them with the states that the system already occupies.

\vskip 4pt
    Critically, the width of the resonance defined by (\ref{eq:resonanceCondition}) is roughly set by $|\Delta \Omega| \sim \eta_{ab}$. Once
     these states go through a resonance, they very quickly decouple and the magnitudes of their coefficients become non-dynamical. Hence, if the resonances are widely separated, we may consider them as a sequence of events and analyze them individually. Associated to each event is thus an S-matrix, which describes how the system evolves through the resonance.

  \vskip 4pt
    This logic can be extended to more general orbital configurations. 
    As we discuss in \S\ref{app:adFloTheo}, each pair of states is generically connected by a perturbation that oscillates at multiple frequencies.  The dressed frame transformation (\ref{eq:multiDressedUnitary}) must then be generalized beyond a simple unitary transformation in order to find a slowly varying dressed frame Hamiltonian like (\ref{eq:multiStateDressedHam}). This can be done, but since it requires a technology that would distract from our main story, we do not describe it here. Instead, we relegate this generalization to \S\ref{app:adFloTheo}. Fortunately, the above discussion also applies there,
     \emph{mutatis mutandis}. To avoid getting bogged down by unnecessary technical details, we thus continue with the simplified setup described by~(\ref{eq:multiStateDefPhase}).

  \subsubsection*{Interaction picture} 

    We must define a scattering matrix with respect to a set of states which do not evolve in the asymptotic past or future. Since the dressed frame's instantaneous eigenstates are stationary (up to a phase) under adiabatic motion, it is natural to define an \emph{interaction frame} by expanding the general state $|\psi(t)\rangle$ as 
    \begin{equation}
      |\psi(t) \rangle = \sum_{i} e_i(t)\, e^{-i \int^t \!\ud t' \, E_i(t')} |E_{i}(t)\rangle\,, \label{eq:instantBasisExpansion}
    \end{equation}
    where the sum runs over all instantaneous eigenstates. From our previous discussion, we know that this expansion is approximately the same as the expansion in terms of dressed states
    \begin{equation}
      |\psi(t) \rangle = \sum_{a} c_a(t)\, e^{-i \int^t \!\ud t' \, (E_a - m_a \dot{\varphi}_*(t'))} |a \rangle\,, \label{eq:dressedBasisExpansion}
    \end{equation}
    at least far from any resonance. We can think of (\ref{eq:dressedBasisExpansion}) as the analog of the interaction picture in textbook quantum field theory in which one works with free particle states, while (\ref{eq:instantBasisExpansion}) is the analog of the  renormalized state basis. While either can be used, our focus will be on (\ref{eq:instantBasisExpansion}) since the dynamics are clearer in that basis.

  \begin{figure}
  		\centering
        \includegraphics{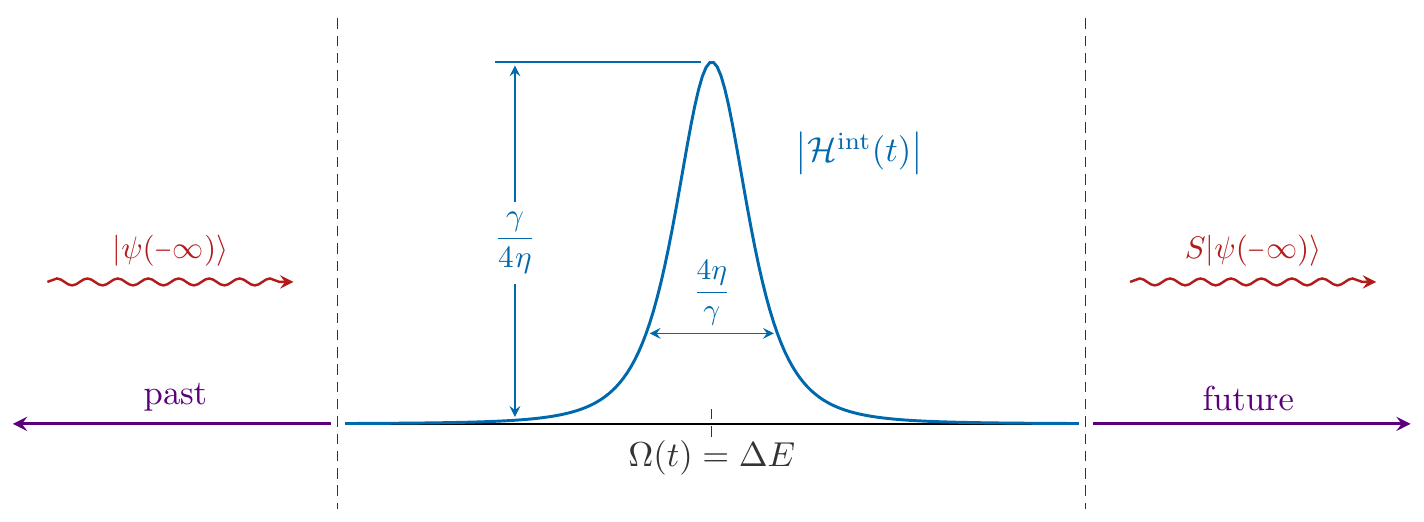}
        \caption{Interaction Hamiltonian for the two-state system as a function of time. We see that the interaction is localized over a width $4 \eta/\gamma$ near the resonance at $\Omega(t) = \Delta E$. \label{fig:interactionPotential}}
    \end{figure}
  
    \vskip 4pt
    The Schr\"{o}dinger equation simplifies in the instantaneous eigenstate basis (\ref{eq:instantBasisExpansion}) to 
    \begin{equation}
      i \frac{\ud e_i}{\ud t} = \mathcal{H}_{ij}^{\lab{int}}(t) \,e_j(t)\,,
    \end{equation}
    where we have defined the \emph{interaction Hamiltonian}\footnote{In general, we should also  include the adiabatic phase factor $e^{i \delta}$ in (\ref{eq:interactionPotential}), with $\delta = i \int^t\!\ud t' \, \langle E_{i}(t') | \partial_{t'} E_{i}(t')\rangle$. However, we can normalize the instantaneous eigenstates $|E_{i}(t')\rangle$ to  be real and set $\delta =0$.}
    \begin{equation}
       \mathcal{H}_{ij}^\lab{int} = \begin{dcases}
           \frac{i \langle E_i(t) | \dot{\mathcal{H}}_\lab{D}(t) | E_j(t) \rangle}{E_i(t) - E_j(t)} \exp\left(i \int^t \!\ud t' \,(E_i(t') - E_j(t'))\right) & i \neq j\, , \\
           \qquad \qquad\,\, 0 & i = j\, .
          \end{dcases} \label{eq:interactionPotential}
    \end{equation}
    Clearly, this term is relevant only when two or more energy levels become nearly degenerate. Furthermore, since eigenvalues repel, this interaction is only relevant for a short amount of time. This can be seen quantitatively for the two-state system from \S\ref{sec:twoState}, where the interaction Hamiltonian is explicitly
     \begin{equation}
      \mathcal{H}_{+-}^\lab{int}(t) = -\frac{i \gamma}{4 \es \eta} \frac{(\tau + \sqrt{1 + \tau^2})^{2 i \eta^2/\gamma}}{1 + \tau^2} \,e^{(2 i \eta^2/\gamma) \tau \sqrt{1 + \tau^2}}\,,\quad \text{with} \quad \tau = \frac{\gamma t}{2 \eta} \,.
    \end{equation}
     As pictured in Fig.~\ref{fig:interactionPotential}, this interaction is well localized in time about the  resonance and becomes irrelevant on a timescale  $\Delta t \sim 2 \eta/\gamma$, cf.~(\ref{eq:lzTimeScale}).

     \begin{figure}
      	\centering
        \includegraphics{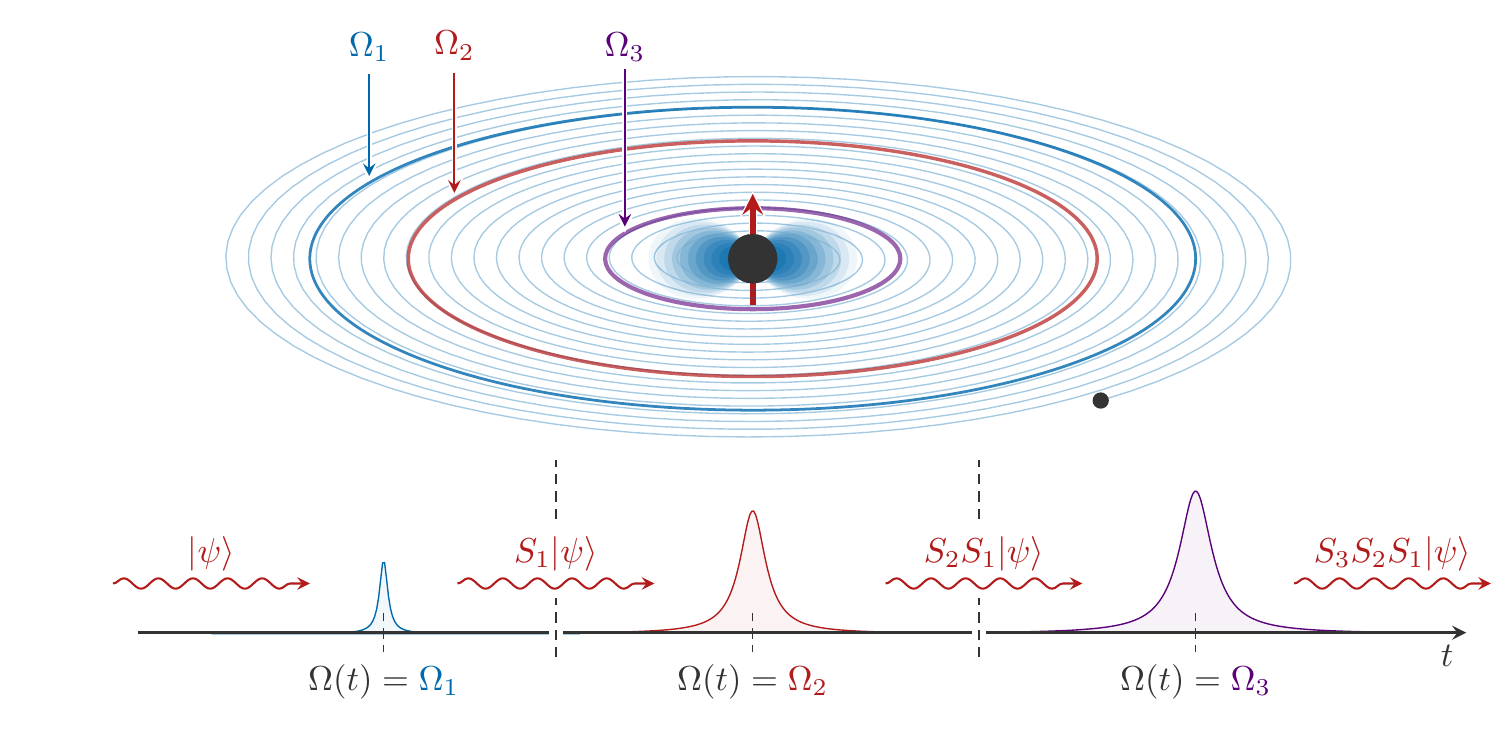}
        \caption{The orbit of a quasi-circular inspiral slowly scans through orbital frequencies $\Omega(t)$. The behavior of the cloud during the inspiral can be decomposed into a set of resonances at frequencies $\Omega_i$. Each resonance is characterized by an S-matrix, $S_i$, which describes the evolution of the system through that event. The S-matrix that describes the entire inspiral is simply the product of all individual S-matrices,~$S_\lab{tot} = \prod_i S_i$.  \label{fig:inspiral}}
    \end{figure}

     \vskip 4pt
    The fact that this interaction quickly turns off allows us to define a notion of stationary asymptotic states---the instantaneous eigenstates with their dynamical phases extracted---and thus a scattering matrix
    \begin{equation}
      S = U^\lab{int}(\infty, -\infty)\,,
    \end{equation}   
    where $U^\lab{int}(t, t_0)$ is the time-evolution operator associated to the interaction Hamiltonian~$\mathcal{H}^{\lab{int}}$. In fact,  since the interaction is negligible between resonances, we can define a scattering matrix associated to each resonance,
    \begin{equation}
      S_i = U^\lab{int}(t_i + \Delta t_\lab{w}, t_i - \Delta t_\lab{w})\,,
    \end{equation}
where $t_i$ is the time at which the $i$-th resonance occurs and $\Delta t_\lab{w}$ is some multiple of the characteristic interaction timescale. We can thus describe the behavior of the cloud throughout the inspiral by combining the $S$-matrices for all previous resonances (see Fig.~\ref{fig:inspiral}),\footnote{We must note that this separation into isolated events discards potentially crucial phase information, especially if we only consider the S-matrix magnitudes in anything other than the interaction picture. This phase information would be needed to characterize possible coherence effects. Fortunately, if the resonances are widely separated, it is very unlikely that any type of coherence can be achieved and it is thus a useful approximation to simply drop this information. However, since such coherent behavior is more likely for nearby resonances, it may be necessary to consider these resonances as a single event.}
    \begin{equation}
      U^\lab{int}(t, -\infty) \approx \prod_{i=1}^{k} S_i\,, \mathrlap{\qquad t_{k+1} > t > t_{k}\,,}
    \end{equation}
    for any time $t$ between the $k$-th and $(k+1)$-th resonance.

    \vskip 4pt
  There is one more simplifying approximation we can make. Note that the instantaneous eigenstates asymptote to dressed frame states far from the resonance, cf.~(\ref{eq:instantBasisExpansion}) and (\ref{eq:dressedBasisExpansion}). This implies that, unless the system occupies multiple instantaneous eigenstates which oscillate among one another at some fixed frequency as in \S\ref{sec:MultiState}, the \emph{magnitudes} of the S-matrix elements in the dressed frame basis will have well-defined limits far from the resonance. In fact, their magnitudes will equal those of the instantaneous eigenstate basis. This is useful, since it then is not necessary to compute the interaction Hamiltonian (\ref{eq:interactionPotential}) at all if we are only interested in the relative populations of the states that the cloud occupies after a resonance. We can instead work entirely in the dressed frame basis, in which the Hamiltonian takes a simple form, and analyze the resonance there.

  \subsubsection*{General S-Matrix} 

    Since we will only need to consider the orbit shortly before and after a resonance, we may take the orbital frequency $\Omega(t)$ to be roughly linear in time (\ref{eq:linearizedOmeg}). This relies on the assumption that the resonance bandwidth, $\Delta \Omega \sim \eta$, is much smaller than the orbital frequency itself, cf.~(\ref{eqn:Omega-Quasi-Circular}), which we argued in \S\ref{sec:TimeDependentTidalMoments} is true for all the transitions we consider. Similarly, we may also assume that the level mixings $\eta_{ab}(t)$ are time-independent.

    \vskip 4pt
Using (\ref{eq:linearizedOmeg}), the dressed frame Hamiltonian (\ref{eq:multiStateDressedHam}) can be written as
    \begin{equation}
      \left(\mathcal{H}_\lab{D}\right)_{ab} = \mathcal{A}_{ab} + \mathcal{B}_a \delta_{ab} t \,,\label{eq:lzHam}
    \end{equation}
    where we have defined 
    \begin{equation}
    \begin{aligned}
      \mathcal{A}_{ab} &\equiv \delta_{ab} \left(E_a \mp m_a \Omega_0\right) + \eta_{ab} \, ,\\
      \mathcal{B}_{a} &\equiv \mp m_a \gamma\,.
      \end{aligned}
    \end{equation}
    This is known as the \emph{generalized Landau-Zener problem} \cite{Brundobler:1993smat}. We will be specifically interested in the magnitudes of the S-matrix in this basis, defined by
    \begin{equation}
      |S_{ab}| = \lim_{t \to \infty} |U_{ab}(t, -t)|\,,
    \end{equation}
    where $U_{ab}$ is the time-evolution operator associated with (\ref{eq:lzHam}). 
    Unfortunately, analytic solutions for this problem are only known in special cases. However, a small amount of progress can be made by noting that the Schr\"{o}dinger equation for this system is invariant under a collection of symmetry transformations \cite{Brundobler:1993smat}, and thus so are the S-matrix elements. The invariant combinations of parameters made from $\mathcal{A}_{ab}$ and $\mathcal{B}_{a}$ are the generalized Landau-Zener parameters
    \begin{equation}
      z_{ab} \equiv \frac{|\mathcal{A}_{ab}|^2}{|\mathcal{B}_{a} - \mathcal{B}_{b}|} = \frac{\eta_{ab}^2}{\gamma |\Delta m_{ab}|}\,, \label{eq:genLZ}
    \end{equation}
    which extends the original definition (\ref{eqn:LZParam}) to the general, multi-state system. The S-matrix elements must then be functions of these parameters and their combinations.

\section{Backreaction on the Orbit} 
\label{sec:Backreaction}

In the previous section, we saw that Landau-Zener transitions can significantly redistribute energy and angular momentum between different states of the cloud.
The mass and spin of the cloud in a general superposition of states, at small $\alpha$, are given by
\beq
\begin{aligned}
M_c(t) &= M_{c, 0} \left(  |c_1|^2 +  |c_2|^2 + \cdots +  |c_N|^2 \right)  , 
\\
\textbf{S}_c(t) &  = S_{c, 0} \left( m_1 |c_1|^2 +  m_2 |c_{2}|^2 + \cdots + m_{N} |c_{N}|^2 \right) \hat{\textbf{z}} \, ,\label{eqn:Scloud}
\end{aligned}
\eeq
where  $S_{c, 0} \equiv M_{c, 0}/ \mu$ and $\hat{\textbf{z}}$ is a unit vector along the spin-axis of the black hole, cf.~Fig.~\ref{fig:BinaryPlane}, and the $c_i$ account for the population of the different states. 
We have ignored the $\hat{\mb{x}}$ and $\hat{\mb{y}}$ components of the spin, as they decouple for equatorial orbits.
Any change in the energy and angular momentum of the cloud must be balanced by an associated change of the binding energy and orbital angular momentum of the binary's orbit. As we will see, this greatly enriches the dynamics of the system and the gravitational-wave signals emitted from the binary. In order to separate the different phenomena, we will split the discussion into the effects occurring during (\S\ref{sec:floatdive}) and after (\S\ref{sec:modulatingfinite}) the resonant transitions.

\subsection{Floating, Sinking, and Kicked Orbits}
\label{sec:floatdive}

To analyze the effect an LZ transition has on the binary orbit, it is most convenient to consider the transfer of angular momentum between the cloud and the binary. 
Ignoring the intrinsic spins of the black holes, conservation of 
angular momentum implies 
\beq
\frac{\d}{\d t} \left(\textbf{L} + \textbf{S}_{c}\right)  = \textbf{T}_{\rm gw} \, , \label{eqn:JConserved}
\eeq
where $\textbf{L}$ is the orbital angular momentum and $\textbf{T}_{\rm gw}$ is the torque due to gravitational-wave emission from the binary. 
Since (\ref{eqn:JConserved}) is a vectorial equation, the dynamics associated to angular momentum transfer can in general be very complicated, and depend on the relative orientations of the different angular momenta.  For simplicity, we will only consider equatorial orbits and ignore precession~\cite{review}. Hence, the magnitudes of the vectors in (\ref{eqn:JConserved}) obey the relation
\beq
 \frac{\d}{\d t}\left(L \pm S_{c} \right) = T_{\rm gw} \, , \label{eqn:JCons1}
\eeq
where, as usual, the upper (lower) sign denotes co-rotating (counter-rotating) orbits. 
Specializing further to quasi-circular orbits, we find that the orbital frequency evolves as (see Appendix~\ref{app:AMTransfer} for a derivation)  
\beq
\frac{\d \Omega}{\d t}  = \gamma \left( \frac{\Omega}{\Or}  \right)^{11/3} \!\pm  3 R_J \,\Or \!\left( \frac{\Omega}{\Or} \right)^{4/3}  \! \frac{\d }{\d t} \left[ m_1 |c_1|^2 + m_{2} |c_{2}|^2 + \cdots + m_{N} |c_{N}|^2  \right]  , \label{eqn:CircularBackreaction}
\eeq
where $\gamma$ was defined in (\ref{eqn:circleRate}), and $R_J$ is the ratio of the spin of the cloud to the orbital angular momentum of the binary at the resonance frequency,
\beq
R_J \equiv \left( \frac{ S_{c, 0}}{M^2} \right) \frac{(1+q)^{1/3}}{q} (M \Or)^{1/3} \, . \label{eq:ratioAngMom}
\eeq  
Far from the transition, the occupation densities in (\ref{eqn:CircularBackreaction}) are constant and the instantaneous frequency during the inspiral is well-approximated by the standard quadrupole formula, 
cf.~(\ref{eqn:Omega-Quasi-Circular}). 

\vskip 4pt
During the transition, on the other hand, the angular momentum transfer between states with different $m$ exerts an additional torque on the binary. Crucially, the direction of the torque depends on various factors, including the orientation of the orbit, the signs of the azimuthal quantum numbers, and whether the occupation densities are growing or decaying in time. For simplicity, we set $m_{2} = m_{3} = \cdots = m_{N} $,\footnote{This turns out to be the case for any transition induced by the gravitational quadrupole $\ell_* = 2$.} such that near the resonance (\ref{eqn:CircularBackreaction}) simplifies to
\beq
\begin{aligned}
\frac{\d \Omega}{\d t}  \simeq \gamma  \mp 3\hskip 1pt \Delta m \hskip 2pt R_J  \, \Or\, \frac{\d |c_1 (t)|^2}{\d t} \, ,  \label{eqn:CircularBackreaction2}
\end{aligned}
\eeq
where we have defined $\Delta m \equiv m_2 - m_1$ and used conservation of the occupation densities during the transitions.

\begin{figure}[t]
\centering
\makebox[\textwidth][c]{ \includegraphics[scale=0.975, trim = 0 0 0 0]{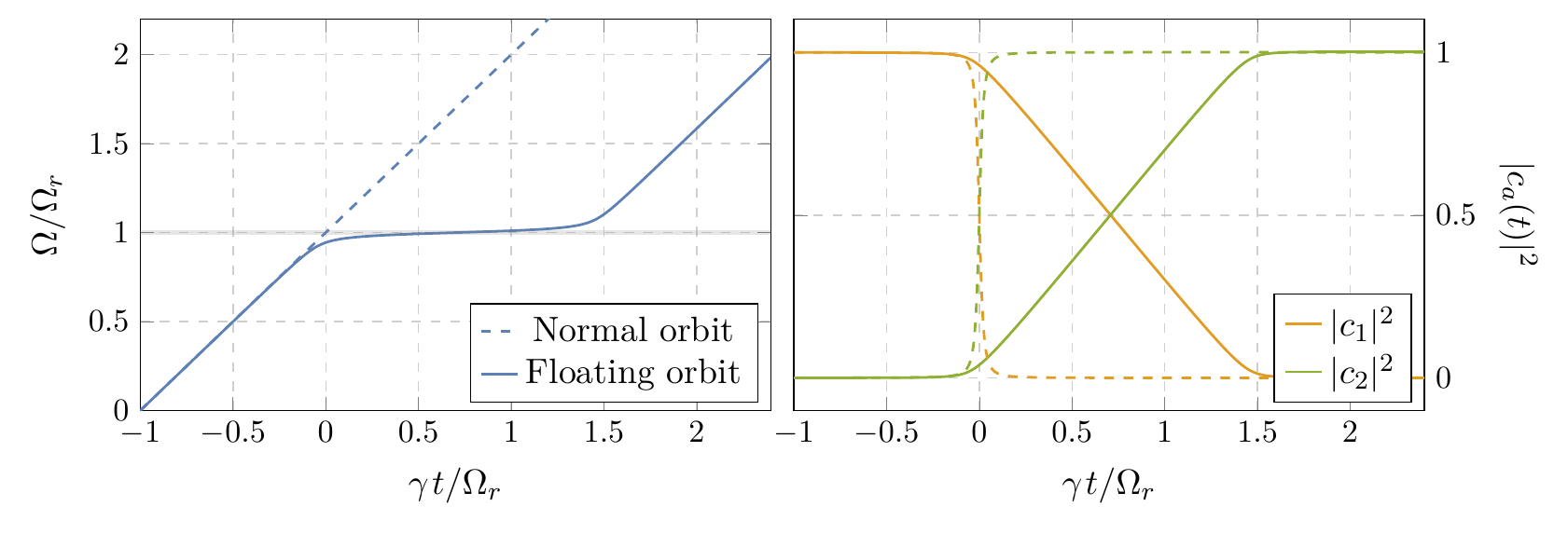}}
\caption{Evolution of the orbital frequencies (\textit{left}) and the occupation densities (\textit{right}) of an unbackreacted orbit ({\it dashed lines}) and a floating orbit ({\it solid lines}). Shown is a transition near the resonance frequency $\Omega_r = 5\mu \alpha^2/144$ with $\alpha = 0.07$ and $q=0.1$.  We also assumed that the parent black hole spun maximally, $S_{c, 0}/M^2 = 1$, before it grew the cloud.
The transition begins
when the binary enters the resonance band, denoted by the thin gray band about $\Omega = \Omega_r$.} 
\label{fig:Floating}
\end{figure}

\vskip 4pt 
By choosing the initial condition $|c_1(-\infty)|^2 = 1$, the time-derivative of $|c_1(t)|^2$ must decrease during the transition.  When $\Delta m < 0$, the cloud loses angular momentum to the orbit, forcing a co-rotating inspiral to stall and counter-rotating inspiral to shrink faster. 
These phenonema are reversed between co-rotating and counter-rotating orbits when $\Delta m > 0$.  To investigate the detailed dynamics of these backreaction effects, we must solve (\ref{eqn:CircularBackreaction2}) numerically. The solution in Figs.~\ref{fig:Floating} and~\ref{fig:Sinking} display the floating and sinking orbits, whose properties we describe below.

\subsubsection*{Floating orbits}

When the torque acts against the shrinking of the orbit, we find that LZ transitions naturally induce \textit{floating orbits}~\cite{Press:1972zz}.\footnote{This floating mechanism is distinct from the one studied in~\cite{Zhang:2018kib}, which assumes $\gamma=0$. It is also different from the analysis in~\cite{Cardoso:2011xi}, which relies on modifications of general relativity.} This is shown in Fig.~\ref{fig:Floating} for a two-state transition, where the orbital frequency increases more slowly in the resonance band.
More precisely, the rate of change in the frequency of gravitational waves emitted by the binary, $f_{\rm gw}$, during the transition satisfies (cf. Appendix~\ref{app:AMTransfer})
\beq
\left( \frac{\d f_{\rm gw}}{\d t} \right)_{\rm float} =  \frac{1}{1 + 3  R_J  \, |\Delta m \Delta E| / (4 \eta)} \left(  \frac{\gamma}{\pi} \right)  , \label{eqn:floatingChirp}
\eeq
which is clearly smaller than the unperturbed rate, $\gamma / \pi$. Interestingly, in the large backreaction limit, (\ref{eqn:floatingChirp}) asymptotes to zero and never turns negative.\footnote{While the chirp rate (\ref{eqn:floatingChirp}) was derived assuming perfect adiabaticity, this observation remains true even in numerical simulations. However, if the system is tuned to be nearly non-adiabatic, there may be small oscillations in the instantaneous frequency. The parameters for which these oscillations occur are not physically realizable and so we do not pursue them further.} This means that the orbit can at most float, but never grows during a transition.
Furthermore, the frequency of the gravitational waves emitted during the floating phase, 
$f_{\rm float}$, is directly related to the resonance frequency
\beq
f_{\lab{float}} = \frac{\Or}{\pi} = \frac{1}{\pi} \left| \frac{\Delta E}{\Delta m} \right|   \, . \label{eqn:floatingMean}
\eeq
A measurement of this approximately monochromatic gravitational-wave signal thus provides a direct probe of the spectral properties of the boson cloud. In Section~\ref{sec:unravel}, we will describe the phenomenological consequences of these floating orbits in more detail. In particular, we will discuss how the duration of the floating depends on the parameters of the system.

\vskip 4pt
The description above remains qualitatively unchanged for multi-level transitions, as they typically satisfy $m_2 = \dots = m_N$.
Moreover, since multi-level transitions necessarily involve nearly degenerate excited states, $E_2 \approx \cdots \approx E_N$, the expression (\ref{eqn:floatingMean}) remains an excellent approximation to the gravitational wave frequency emitted by the corresponding floating orbits. The only quantitative difference is that $\eta$ must be replaced by an effective coupling $\eta_\lab{eff}$ that characterizes the multi-state transition, which is well-approximated by the Pythagorean sum of the diagonalized couplings $\tilde{\eta}_{1a}$, cf.~(\ref{eq:diagDressFrameHam}). We discuss its precise definition in Appendix~\ref{app:AMTransfer}.

\vskip 4pt
Another key feature of these floating orbits is that the adiabaticity of the transition is \textit{enhanced}. 
In  analogy to the LZ parameter (\ref{eqn:LZParam}), the degree of adiabaticity of the evolution can now be quantified by the parameter $z^\prime \equiv \eta^2/\dot{\Omega}$. 
Floating orbits, with $\dot{\Omega} < \gamma$, thus enhance the adiabaticity of the transition, $z^\prime > z$.  The predictions made by adiabatically following the instantaneous eigenstates of the system, such as the final-state occupation densities, are therefore robust.

\subsubsection*{Sinking and kicked orbits}

\begin{figure}[t]
\centering
\makebox[\textwidth][c]{\includegraphics[scale=0.975]{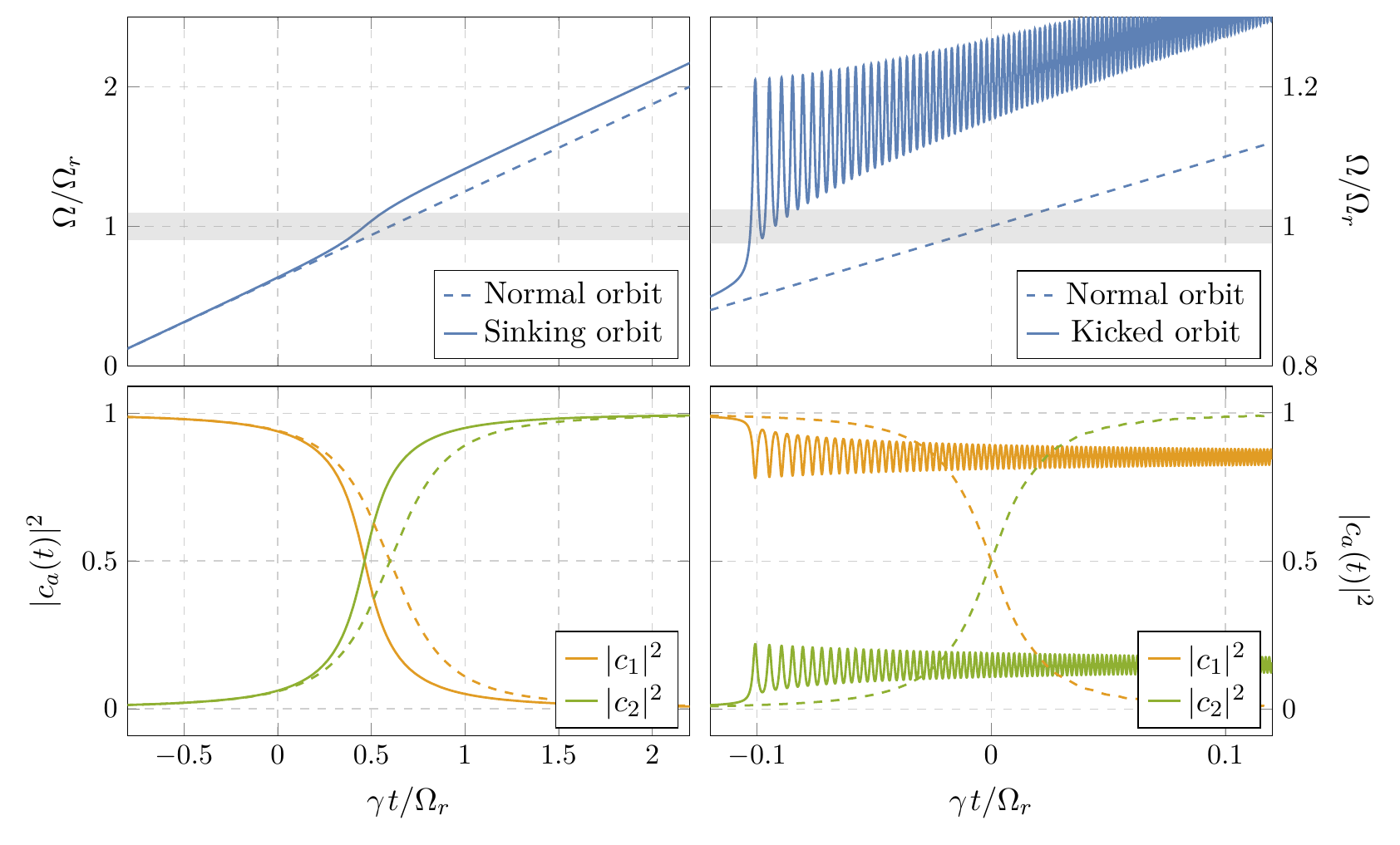}}
\caption{ Sinking (\emph{left}) and kicked (\emph{right}) orbits, with their corresponding occupation densities below. As in Figure~\ref{fig:Floating}, this transition is near the resonance $\Omega_r = 5\mu \alpha^2/144$ with $\alpha = 0.07$.  However, we flip the sign of~(\ref{eqn:CircularBackreaction2}) so that backreaction causes the orbit to either sink ($q =1$) or kick ($q = 0.1$).} 
\label{fig:Sinking}
\end{figure}

The transition can also exert a torque that increases the orbital frequency of the binary, and therefore temporarily accelerates the shrinking of the orbit.
If the effect is weak, we call it a {\it sinking orbit} (see the left panel in  Fig.~\ref{fig:Sinking}). As long as the adiabaticity of the transition is preserved, the chirp rate
of the emitted gravitational waves is\hskip 1pt\footnote{While (\ref{eqn:KickingChirp}) naively admits solutions that are divergent and negative, this simply indicates the breakdown of the adiabatic approximation in those corresponding regions of parameter space, cf.~(\ref{eqn:Non-adiabaticity}). \label{footnote:non-adiabatic}}  
\beq
 \left( \frac{\d f_{\rm gw}}{\d t} \right)_{\rm sink}  = \frac{1}{1 - 3  R_J  |\Delta m \Delta E| / (4 \eta)} \left( \frac{\gamma}{\pi} \right)   , \label{eqn:KickingChirp}
\eeq
which differs from (\ref{eqn:floatingChirp}) by an important sign. 
Compared to the unperturbed orbits, these sinking orbits spend a much shorter amount of time in the region where the LZ transition is efficient.

\vskip 4pt
A sufficiently strong ``kick,'' on the other hand, can destroy any initial adiabaticity. Using $ \dot{\Omega} \lesssim \eta^2$ as a diagnostic, we estimate that adiabaticity is preserved whenever
\beq
R_J \lesssim \frac{4 \eta \left( 1 - \gamma/\eta^2 \right)}{3  \left| \Delta m \Delta E \right|  } \simeq \frac{4 \hskip 1pt  \eta }{3  \left| \Delta m \Delta E \right| } \, , \label{eqn:Non-adiabaticity}
\eeq
where the second relation
 assumes an initial adiabatic transition, $\gamma \ll \eta^2$. As we discussed in Section~\ref{sec:gcollider}, the resulting non-adiabatic transition yields drastically different final states, most prominently through the presence of oscillations in the occupation densities of final state. These oscillations transfer angular momentum between different states of the cloud, even after the transition. As shown in bottom panel of Fig.~\ref{fig:Sinking}, this effect also modulates the quasi-circular motion of the binary, leaving dramatic oscillatory features in the binary's gravitational-wave signal. Having said that, we caution that (\ref{eqn:CircularBackreaction2}) assumes that the binary motion is always well-described by a quasi-circular orbit. This need not be the case---strong non-adiabatic transitions in which $\dot{\Omega} \gtrsim \Omega^2$ can induce significant eccentricity (or even unbind the orbit). The full phenomenology of the kicked orbits can therefore be much richer than what was described above, and deserves a more detailed investigation.  We will focus thus primarily on sinking orbits, where this backreaction is weak enough to preserve both adiabaticity and the quasi-circularity of the initial orbit.

\subsection{Time-Dependent Finite-Size Effects}
\label{sec:modulatingfinite}

As discussed in \cite{Baumann:2018vus}, finite-size effects can be large for boson clouds and feature  time-dependent behavior, such as rapid depletion.  After a brief recap of these finite-size effects, we sketch how LZ transitions produce additional time-dependent signatures. 

\vskip 4pt
We denote the intrinsic quadrupole moment of the cloud by $Q_c$ and introduce the dimensionless parameter,
\beq
\kappa_{c} \equiv - \frac{Q_{c} M}{J^2} \, , \label{eqn:kappa}
\eeq
where $M$ and $J$ are the total mass and angular momentum of the combined black hole-cloud system. For comparison, isolated Kerr black holes have $\kappa=1$. Parametrically, the quadrupole moment scales with the mass and the Bohr radius of the cloud as $Q_c \sim M_c \hskip 1pt  r_c^2$. 
The tidal force exerted by the binary companion can also deform the cloud. The linear response to tidal fields is
\beq
\delta Q_{c, ij} \equiv \Lambda_{c} \, r_{c}^5 \, \mathcal{E}_{ij} \, , \label{eqn:TLN}
\eeq
where $\delta Q_{c, ij}$ is the correction to the quadrupole moments and $\mathcal{E}_{ij}$ is the tidal tensor.
The parameter $\Lambda_{c}$ is the ``tidal Love number," and we have made it dimensionless by extracting factors of the Bohr radius.
Since $r_c/r_g \gg 1$ for small~$\alpha$, these finite-size effects may be very large for extended boson clouds~\cite{Baumann:2018vus}. 
This can compensate for the fact
that $\kappa_c$ and $\Lambda_c$ affect the dynamics at 2PN and 5PN orders, respectively~\cite{review,Poisson:1997ha, Flanagan:2007ix, Hinderer:2007mb, nrgr,nrgrs,nrgrs2}, and so these enhanced finite-size effects may impact the waveforms even during the early inspiral phase.

\subsubsection*{Superpositions and oscillations}

A resonant transition will force the cloud's spatial profile to change dramatically in time. This profile can be characterized by a set of time-dependent mass multipole moments, 
\begin{equation}
  Q_{\ell m}(t) \equiv  \sqrt{\frac{4\pi}{2\ell+1}} \int \d^3 r \, \left[ - {T^0}_0(t) \right]  r^{\ell}  \, Y_{\ell m}(\theta, \phi) \,, \label{eqn:EnergyDensityMultipole}
\end{equation}
where $T_{\mu \nu}$ is the cloud's energy-momentum tensor.  The goal of this section is to relate the occupation densities $|c_a(t)|^2$ to these multipole moments, and to discuss their time dependences. 

\vskip 4pt
Let us consider a cloud that occupies a \emph{single} eigenstate. Its mass quadrupole moments then experience $\alpha$-suppressed oscillations with frequency $\mu$, which are responsible for the cloud's well-studied decay via gravitational-wave emission \cite{Arvanitaki:2010sy,Yoshino:2014}.  For example, the axisymmetric and non-axisymmetric quadrupole moments of the scalar $|2 \es 1 \es 1 \rangle$ state  are 
\beq
Q_c \equiv Q_{20} = -6 M_c r_c^2 \quad \text{and} \quad Q_{2, \pm 2} \sim \, \alpha^2 Q_c \hskip 1pt e^{\pm 2 i \mu t} \, . \label{eqn:QcSingle}
\eeq
Since $\mu \gg \Omega(t)$, these oscillations average out over an orbital period and therefore do not affect the dynamics of the binary significantly. This is the generic behavior of all clouds occupying a single eigenstate---finite-size effects in these cases are dominated by their time-independent component.

\vskip 4pt
For scalar clouds, a typical adiabatic transition forces an initial state to fully transfer its population to another state with a different shape. As a result, the effective parameters (\ref{eqn:kappa}) and~(\ref{eqn:TLN}) evolve in time. Interesting multipolar time dependences are possible if the cloud occupies \emph{multiple} eigenstates, which is typical for vector clouds. In a general time-dependent superposition, the axisymmetric quadrupole moment is approximately
\beq
Q_c(t) = \sum_{a} |c_a (t)|^2 \, Q_{c, a} + \sum_{\substack{a \neq b}}  \delta_{m_a, m_b}\,|c_a(t) c_b(t)|  \, Q_{ab}  \cos ( \Delta E_{ab} \,t) \,  , \label{eqn:QAxisymmetric}
\eeq
where we have discarded terms that are either subleading in $\alpha$ or rapidly oscillating. The first term represents the weighted sum of the individual axisymmetric moments $Q_{c, a}$ of each state~$|a \rangle$, while the second represents additional contributions that arise from \textit{interference effects}, which occur between different states with the same azimuthal angular momenta. Crucially, the $Q_{ab}$'s are not $\alpha$-suppressed and may lead to large oscillations in $Q_c(t)$ at frequencies set by the energy differences~$\Delta E_{ab}$. Furthermore, there can be similar interference effects in the non-axisymmetric moments if the superposition involves states with different azimuthal quantum numbers, as may happen after a kick-induced non-adiabatic transition. 

\vskip 4pt
There are, in fact, two sources of oscillatory behavior in (\ref{eqn:QAxisymmetric}). As we discussed in \S\ref{sec:MultiState}, a mildly non-adiabatic multi-state transition can excite long-lived oscillations in the occupation densities $|c_{a}(t)|$, with frequencies set by the gravitational perturbation $\eta_{ab}$. It is clear from (\ref{eqn:QAxisymmetric}) that the multipole moment will inherit these oscillations. On the other hand, even if the $|c_a(t)|$ are constant, there are still oscillations in the multipole moment since the system occupies multiple, non-degenerate energy eigenstates, each with their own spatial profile. These are similar to neutrino oscillations,\footnote{The oscillations are neutrino-like in the following sense: after the resonance, the system occupies an instantaneous eigenstate (like a ``flavor eigenstate" for neutrinos) which, away from the interaction, becomes a superposition of energy eigenstates (like ``mass eigenstates" for neutrinos).} 
and are present in the final states of both adiabatic and non-adiabatic multi-state transitions. Notice that their frequencies depend only on the energy differences $\Delta E_{ab}$, which have been calculated~\cite{Baumann:2019eav}.

\vskip 4pt
In summary, whenever the cloud occupies a superposition of energy eigenstates, the mass multipole moments oscillate with unsuppressed amplitude at frequencies that are slow compared to the orbital frequency.\footnote{In principle, these oscillating quadrupoles are also new sources of gravitational waves emitted by the cloud. While we leave the study of their emission rates to future work, we note that their frequencies are typically very small and so we do not expect that this is an efficient decay channel for the cloud.}   A detailed study of the impact that  these dynamical finite-size effects have on the orbit, and ultimately on the waveform, is beyond the scope of this work. Yet, it is clear that they serve as important relics that can help us decode the nature of an LZ transition that occurred in the past.  As we will discuss in \S\ref{sec:final}, correlating these oscillatory finite-size effects with the dephasing induced by floating and sinking orbits will serve as a powerful probe of the mass and spin of the ultralight particles.

\subsubsection*{Cloud depletion}

Since a transition can populate decaying states, the energy and angular momentum of the cloud
 can also be reabsorbed by the central black hole over the typical decay timescales of these modes.
The mass of the cloud approximately evolves as~\cite{Baumann:2018vus} 
\beq
\begin{aligned}
M_{c} (t) &= M_{c, 0} \exp \left( \sum_a 2 \, \Gamma_a \!\int^t_{0}\! \d t^\prime \,|c_a(t^\prime)|^2 \right)  . \label{eqn:CloudMassRatio}
\end{aligned}
\eeq
Since the quadrupole moment depends on the mass of the cloud, $Q_c \sim M_c r_c^2$, this decay can introduce additional time dependence in the various effects discussed above, which in turn also affect the waveforms.  
As we have seen in \S\ref{sec:ScalarVector}, the depletion is generally slower than the oscillation timescale of the effective quadrupole.
During the transition, they may therefore appear as modulations of the dominant finite-size effects. However, unlike the resonant depletions that were considered in~\cite{Baumann:2018vus}, those induced by LZ transitions persist even after the cloud has passed the resonance.\footnote{The total amount of depletion during a resonance was estimated in \cite{Baumann:2018vus} and \cite{Berti:2019wnn}, assuming a constant resonance frequency. As we show in this paper, LZ transitions lead to a much longer depletion time which overwhelms the results in \cite{Baumann:2018vus, Berti:2019wnn} in almost all scenarios.} This depletion can then be observed even away from resonances and is another unique signature of boson clouds.

\newpage
\section{Unraveling the Atomic Structure}
\label{sec:unravel}

So far, we have studied simplified models for the evolution of boson clouds in black hole binaries. This allowed us to identify a number of interesting dynamical effects that, in principle, can have important observational consequences. In this section, we investigate under which conditions these effects are accessible to current and future gravitational-wave observations.  Although a comprehensive study of the rich phenomenology is beyond the scope of this paper, we will identify a few robust observational signatures that are worth exploring further.

\vskip 4pt
We will begin, in \S\ref{sec:initial}, with a discussion of the most likely state of the cloud {\it before} 
it experiences any resonances. 
This initial state is prepared via superradiance and is sufficiently long-lived to be subsequently observed with ground-based and space-based observatories. We will entertain the possibility that all of the interesting physics of the resonances and the associated backreaction on the orbit occurs within the sensitivity of gravitational-wave experiments. In \S\ref{sec:single}, we start by determining the regions in parameter space for which resonant transitions can occur within the frequency bands of present and future detectors. 
For this range of parameters, we will then determine which effects can be observed due to the existence of one or several resonant transitions. 
In~\S\ref{sec:final}, we discuss the states of the boson cloud {\it after} each resonance transition, and explain how they can be probed during the inspiral.

\subsection{Initial States of the Cloud}
\label{sec:initial}

Superradiant growth occurs when the angular velocity of the black hole is larger than the angular phase velocity of a quasi-bound state, 
\beq
\Omega_\lab{H} > \frac{\omega}{m} \, . \label{eqn:superradiant}
\eeq
The resulting boson cloud will first populate the dominant growing mode, with $m=1$, until the superradiance condition saturates at $\Omega_\lab{H} = \omega/m \approx \mu$. 
This mode remains stable for a long time, as it only
decays via gravitational-wave emission on a much longer timescale. After that, the fastest growing $m=2$ mode will continue to superradiantly drain mass and angular momentum from the black hole, until it too becomes stable when $\Omega_\lab{H} = \omega/2$.
While this process can, in principle, continue towards larger values of $m$, each subsequent growth
occurs over timescales that are much longer than the previous one, cf.~(\ref{eqn:ScalarRate}) and (\ref{eqn:VectorRates}), so that 
only the first few $m$ modes can be produced on astrophysical timescales. 

\vskip 4pt
Both the mass $ M_{c, 0}$ and the angular momentum $S_{c} \equiv m \hskip 1pt S_{c, 0}$  contained in each growing mode increase during superradiance, until  $\Omega_\lab{H} = \omega/m$. The spin of the black hole at saturation is 
\beq
\frac{a}{M} = \frac{4 m (M \omega)}{m^2 + 4 (M \omega)^2} = \frac{4 \alpha}{m} + \mathcal{O}(\alpha^3) \, . \label{eqn:BHspinSat}
\eeq
Using angular momentum conservation and the relation $M_{c, 0} = \omega \hskip 1pt S_{c, 0}$~\cite{Bekenstein:1973mi}, we can estimate the final mass and angular momentum stored in each mode. The results are summarized in Table~\ref{table:Superrad}. These estimates are rather conservative, as they ignore the re-absorption of the lower $m$ modes, which could spin up the black hole and further enhance the amplitude of the cloud. Within this approximation, we find that $M_{c,0}$ and $S_{c, 0}$ are suppressed by a power of $\alpha$ for modes with $m > 1$.  In the following, we will refer to the $m =1$ modes as ``ground states,'' and call the $m >1$ modes ``excited states.'' However, we stress that this terminology is only meant to distinguish between the growing modes, and that the ground state is not the lowest frequency mode of the cloud (see Figs.~\ref{fig:ScalarSpectra} and \ref{fig:VectorSpectra}). 

\begin{table}[t!]
\begin{center}
\begin{tabular}{ | m{6.7em} || m{4em} | m{4em}| m{4em} |  m{4em} | } 
\hline
 & $m=1$ & $m=2$ & $m=3$ & $\cdots$ \\ 
\hline
Scalar: \hskip 1pt$ | n \es \ell \es m \rangle $ & $| 2 \es 1 \es 1\rangle$ & $|3 \es 2 \es 2\rangle$ & $|4 \es 3 \es 3\rangle$ & $\cdots$ \\ 
\hline
Vector: $|n \es \ell \es j\es  m\rangle $  & $|1 \es 0\es 1\es 1\rangle$ & $|2 \es 1 \es 2 \es 2\rangle$ &$|3\es 2 \es 3 \es 3\rangle$ & $\cdots$\\ 
\hline
$M_{c, 0}/M$ & $\alpha - 4 \alpha^2$ & $ \alpha^2$ & $2\alpha^2/9$ & $\cdots$ \\ 
\hline
$S_{c, 0}/M^2$ & $1- 4\alpha$ & $ \alpha$ & $2\alpha/9 $ & $\cdots$ \\ 
\hline 
\end{tabular}
\caption{ List of the dominant growing modes for scalar and vector fields. Shown also are the mass $M_{c, 0}$ and angular momentum $S_{c, 0}$ extracted by each mode in the  limit $\alpha \ll 1$. For the $m=1$ modes, we assumed that the black hole was initially maximally spinning. 
} \label{table:Superrad}
\end{center} 
\end{table}

\subsubsection*{Ground states}

It is natural to first consider the fastest growing modes, with $m=1$.
These are the states $|2\es 1\es 1\rangle$ and $|1 \es 0 \es 1 \es 1\rangle$ for the scalar and vector clouds, respectively. In the  limit $\alpha \ll 1$, the typical growth timescales of these states are 
\beq
\begin{aligned}
\Gamma^{-1}_{211} & \simeq \frac{10^6 \,\es \lab{yrs}}{\tilde a} \left( \frac{M}{60 M_\odot} \right) \left( \frac{0.019}{\alpha} \right)^9 \mathrlap{\qquad\,\,\, \text{(scalar)}\,,} \\ 
\Gamma^{-1}_{1011} & \simeq \frac{10^6 \,\es \lab{yrs}}{\tilde a} \left( \frac{M}{60 M_\odot} \right)  \left( \frac{0.0033}{\alpha} \right)^7 \mathrlap{\qquad \text{(vector)}\,,} \label{eqn:m=1GrowthRates}
\end{aligned}
\eeq
where $\tilde{a} \equiv a/M \leq 1$ is the dimensionless spin of the black hole. 
We see that for $\alpha \gtrsim 0.019$ (scalar) and $\alpha \gtrsim 0.0033$ (vector) the clouds grow quickly on the timescale that is shorter than the typical merger time ($\sim 10^{6}\,\es \lab{yrs}$). 
Since these clouds emit continuous, monochromatic gravitational waves, they gradually deplete over timescales of order~\cite{Arvanitaki:2010sy, Baumann:2018vus, Yoshino:2014, Baryakhtar:2017ngi, East:2017mrj, East:2018glu} 
\beq
\begin{aligned} 
T_{211} & \simeq 10^8 \, \text{yrs} 
\left( \frac{M}{60 M_\odot} \right) \left( \frac{0.07}{\alpha} \right)^{15} \mathrlap{\qquad \ \text{(scalar)}\, ,} \\[2pt] 
T_{1011} & \simeq 10^8 \, \text{yrs} 
\left( \frac{M}{60 M_\odot} \right) \left( \frac{0.01}{\alpha} \right)^{11} \mathrlap{\,\,\qquad \text{(vector)}\, ,} \label{eqn:m=1Lifetimes}
\end{aligned}
\eeq
where we have assumed that $M_c \simeq \alpha M$.
Requiring these modes to be stable on astrophysical timescales, $T \gtrsim 10^8$ years, puts upper bounds on the values of $\alpha$. For stellar-mass black holes, $M \sim 60 M_\odot$, we impose $\alpha \lesssim 0.07 $ (scalar) and $ \alpha \lesssim 0.01$ (vector), while for supermassive black holes, $M \sim 10^6 M_\odot$, the range can increase to $\alpha \lesssim 0.18 $ (scalar) and  $\alpha \lesssim 0.04$ (vector).

\begin{figure}
		\centering
		\includegraphics{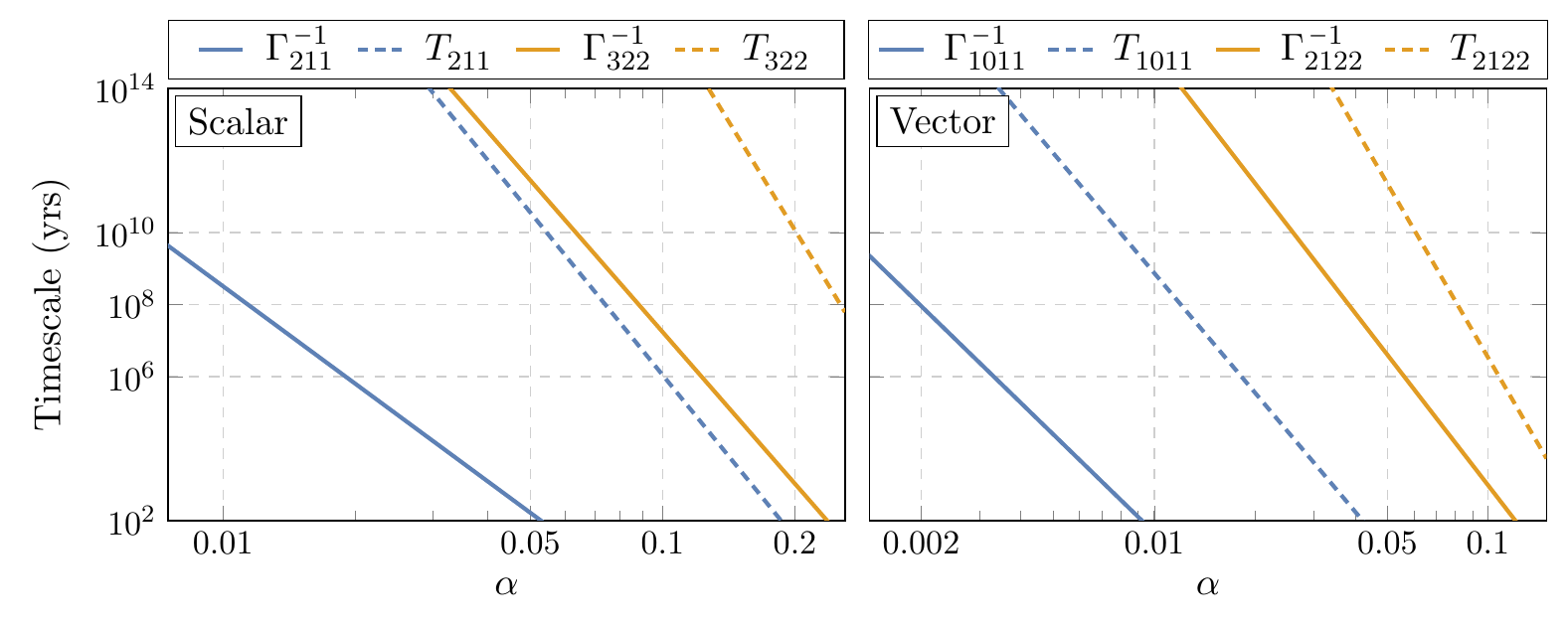}
		\caption{Relevant timescales for the ground states and first excited states of the scalar (\emph{left}) and vector (\emph{right}), as functions of $\alpha$ and for fixed $M = 60 M_\odot$. Note that both the growth time and the lifetime scale linearly with $M$. We have indicated the timescales that are shorter than typical merger times ($\sim 10^{6}$ years) 
and general astrophysical processes ($\sim 10^8$ years), as well as the age of the universe ($\sim 10^{10}$ years).    \label{fig:growthDecay}}
\end{figure}

\subsubsection*{Excited states}

To probe larger values of $\alpha$, we must instead consider excited states with $m\geq 2$, which are much longer lived. These excited states also take much longer to form than the ground states. For example, the typical growth timescales for 
the scalar $|3 \es 2\es 2\rangle$ and vector $|2\es 1\es 2\es 2\rangle$ modes are \beq
\begin{aligned}
\Gamma^{-1}_{322} & \simeq   \frac{10^6 \,\text{yrs}}{\tilde a} \left( \frac{M}{60 M_\odot} \right) \left( \frac{0.11}{\alpha} \right)^{13}  \mathrlap{\qquad \ \ \,  \text{(scalar)}\, ,}   \\ 
\Gamma^{-1}_{2122} & \simeq \frac{ 10^{6} \, \text{yrs}}{\tilde a} \left( \frac{M}{60 M_\odot} \right)  \left( \frac{0.046}{\alpha} \right)^{11}   \mathrlap{\,\,\qquad \text{(vector)} \, .} \label{eqn:GrowthRates}
\end{aligned}
\eeq
At the same time, these excited states are also much more stable than the ground states, depleting via gravitational-wave emission over the timescales\hskip 1pt\footnote{While the gravitational-wave emission rate of the scalar $|3 \es 2\es 2\rangle$ mode is known~\cite{Yoshino:2014}, the analogous rate for the vector $|2 \es 1 \es 2 \es 2 \rangle$ mode is not. Nevertheless, since the emission rate for the $\ell = 0$ modes of both the scalar and vector states share the same $\alpha$-scaling, we assume that a similar relation also applies for the scalar and vector excited states. In other words, we assume that $|2 \es 1 \es 2 \es 2 \rangle$ decays at the same rate as $|2 \es 1 \es 1 \rangle$. The only difference between the lifetimes $T_{211}$ and $T_{2122}$ in (\ref{eqn:m=1Lifetimes}) and (\ref{eqn:m=2Lifetimes}) is then that we take the initial mass of the cloud to be $M_{c,0} = \alpha M$ in the former and $M_{c, 0} = \alpha^2 M$ in the latter.}
\beq
	\begin{aligned}
		T_{322} &\simeq 10^{8}\,\lab{yrs} \left(\frac{M}{60 M_\odot}\right) \left(\frac{0.22}{\alpha}\right)^{20}  \mathrlap{\,\,\,\,\,\qquad \text{(scalar)}\,,}\\
		T_{2122} &\simeq 10^8 \, \lab{yrs} \left(\frac{M}{60 M_\odot}\right)\left(\frac{0.08}{\alpha}\right)^{16} \mathrlap{\,\,\qquad\,\,\, \text{(vector)}\,.}
\end{aligned} \label{eqn:m=2Lifetimes}
\eeq
Demanding the cloud to be stable on astrophysical timescales, $T \gtrsim 10^8 \, \lab{years}$, leads to  $\alpha \lesssim 0.22$ (scalar) and $\alpha \lesssim 0.08$ (vector) for $M \sim 60 M_\odot$, and  $\alpha \lesssim 0.44$ (scalar) and 
 $\alpha \lesssim 0.19$ (vector) for $M \sim 10^6 M_\odot$. 
Beyond these limits, we must consider excited states with  larger $m$.  However, we do not expect these higher excited states to be qualitatively different, and so we will only focus on the phenomenology of the $m=1$ and $m=2$ states.

\vskip 4pt
The cloud's history is summarized in Figure~\ref{fig:growthDecay}. The black hole will first grow the ground state (blue, solid). We require that this process takes place on a
short enough timescale ($\lesssim 10^{6}\,\lab{yrs}$) to be observable, which sets a lower bound on $\alpha$.   This ground state will then decay via gravitational-wave emission (blue, dashed). We find an upper bound on $\alpha$ by demanding that this state is suitably long-lived ($\gtrsim 10^{8} \, \lab{yrs}$). For larger $\alpha$, we are more likely to observe the cloud in an excited state, which grows (yellow, solid) and decays (yellow, dashed) via the same mechanisms, but on longer timescales.

\subsection{Resonant Transitions }
\label{sec:ResonanceSignals}

The states presented above can live long enough to be accessible to gravitational-wave observatories. Resonant transitions can then occur {\it in band} during the binary's inspiral phase.
As a consequence, there will be a series of distinct signatures which can reveal the nature of the gravitational~atom.  
We will first describe the observational characteristics of a single transition, and then discuss how observing multiple, sequential transitions can be used to further elucidate the properties of the boson clouds.

\subsubsection*{Single transition}
\label{sec:single}

\begin{figure}
		\centering
		\includegraphics[scale=1]{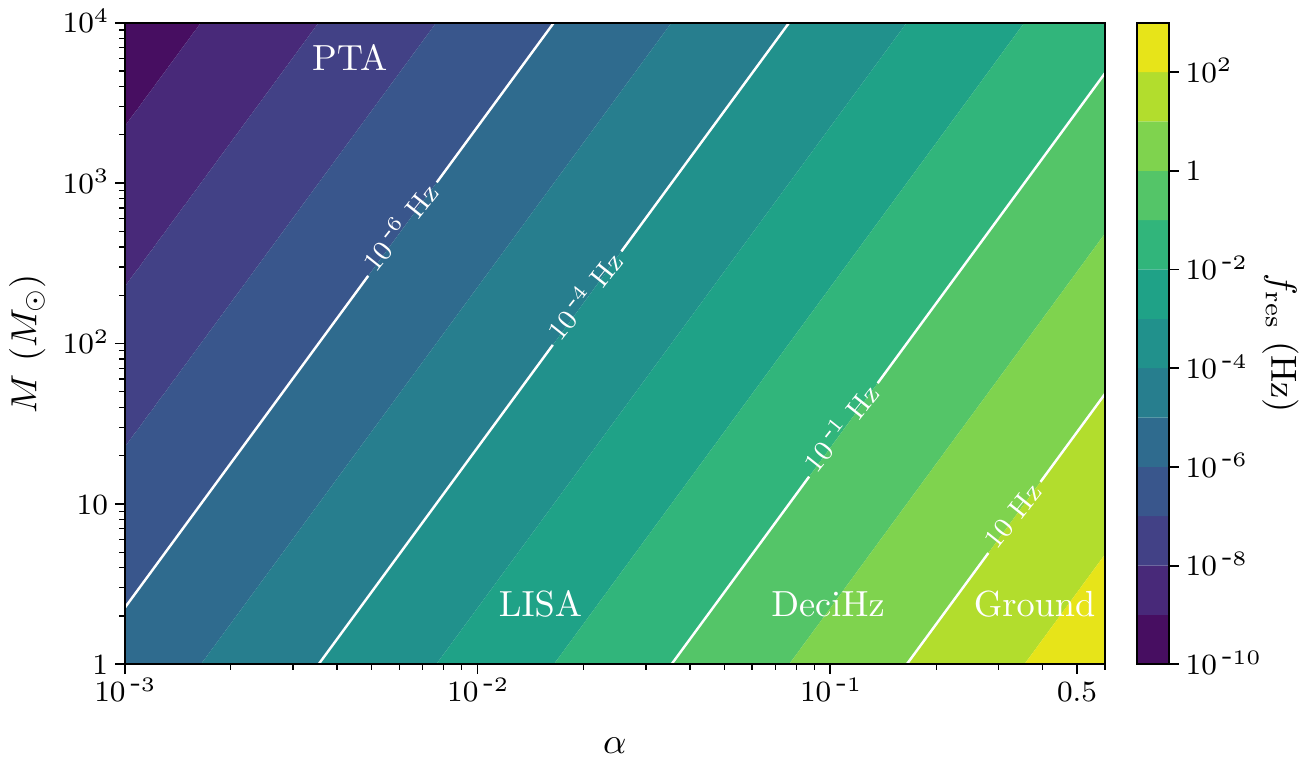}
		\caption{Resonance frequency of the Bohr transition $|2\es 1 \es 1 \rangle \to |3 \es 1 \es\es {\protect \minus 1}\rangle$, as a function of $M$ and~$\alpha$. This is representative for all Bohr transitions, since the frequency (\ref{eq:resonanceFrequencies}) is relatively insensitive to the choice of  transition, 	
under which $f_\lab{res} \to \varepsilon_\lab{B} f_\lab{res}$. Depending on the values of $M$ and $\alpha$, the resonance may fall in the band of pulsar timing arrays (PTA), the LISA observatory, proposed deciHertz experiments, or currently operating ground-based LIGO/Virgo detectors.   \label{fig:resonanceFrequencies}}
\end{figure}

Consider a transition between states with principal quantum numbers $n_a$ and $n_b$. 
The orbital frequency at the resonance, $\Omega_{r}$, implies the following frequency of the emitted gravitational waves
\beq
f_{\rm res} = \frac{\Omega_{r}}{\pi} = 1.3 \times 10^{-2} \, \text{Hz} \,\left( \frac{60 M_\odot}{M} \right) \left( \frac{\alpha}{0.07} \right)^3 \varepsilon_{\lab{B}} \, , \label{eq:resonanceFrequencies}
\eeq
where we have introduced the transition-dependent quantity
\beq
	\varepsilon_{\lab{B}} \equiv  \frac{36}{5} \frac{2}{|\Delta m|} \!\left|\frac{1}{n_
	a^2} - \smash{\frac{1}{n_{b}^{2}}} \right| ,  \label{eqn:VarepsilonBohr}
\eeq
and chosen the normalization such that
$\varepsilon_{\lab{B}} \equiv1$ for the first Bohr transition of the scalar ground state, $|2 \es 1 \es 1 \rangle \to |3 \es 1 \, \minus{1}\rangle$. The resonance frequency for this specific transition, as a function of $M$ and $\alpha$, is shown in Fig.~\ref{fig:resonanceFrequencies}. Indicated are the ranges for which the signal falls into the frequency bands of current and future gravitational-wave observations.  

   \begin{figure}[]
  	\centering
    \includegraphics[scale=1]{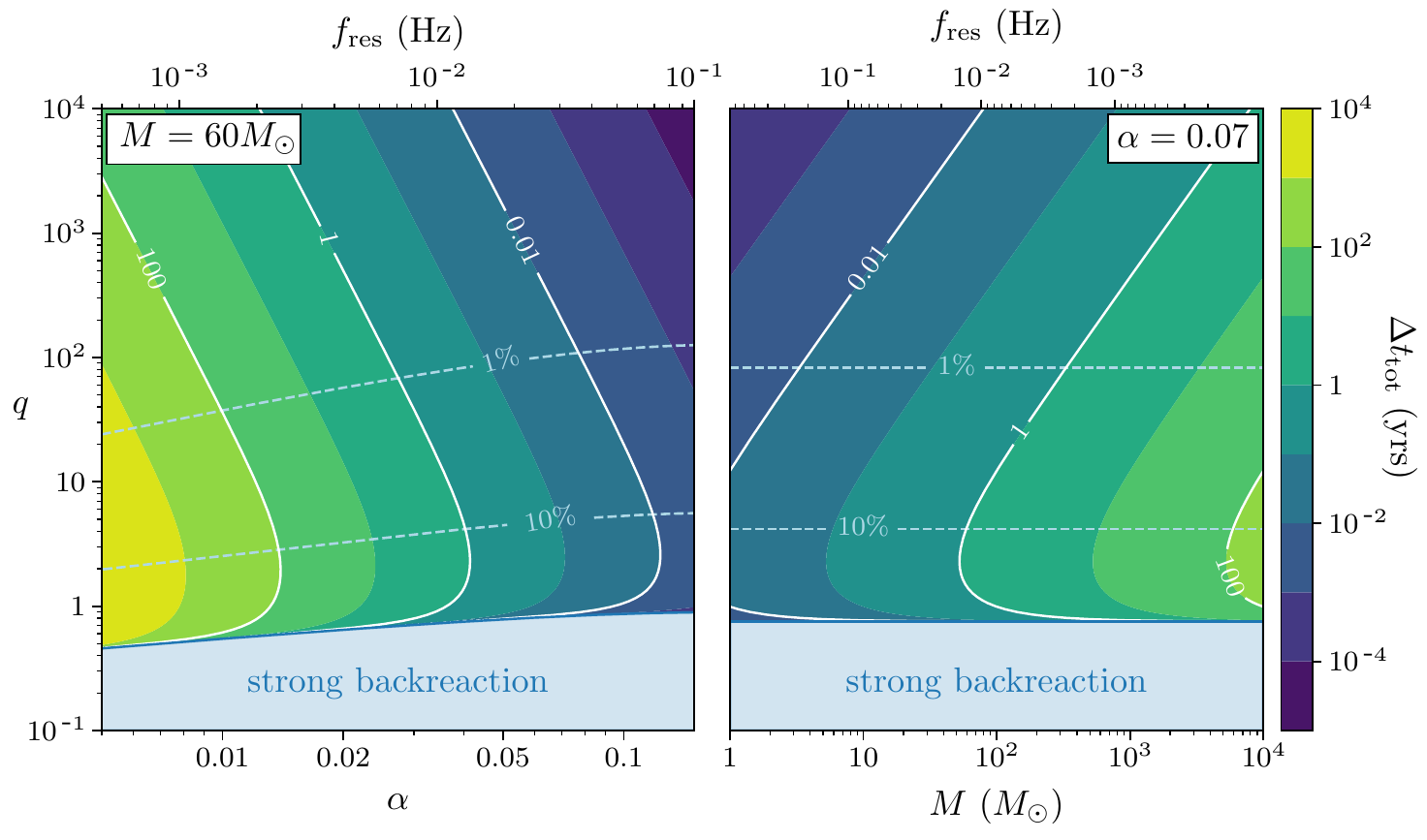}
    \caption{Total time $\Delta t_\lab{tot}$ spent in the sinking Bohr transition $|2 \es 1 \es 1 \rangle \to |3 \es 1 \es\es {\protect \minus} 1\rangle$. On the left, we plot this as function of $\alpha$ and $q$, for fixed mass $M = 60 M_\odot$.  On the right, we instead fix $\alpha = 0.07$ and plot the total time as a function of $M$ and $q$. For both, we provide the corresponding resonance frequency $f_\lab{res}$ on the top axis. 
 We plot constant contours of the ratio $|\Delta t_\lab{c}/\Delta t|$ as dashed light blue 
  lines. The cloud's backreaction on the orbit becomes stronger as $q$ decreases, shortening the time it takes for the binary to move through the transition. In blue, we indicate the region in parameter space where the adiabatic condition (\ref{eqn:Non-adiabaticity}) is violated and the backreaction is strong enough that we lose predictive control and the estimation (\ref{eq:totalTime}) is inapplicable. 
    \label{fig:211Kick}}
\end{figure}

  \vskip 4pt
An important additional characteristic is the time spent within the resonance band.
Given the resonance bandwidth of $\Delta \Omega \sim 2 \eta$, and using (\ref{eqn:floatingChirp}) and (\ref{eqn:KickingChirp}), we can estimate this as  
 \beq
 \Delta t_{\rm tot} \,\simeq\, \Delta t \pm \Delta t_c\, ,\label{eq:totalTime}
 \eeq 
 where $+/-$ represents floating/sinking orbits, respectively. Here, $\Delta t$ is the time it takes the inspiral to move through the resonance in the absence of backreaction, whereas $\Delta t_c$ is the additional contribution from the presence of the cloud,  \begin{align}
 \Delta t  &\simeq \frac{2 \eta }{\gamma}   \, \simeq\, 4 \, \text{yrs} \,  \left(\frac{M }{60 M_\odot} \right) \frac{1}{(1+q)^{2/3}} \left( \frac{0.07}{\alpha}\right)^8 \left(\frac{R_{ab}}{0.3} \right)  \varepsilon_\lab{B}^{-8/3} \, , \label{eqn:NoBackreactionTime} \\[4pt]
 \Delta t_{c}  &\simeq\frac{3R_J  | \Delta m \Delta E | }{2\gamma} \, \simeq \, 1 \, \text{yr} \, \left(\frac{M }{60 M_\odot} \right) \frac{(1+q)^{2/3}}{q^2}   \left( \frac{S_{c, 0}}{M^2} \right)  \left( \frac{0.07}{\alpha}\right)^7  \left(\frac{|\Delta m|}{2} \right)^2 \varepsilon_\lab{B}^{-7/3}\, . \label{eqn:BackreactionTime}
 \end{align}
 In (\ref{eqn:NoBackreactionTime}), we used (\ref{eqn:Eta}) to extract the value of $\eta_{ab}$ as a function of $R_{ab}$. To assess the strength of the backreaction, it is useful to take the ratio between these two timescales, 
\beq
\frac{\Delta t_{c}}{\Delta t} \simeq \frac{1}{4} \frac{(1+q)^{4/3}}{q^2} \left(\frac{\alpha}{0.07}\right)  \left( \frac{S_{c, 0}}{M^2} \right) . \label{eq:timeRatio}
\eeq 
In all cases, we find that the backreaction crucially depends on the angular momentum stored in the cloud, $S_{c,0}$, as well as the mass ratio, $q$. Notice that, for excited states, $S_{c,0}$ is suppressed by a power of $\alpha$, and the backreaction is typically weaker.  
Alternatively, we can also estimate the size of the backreaction through the difference in the number of orbital cycles spent in the resonance band with and without the cloud 
\beq
\begin{aligned}
\Delta N_c &=  f_{\rm res} \Delta t_{c}  \,\simeq\, 10^5 \, \frac{(1+q)^{2/3}}{q^2}  \left(\frac{S_{c, 0}}{M^2}\right)  \left( \frac{0.07}{\alpha} \right)^{4}  \left(\frac{|\Delta m|}{2} \right) \varepsilon^{-4/3}_\lab{B} \, . \label{eqn:DeltaN}
\end{aligned}
\eeq
Since binary searches are sensitive to $\gtrsim \mathcal{O}(1)$ difference in the number of orbital cycles between signal and template waveforms, the estimate (\ref{eqn:DeltaN}) suggests that backreaction on the orbit can dramatically alter the gravitational-wave signal, introducing a substantial dephasing with respect to the waveform model without the cloud. The effect is especially prominent for $q \ll 1$, yielding a significant dephasing for intermediate (IMRIs) and extreme mass ratio inspirals (EMRIs), where a small black hole perturbs a larger cloud.\footnote{Requiring these resonant transitions to be adiabatic, cf.~(\ref{eqn:HigherOrderLZ}), weakly constrains the mass ratio $q \gtrsim \alpha^5$.} 

\begin{figure}[]
  	\centering
    \includegraphics[scale=1]{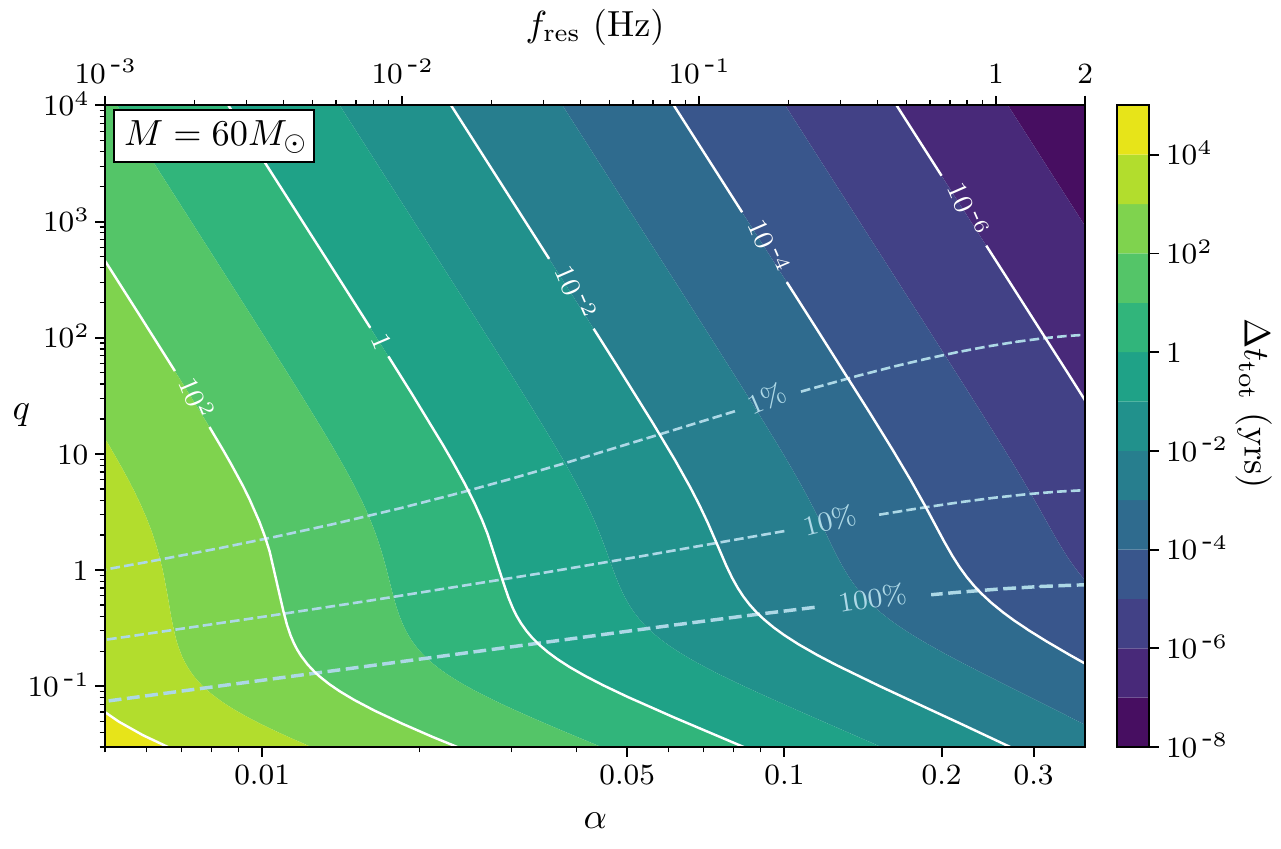}
    \caption{Total time $\Delta t_\lab{tot}$ spent in the floating Bohr transition $|3 \es 2 \es  2 \rangle \to |2 \es 0 \es 0 \rangle$, for a parent black hole of mass $M = 60 M_\odot$, as a function of $\alpha$, $f_\lab{res}$, and $q$. Increasing the fiducial mass $M$ reduces the resonance frequency and enhances the floating time, according to $f_\lab{res} \to (60 M_\odot/M) f_\lab{res}$  and ${\Delta t_\lab{tot} \to (M/60 M_\odot) \Delta t_\lab{tot}}$, for fixed $\alpha$. We plot constant contours of the ratio $\Delta t_\lab{c}/\Delta t$ as dashed light blue lines, 
    indicating where the inspiral spends an additional $1\%$, $10\%$, and $100\%$ amount of time in the transition. \label{fig:322Float}}
\end{figure}

\vskip 4pt 
The dephasing due to the backreaction on the orbit is a robust signature of boson clouds in binary systems. 
However, whether the orbit floats or sinks depends on its orientation (co-rotating or counter-rotating with respect to the black hole's spin), and the nature of the transition; see the discussion below (\ref{eqn:CircularBackreaction2}). For example, for the transition $|2\es 1 \es 1 \rangle \to |3 \es 1 \es\es {\protect \minus 1}\rangle$, which occurs for counter-rotating orbits, the cloud absorbs angular momentum from the orbit, which therefore shrinks faster, thus reducing
the time it takes for the binary to move through the transition. We plot the total time it takes for the inspiral to cross this resonance in Fig.~\ref{fig:211Kick}. Notice that backreaction dominates for $q \ll 1$, when the cloud contains a large fraction of the total angular momentum. Especially in this limit, there is a danger that the system moves through the transition too quickly, violating the condition (\ref{eqn:Non-adiabaticity}) and hence destroying the validity of~(\ref{eq:totalTime}). In that case, the binary's orbit can receive a significant kick from the cloud (cf.~Fig.~\ref{fig:Sinking}), potentially making the orbit highly eccentric. As discussed in Section~\ref{sec:Backreaction}, the cloud's dynamics then becomes highly non-adiabatic and depends sensitively on the backreacted dynamics of the orbit. A precise characterization of this interesting region in parameter space thus requires either different analytic techniques or direct numerical simulations.

\vskip 4pt
For excited states, on the other hand, Bohr transitions can also occur for co-rotating orbits. For example, during the transition $|3 \es 2 \es 2 \rangle \to |2 \es 0 \es 0 \rangle$,\footnote{Since we only focus on Bohr transitions, we will assume that the initial frequency of this binary is higher than the resonance frequencies of the hyperfine $|3 \es 2 \es 2\rangle \to |3\es 2 \es 0\rangle$ and fine $| 3\es 2 \es 2\rangle \to |3\es 0\es 0\rangle$ transitions, such that they were missed by the binary inspiral.} the cloud loses angular momentum, thereby producing a floating orbit. Fig.~\ref{fig:322Float} illustrates the total time it takes for the binary inspiral to move through this transition.
 As a reference we choose the fiducial value $M=60M_\odot$, but the plot can be easily scaled to any value of $M$. 
 The thick dashed line (labeled 100\%) denotes parameters for which the presence of the cloud doubles the time spent in the resonance region. During this transition, the inspiral floats and the orbital frequency $\Omega(t)$ remains roughly constant at the resonance value~(\ref{eqn:floatingMean}). The backreaction time again dominates for $q \ll 1$ but, unlike for the sinking orbits, this regime is still under analytic control.
 As before, this effect is particularly relevant for IMRIs and EMRIs, in which case the float can be extremely long-lived---much longer than the typical timescale of gravitational-wave observatories ($\sim 10$ years). They may thus serve as another source of continuous monochromatic gravitational waves.

\vskip 4pt
As we have demonstrated, for favorable values of $M$ and $M_*$, 
the cloud provides large corrections to the inspiral at a frequency that falls within the bands of 
future gravitational-wave observatories. A measurement of this resonance frequency, together with a measurement of $q$ and $M$, would then provide a strong constraint on the allowed values of $\alpha$, and thus the boson's mass~$\mu$. However, it is important to note that the determination of a single resonance frequency does not uniquely fix the boson mass, since additional information is needed to identify which specific transition occurred. This degeneracy is broken if we observe multiple transitions.

\subsubsection*{Multiple transitions}

\begin{figure}[t!]
\centering
	\includegraphics[scale=1]{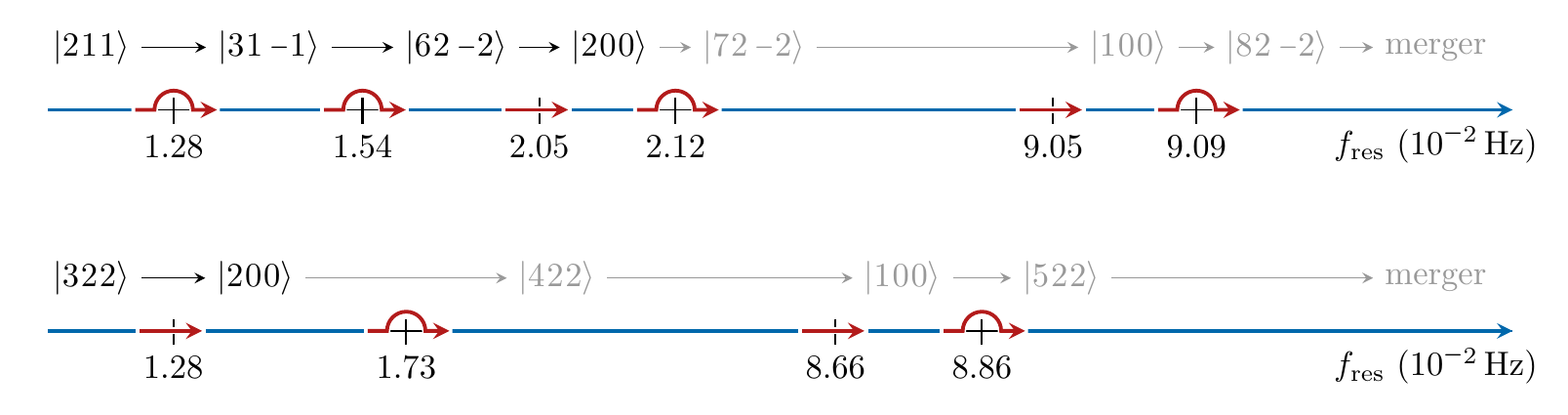}
\caption{Evolution of the $|2 \es 1 \es 1\rangle$ ({\it top}) and $|3 \es 2 \es 2 \rangle$ ({\it bottom}) states, during a counter-rotating and co-rotating inspiral, respectively (for $\alpha=0.07$, $q=1$ and $M=60M_\odot$).  Each history contains a series of floating ({\protect\tikz[baseline=-3pt] \protect\draw[cornellRed, line width=1.1, >=stealth, ->] (-0.3, 0) -- (0.2, 0);}) and sinking ({\protect\tikz[baseline=-1pt] \protect\draw[cornellRed, line width=1.1, >=stealth, ->] (-0.4, 0)--(-0.2,0) arc (180 : 0: 0.2)--(0.45, 0);}) orbits, separated by periods of ``normal'' inspiral evolution ({\protect\tikz[baseline=-3pt] \protect\draw[cornellBlue, line width=1.1, >=stealth] (-0.2, 0) -- (0.2, 0);}), with only weak perturbative mixing. As we discuss in the main text, the $|2 \es 0 \es 0 \rangle$ state has a large decay width. Unless the binary moves quickly to the next transition, which occurs only for $q \gg 1$, this forces the cloud to deplete before experiencing the next resonance.  We indicate this by the reduced opacity of the states after $|2 \es 0\es 0 \rangle$.}
\label{fig:scalarTree}
\end{figure}

As seen in Figs.~\ref{fig:211Kick} and~\ref{fig:322Float}, the binary passes through the transition relatively quickly for moderate values of $\alpha$ and $q$.  
Since subsequent resonance frequencies may be nearby, this region of parameter space suggests that we may observe \emph{multiple} resonant transitions in a given observational window.
In Fig.~\ref{fig:scalarTree}, we show two representative
 examples  for the evolution of the scalar cloud, starting from the ground state $|2 \es 1 \es 1\rangle$ and first excited state $|3 \es 2 \es 2 \rangle$, respectively. Notice that, because the gravitational atom has a finite \emph{ionization energy}, cf. (\ref{eqn:scalarspectrum}) and (\ref{eqn:vectorspectrum}), there is a maximum frequency at which resonant transitions can occur,
	\begin{equation}
		f_\lab{max} \simeq  0.2\,\, \lab{Hz} \left(\frac{60 M_\odot}{M}\right) \left(\frac{\alpha}{0.07}\right)^3. \label{eqn:fResMax}
	\end{equation}
Together with the selection rules discussed in \S\ref{sec:gravlevelmix}, this explains why the transition trees\footnote{The dominant $\ell_*=2$ perturbation mediates all of the resonances displayed in these transition trees, except for the $|3 \es 1\, \es \minus 1\rangle \to |6 \es 2 \, \es \minus 2\rangle$ and $|3100\rangle \to | 62j\, \es \minus 1 \rangle$ resonances, which are instead induced by the weaker $\ell_*=3$ perturbation. This is because, after experiencing an earlier resonance in these cases, the binary has an orbital frequency that already exceeds all resonance frequencies that can be excited from the newly prepared $| n=3, \ell=1 \rangle$ states by $\ell_*=2$. A further transition is nevertheless still possible through $\ell_*=3$, which supports perturbations with  $|\Delta m|=1$ (see \S\ref{app:TidalMoments}) and hence induces transitions at higher frequencies, cf. (\ref{eqn:VarepsilonBohr}).}    pictured in Figs.~\ref{fig:scalarTree} and~\ref{fig:vectorTree} terminate at a particular \emph{end state}.\footnote{In principle, there is also a continuum of states above this ionization energy where the cloud becomes unbound from the black hole. However,  since the gravitational perturbation is very weak, it is unlikely these states can be appreciably occupied by the sort of resonant transitions we described in \S\ref{sec:ResonanceSignals}. } A more detailed numerical analysis may then be needed to fully incorporate all of the relevant physics.

\vskip 4pt
As a rough guide as to whether or not multiple transitions can be observed, we estimate the time it takes the inspiral to move from one resonance frequency $f_{{\rm res}}^{(i)}$ to another~$f_{{\rm res}}^{(j)}$. Assuming quasi-circular adiabatic evolution, we find  
\beq
\Delta T_{i \to j} = 2.5 \, \text{yrs} \, \left( \frac{M}{60M_\odot} \right) \frac{(1+q)^{1/3}}{q} \left( \frac{0.07}{\alpha} \right)^8 \left( \varepsilon_{\lab{B}, (i)}^{-8/3} - \varepsilon_{\lab{B},(j)}^{-8/3} \right) , \label{eq:timeSpent}
\eeq
where $\varepsilon_{\lab{B}, (i)}$ represents the parameter (\ref{eqn:VarepsilonBohr}) associated to the $i$-th transition. For example, we have $\varepsilon_{\lab{B},(1)} = 1$ and  
$\varepsilon_{\lab{B},(2)} = 6/5$ for the transitions $|2 \es 1 \es 1 \rangle \to |3 \es 1\, \es \minus 1\rangle$ and  $|3 \es 1\, \es \minus 1\rangle \to |6 \es 2 \, \es \minus 2\rangle$, respectively. Comparing \eqref{eq:timeSpent} with (\ref{eqn:NoBackreactionTime}), we find that, as long as the binary moves through a single Bohr transition in a reasonable amount of time (say $1$ year), we should expect to observe a second Bohr transition on a similar timescale. Furthermore, since the rate at which the binary sweeps through the frequency accelerates, we should generically expect to observe several transitions. 

\vskip 4pt
Observing the dephasing of sequential Bohr transitions allows us to break degeneracies among the different parameters.
 For instance, let us consider successive Bohr transitions, labeled $i$ and $i+1$, between states with principal quantum numbers $n_a$, $n_b$, and $n_c$. The azimuthal angular momentum differences between the states are $\Delta m_{ab}$ and $\Delta m_{bc}$. The ratio of the resonance frequencies then is
\begin{equation}
\frac{f_{\lab{res}}^{(i+1)}}{f_{\lab{res}}^{(i)} }= \frac{\varepsilon_{\lab{B},(i+1)}}{\varepsilon_{\lab{B},(i)} }= \left| \frac{\Delta m_{ab}}{\Delta m_{bc}} \right|  \left(\frac{n_a}{n_c}\right)^2 \left| \frac{n^2_c - n_b^2}{n_b^2 - n_a^2} \right| , \label{eqn:ratiosFres}
\end{equation}
which crucially depends only on integer quantities.\footnote{Note that the presence of fine or hyperfine corrections does not affect our conclusions.} An observed sequence of these ratios then provides a fingerprint with which a particular transition history can be identified. For instance, the sequence of successive frequency ratios \{1.2, 1.33, 1.03, 4.27, 1.005\} of the counter-rotating history in Fig.~\ref{fig:scalarTree}  
 is clearly different from that of the co-rotating history, 
 \{1.35, 5, 1.02\}. Even though some of the ratios in these two histories are nearly equal, they never appear in the same order, and so we may use this sequence as a unique identifier for each history.  
This, combined with a measurement of the black hole mass $M$, can then be used to infer the boson mass $\mu$. However, there is a caveat. As illustrated in Figs.~\ref{fig:scalarTree} and \ref{fig:vectorTree}, the cloud may deplete considerably before reaching the next resonance. Fortunately, as we discuss next, this can also be used as a unique signature of the boson cloud.

\subsection*{Cloud depletion}

Throughout the above discussion, we have implicitly assumed that the cloud does not appreciably evolve away from the resonant transitions. 
However, in the later stages of the evolution, the cloud may occupy a rapidly decaying mode and is quickly reabsorbed back into the black hole, before arriving at the next resonance. 
After that, the effects of the cloud on the gravitational-wave signal become negligible.

\vskip 4pt
 Using \eqref{eqn:CloudMassRatio}, we can estimate the number of $e$-folds of decay between two successive transitions. At leading-order in $\alpha$, we find\hskip 1pt\footnote{In order to obtain the $\alpha$-scaling in \eqref{eqn:efold} , we also used the fact that the central black hole is slowly spinning after experiencing superradiance, cf. (\ref{eqn:BHspinSat}).}
\begin{equation}
  |\Gamma| \Delta T_{i \to j} = \mathcal{C}\, \frac{(1+q)^{1/3}}{q} \alpha^{2j + 2 \ell - 2} \left(\varepsilon_{\lab{B}, (i)}^{-8/3} - \varepsilon_{\lab{B}, (j)}^{-8/3}\right)  , \label{eqn:efold}
\end{equation}
where $\Gamma$ is the decay rate of the state that is occupied after the $i$-th transition. Typically, the dimensionless coefficient $\mathcal{C} \lesssim \mathcal{O}(0.1)$ for the states of interest, and can depend sensitively on the state's quantum numbers.
Depletion is significant when the ratio in (\ref{eqn:efold}) is greater than $1$.

\begin{figure}[t!]
	\centering
		\includegraphics{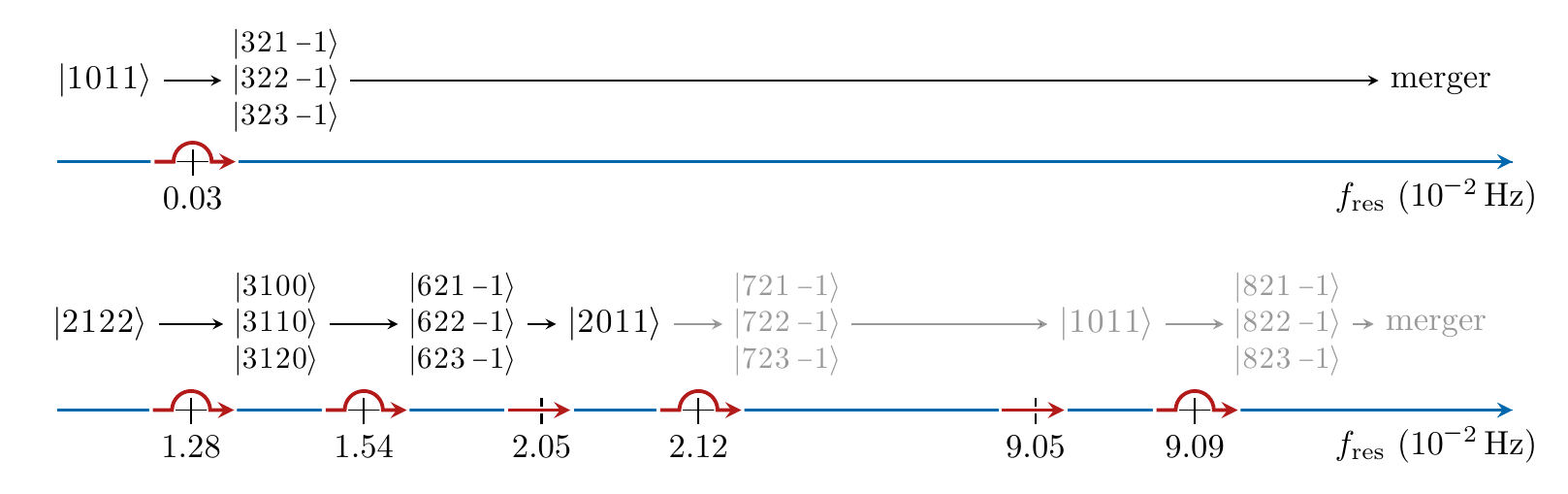}
		\caption{
		Evolution of the $|1 \es 0 \es 1 \es 1\rangle$ ({\it top}) and $|2 \es 1 \es 2 \es 2\rangle$ ({\it bottom}) states during a counter-rotating inspiral (for $\alpha = 0.01$ and $\alpha=0.07$, respectively). In both cases, $q=1$ and $M=60M_\odot$. Notice that the ground state of the vector cloud $|1 \es 0 \es 1 \es 1 \rangle$ only experiences a single transition. The excited state $|2 \es 1 \es 2 \es 2 \rangle$, on the other hand, mimics the history of the scalar ground state $|2 \es 1 \es 1 \rangle$. Yet, we see that vector transitions involve superpositions of many  states. 
The large decay width of the state  $|2 \es 0 \es 1 \es 1 \rangle$ makes it unlikely for the cloud to survive after that point, unless $q \gg 1$. As in Fig.~\ref{fig:scalarTree}, we indicate this by reducing the opacity of the states after the $|2 \es 0 \es 1 \es 1 \rangle$ state.}
		\label{fig:vectorTree}
\end{figure}

\vskip 4pt
For scalar fields, the fastest decaying modes have $\ell=0$, in which case the estimator in \eqref{eqn:efold} scales inversely with $\alpha$. A cloud that populates these states will therefore almost always deplete before it reaches the next resonance. The cloud only survives when $q \gtrsim \alpha^{-3}$, namely when the cloud is on the smaller black hole, in which case the system moves quickly enough to the next transition. Similarly, for vector clouds, a fast decaying state with $\ell=0$ and  $j=1$ becomes occupied along the transition tree in Fig.~\ref{fig:vectorTree}.\footnote{Notice that, because we are assuming that the central black hole has a spin that saturates at the superradiance condition, the  state $|2011 \rangle$ turns into a decaying mode, despite having $m>0$.} Unless $q \gg 1$, it therefore suffers the same fate as the scalar cloud. However, a vector cloud still experiences more resonances over a wider range of mass ratios than a scalar cloud.

\vskip 4pt
Although a rapid depletion of the cloud prevents us from exploring later resonant transitions, it is a unique feature of gravitational atoms in binary systems~\cite{Baumann:2018vus} that helps to distinguish them from other exotic compact objects like boson stars, e.g.~\cite{Sennett:2017etc}. Using~(\ref{eqn:efold}), we see that, for $q \ll 1$, a cloud that populates a fast-decaying state will typically deplete before it reaches the next resonance. Nevertheless, these IMRIs and EMRIs are precisely the binaries for which we only expect to see a single transition in a reasonable observational period (\ref{eq:timeSpent}). Furthermore, they are also the binaries for which backreaction is most significant. This suggests that, to probe these types of binary systems, we will typically need to hunt for long floats or strongly kicked orbits, rather than multiple correlated transitions.

\subsection{Post-Resonance Evolution}
\label{sec:final}

As argued above, observing multiple successive resonances provides a detailed mapping of the spectral properties of the cloud, allowing us to extract the mass $\mu$ of the ultralight boson. However, since this fingerprint depends only on the energy differences between states, it can fail to distinguish between scalar and vector clouds, which mainly differ by the number of states involved in a transition. Fig.~\ref{fig:vectorTree} shows transitions of the vector cloud, starting from the ground state $|1 \es 0 \es 1 \es 1\rangle$  and the first excited state $|2 \es 1 \es 2 \es 2\rangle$, respectively.
While the dynamics of the vector ground state is rather unique (involving only a single transition), distinguishing between scalar and vector clouds for other histories is more difficult.  In particular, the vector state $|2 \es 1 \es 2 \es 2 \rangle$ in Fig.~\ref{fig:vectorTree} has a  similar transition history as the scalar state $|2 \es 1 \es 1 \rangle$ in Fig.~\ref{fig:scalarTree}. 
It is thus non-trivial  to disentangle these two histories using only information about their resonance frequencies.  As we discuss here, studying finite-size effects can help us break this degeneracy.\footnote{In principle, this degeneracy can also be lifted through measurements of $\partial_t \Omega$ and $\Delta t_{\rm tot}$ across a resonance, as the mode $|2 \es 1 \es 2 \es 2 \rangle$ carries less angular momentum than the mode $|2 \es 1 \es 1 \rangle$ (for the same value of $\alpha$), and this should have a weaker effect on the orbit.}

\subsection*{Finite-size effects}

 There is a major qualitative difference between the evolution of scalar and vector clouds that we can exploit to distinguish them: vector transitions typically involve more than two states, while scalar transitions do not. As discussed in \S\ref{sec:modulatingfinite}, the result of these multi-state vector transitions is a superposition of states, each with their own shape and characteristic frequency. 
 There are then large oscillations in the shape of the vector cloud, at frequencies which depend on the energy differences $\Delta E_{ab}$ between the occupied states. Detecting these oscillatory finite-size effects is thus a smoking gun of particles with spin.\footnote{We expect the same phenomena to apply also to higher-spin particles. Namely, for a given $n$, $\ell$ and $m$, there will be a larger number of  states, and the finite-size effects can oscillate at a larger number of frequencies.}

  \vskip 4pt
  We can illustrate this effect quantitatively using  the first Bohr transition of the vector excited state, $|2 \es 1 \es 2 \es 2 \rangle \to |3 \es 1 \es j \es 0 \rangle$. After the transition, the state evolves into a superposition with has negligible population in the $|3 \es 1 \es 1 \es 0 \rangle$ mode:\hskip 1pt\footnote{The gravitational perturbation only mediates degenerate subspace transitions between $|3 \es 1 \es 0 \es 0 \rangle$ and $|3 \es 1 \es 2 \es 0 \rangle$. As a consequence, the final state is an equally weighted sum of these two states, independent of both $q$ and $\alpha$.}
  \begin{equation}
    |\psi(t) \rangle \approx \frac{1}{\sqrt{2}}\hskip 2pt e^{-i E_{3 1 0 0} t} \left(|3 \es 1 \es 0 \es 0 \rangle +  e^{-i \Delta E_{2 0} t } |3 \es 1 \es 2 \es 0 \rangle\right) ,
  \end{equation}
where we have defined $\Delta E_{20} \equiv E_{3120} - E_{3100}$.
  The dominant energy difference is set by the fine-structure splitting, cf.~(\ref{eqn:vectorspectrum}) and~(\ref{eq:fineStructure}), and so the frequency of these oscillations is roughly
    \begin{equation} 
    \frac{\Delta E_{20}}{2 \pi} \approx  6 \times 10^{-5} \, \lab{Hz} \, \left(\frac{60 M_\odot}{M}\right) \left(\frac{\alpha}{0.07}\right)^5\,. \label{eq:firstBohrSplit}
  \end{equation}
At the same time, the cloud's axisymmetric quadrupole moment inherits these oscillations, which can be parameterized as
  \begin{equation}
    \kappa_c(t) = \kappa_0 \left(1 - 2 \sqrt{2} \cos(\Delta E_{2 0} \,t)\right)\,,
  \end{equation}
  with $\kappa_c$ defined in (\ref{eqn:kappa}) and we have ignored corrections that are suppressed by $\alpha$. The $\kappa_0$ value represents the quadrupole moment of the state $|3\es 1\es 0 \es 0 \rangle + i |3 \es 1 \es 2 \es 0 \rangle$. For $\alpha = 0.07$ and $M = 60 M_\odot$, the cloud's mass distribution oscillates between bulging at the black hole's spin axis and bulging along its equatorial plane at a period of roughly $4.6$ hours. Since it then takes $1.2$ years to evolve towards the next transition, cf.~(\ref{eq:timeSpent}), we could potentially observe $2.2 \times 10^3$ of these cycles. 
If, in addition, we include the relatively fast depletion of the $|3 \es 1 \es 0 \es  0 \rangle$ mode, these oscillations decay on a timescale of about a month.

  \vskip 4pt
  Because the frequency of these oscillations is so heavily $\alpha$-suppressed, observing them is a challenge. For example, a single oscillation of the $|3\es 2 \es j \, \es \minus 1 \rangle$ superposition produced by the vector ground state can take decades for $\alpha = 0.01$ and $M = 60 M_\odot$. This is far outside present observational timescales.   Similarly, the binary might merge or encounter another resonance before the cloud has a chance to complete a single cycle. This is typically the case for the end state of a transition tree, like the $|8 \es 2 \es j \, \es \minus 1 \rangle$ superposition for the vector excited state in Fig.~\ref{fig:vectorTree}. Though we can avoid these issues by considering supermassive black holes and/or higher excited states where larger values of $\alpha$ are allowed, we are more likely to detect
  these oscillations for the earlier Bohr transitions. For instance, while the superpositions $|6 \es 2 \es j \, \es \minus 1 \rangle$ and $|7 \es 2 \es j \, \es \minus 1 \rangle$ oscillate $\sim \!40$ times slower than the $|3 \es 1 \es j \es 0 \rangle$ with frequency (\ref{eq:firstBohrSplit}), they still execute hundreds of cycles, which may be detected with high-precision templates.

  \vskip 4pt
Since finite-size effects enter at higher post-Newtonian orders, they are difficult to measure during the early stages of the inspiral when the binary's relative velocity is small \cite{review}. However, boson clouds can have much larger multipole moments compared to black holes in isolation \cite{Baumann:2018vus}, which can greatly enhance our chances of detecting them through gravitational-wave precision measurements, even before the system achieves high velocities in the merger phase. As we argued in \S\ref{sec:modulatingfinite}, the axisymmetric moments of the cloud roughly scale as $Q_c \sim M_c r_c^2$, where $M_c$ and $r_c$ are the mass and Bohr radius of the cloud, respectively. For an excited initial state, the dimensionless quadrupole moment (\ref{eqn:kappa}) is then of the order $\kappa_c \sim (M_c/M)\tilde{a}^{-2} \alpha^{-4}$, which is much larger than the corresponding moment for the pure Kerr black hole, $\kappa = 1$. The effects of these large, fluffy clouds are amplified near the merger, and we might expect that we can infer detailed information about the shape of the end states in a transition tree  by accurately modeling the finite-size effects in this phase.

\subsection*{Decaying shapes}

In \S\ref{sec:modulatingfinite}, we described how finite-size effects can receive further time-dependent changes when the cloud occupies decaying states. While this depletion typically occurs over much longer timescales than the orbital period, it can be significant when the decaying states have small angular momentum. 
For examples, the decay times of the $|2\es 0\es 0\rangle$ and $|2\es 0\es 1\es 1\rangle$ modes---the main depletion channels for the histories depicted in Figs.~\ref{fig:scalarTree} and \ref{fig:vectorTree}---are
\beq
\begin{aligned}
\left| \Gamma_{200}^{-1} \right| & \simeq 1 \, \text{yr} \left( \frac{M}{60 M_\odot} \right) \left( \frac{0.014}{\alpha} \right)^6 \mathrlap{\qquad\,\,\, \text{(scalar)}\,,} \\ 
\left| \Gamma_{2011}^{-1} \right| & \simeq 1 \, \text{yr} \left( \frac{M}{60 M_\odot} \right) \left( \frac{0.045}{\alpha} \right)^8 \mathrlap{\qquad\,\,\, \text{(vector)}\,.}
\end{aligned}
\eeq
We see that, even for moderate values of $\alpha$, these depletion effects can be significant and are observable within the typical lifetimes of gravitational-wave observatories. As pointed out in~\cite{Baumann:2018vus}, this change in the contribution from finite-size effects can also strongly indicate the presence of a boson cloud.

\section{Conclusions and Outlook}
\label{sec:Conclusions}

The detection of gravitational waves~\cite{Abbott:2016blz, Abbott:2016nmj, Abbott:2017vtc, Abbott:2017oio, TheLIGOScientific:2017qsa} 
marked the beginning of a new era for multi-messenger astronomy~\cite{GBM:2017lvd, Sathyaprakash:2019rom}. It also raises the interesting question whether precision 
gravitational-wave observations can become a new tool for fundamental physics~\cite{review,Porto:2016zng,Porto:2017lrn, Barack:2018yly, Sathyaprakash:2019yqt,Bertone:2019irm}.  We often associate new physics with short-distance (or high-energy) modifications of the Standard Model.  The decoupling between physics at short and long distances then provides an immediate challenge for using the long-wavelength gravitational waves produced by the dynamics of macroscopic objects to probe physics at shorter distances~\cite{review}.\footnote{This is particularly challenging in gravity, since the equivalence principle implies that the inner structure of a compact body manifests itself only at higher orders in the perturbative expansion~\cite{review, Porto:2016zng,Porto:2017lrn}.} This, however, ignores the possibility that new physics can be both very light and weakly coupled, which allows for coherent effects on astrophysical length scales. 
As it was highlighted in~\cite{Baumann:2018vus}, this is the case for ultralight bosons in  black hole binaries, whose Compton wavelengths are larger than the typical sizes of the constituents. The extended nature of the associated boson clouds enhances the effects due to the internal structure of the compact objects, 
mitigating the decoupling challenge. 
As a consequence, the observation of gravitational waves from binary black holes has also opened a new window into physics beyond the Standard Model at the weak-coupling frontier.

\vskip 4pt
In this paper, we have studied how the presence of boson clouds leads 
to novel dynamical effects when they are part of a binary system. During the inspiral, the clouds are strongly deformed at characteristic resonance frequencies that depend sensitively on their spectral properties.  The
transfer of angular momentum between the cloud and the orbit during each resonance can cause large corrections to the gravitational-wave signal. Notably, there is a dephasing with respect to the frequency evolution without a cloud, arising from transient floating and sinking orbits. 
Furthermore, time-dependent finite-size effects provide additional information about the available states in the gravitational atom which, due to the extended nature of the cloud, may be observed during the early inspiral phase. Similarly, as was emphasized in \cite{Baumann:2018vus}, strong mixing with decaying modes during the resonant transition can also deplete the cloud as it approaches merger. \\

\vskip 4pt
In an ordinary collider, a particle's spin is measured via the angular dependence of the final state. Similarly, in the gravitational collider the position of the resonances determine the boson's mass, and we must observe the final state of a transition to distinguish particles of different spin. 
 Fortunately, the properties of this state are accessible through the imprint of finite-size effects on the waveform.  Hence, a precise reconstruction of a gravitational-wave signal can help us not only to detect new ultralight bosons, but also determine their mass and intrinsic spin. The discovery potential of gravitational-wave observations thus necessitates the development of sufficiently accurate template waveforms, which include the characteristic features of boson clouds in binary systems that we have uncovered.
 
 \vskip 4pt The results of this paper also raise a number of interesting follow-up questions:
\begin{itemize}
\item For simplicity, we have mostly studied circular equatorial orbits.  It would be interesting to extend our analysis to general inclined and elliptical orbits, in which case the quasi-periodic driving force provided by the companion has additional frequency components; see e.g.~\cite{Zhang:2019eid, Berti:2019wnn}, \S\ref{app:TidalMoments}, and \S\ref{app:adFloTheo}. This will affect the allowed transitions of the cloud during the~inspiral. 

\item To maintain maximal theoretical control, we have focused on the early inspiral phase of the binary's evolution. After a sequence of resonant transitions, the system evolves into a final state with distinctive properties, which can be probed most sensitively during the late stages of the inspiral. 
It is unclear to what degree this regime can still be describe analytically, or if we have to resort to numerical simulations. For instance, additional effects such as dynamical friction~\cite{Zhang:2019eid} can become important when the binary companion enters the cloud.

\item Most of our quantitative analysis has been restricted to the regime of weak backreaction.  In these cases, 
the effects of the cloud on the orbital dynamics can be treated perturbatively and the 
dephasing 
during resonances can be predicted analytically. This regime of weak backreaction is most relevant for black holes of roughly equal masses.  There is, however, also great interest in the case of extreme mass ratio inspirals. If the cloud is around the large black hole, the backreaction on the dynamics of the small companion will be very large. The companion will either float for a very long time, or receive a very strong kick that is likely to induce significant eccentricity of the orbit. While the floating behavior is still under analytical control, our approximations break down for strong kicks. These very interesting cases deserve a more dedicated analysis.

\item We have ignored the backreaction of the cloud on the central black hole. 
This may become important when quickly decaying states are populated, where the depletion of the cloud may significantly change the black hole's spin, triggering more superradiant growths or depletions; see e.g.~\cite{Ficarra:2018rfu}. This can further enrich the time dependence of the finite-size effects, and even alter the sequence of transitions if new growing modes are excited.

\item We did not compute the explicit waveforms for the signals described in Section~\ref{sec:unravel}. This will be important for actually detecting these effects, as they are not yet captured by the available templates for standard binary black hole mergers. 
Because both the resonance signals and the time-dependent finite-size effects are intrinsically correlated, one must
incorporate all of the effects described here into a unified framework
to obtain reliable template waveforms. 

\item 
Neither did we 
estimate the event rates and strengths of these signals.  This would require astrophysical models for the merger rates of black holes that are rotating rapidly enough to support the creation of boson clouds; see e.g.~\cite{Arvanitaki:2014wva, Baryakhtar:2017ngi, Brito:2017zvb, Brito:2017wnc}. 
It would also need more accurate computations of the multipole moments associated with the clouds, and their associated gravitational radiation rates. 

\item The ultralight bosons in our analysis only coupled via gravity. In principle, there can also be self-interactions of the field, couplings to other ultralight bosons, or direct couplings to ordinary matter. It would be interesting to explore what additional signatures can arise from these extra interactions.
\end{itemize}
We hope to return to some of these issues in future work.

	\vskip23pt
	\subsection*{Acknowledgements}
	We thank Mina Arvanitaki, Gianfranco Bertone, Vitor Cardoso, Liang Dai, Savas Dimopoulos, Tanja Hinderer, Junwu Huang, Luis Lehner, David Nichols, Samaya Nissanke, Huan Yang, and Jun Zhang for helpful discussions.
	DB~and JS are supported by a Vidi grant of the Netherlands Organisation for Scientific Research~(NWO) that is funded by the Dutch Ministry of Education, Culture and Science~(OCW). The work of DB and JS is part of the Delta-ITP consortium, and the work of HSC is supported by NWO. R.A.P. acknowledges financial support from the ERC Consolidator Grant ``Precision Gravity: From the LHC to LISA"  provided by the European Research Council (ERC) under the European Union's H2020 research and innovation programme (grant agreement No. 817791), as well as from the Deutsche Forschungsgemeinschaft (DFG, German Research Foundation) under Germany's Excellence Strategy (EXC 2121) ``Quantum Universe'' (390833306). HSC would like to thank Perimeter Institute and the Stanford Institute for Theoretical Physics for their hospitality while some of this work was completed. R.A.P. would like to thank the participants of the workshop ``The Science of Third-Generation Gravitational Wave Detectors" in Berlin, for discussions while some of the results in this paper were presented.\footnote{\url{https://indico.desy.de/indico/event/23425/}} Portions of this work were completed by JS at the Aspen Center for Physics, which is supported by National Science Foundation grant PHY-1607611. 
	
\clearpage

\appendix

\section{Gravitational Perturbations}
\label{app:GP}

In this appendix, we provide further details of the gravitational perturbations introduced in  \S\ref{sec:gravlevelmix}. We generalize the tidal moments to inclined orbits (\S\ref{app:TidalMoments}) and present the next-to-leading order corrections to the gravitational potential (\S\ref{app:HigherOrder}).

\subsection{Tidal Moments of Inclined Orbits} 
\label{app:TidalMoments}

We wish to describe the tidal moments $\mathcal{E}_{\ell_* m_*}$ in (\ref{eqn:h00Expansion}) for general orbits.  
In the usual angular spherical coordinates $\{\Theta_*, \Phi_*\}$ (see Fig.~\ref{fig:BinaryPlane}), they are
\beq
\mathcal{E}_{\ell_* m_*}(\Theta_*, \Phi_* ) = \frac{4\pi}{2\ell_* + 1} Y^*_{\ell_* m_*} (\Theta_* , \Phi_* ) \, , \label{eqn:QuadrupoleTidalMomentOld0}
\eeq
where have suppress all explicit dependences on time $t$.  
Although (\ref{eqn:QuadrupoleTidalMomentOld0}) provides a closed-form expression for the tidal moments for all $\ell_*$, these angular coordinates do not naturally adapt to the orbital motion of the binary. Instead, it is more convenient to express the tidal moments in terms of the inclination angle and the true anomaly $\{ \iota_*, \varphi_*\}$ described in \S\ref{sec:gravlevelmix}: 
\beq
\mathcal{E}_{\ell_* m_* }(\iota_*, \varphi_*) = \sum_{ m_\varphi=-\ell_*}^{\ell_*} \varepsilon^{(m_\varphi)}_{\ell_* m_*}(\iota_* ) \, e^{-i m_\varphi \varphi_* } \, , \label{eqn:mgammaDef}
\eeq
where the reduced tidal moments, $\varepsilon_{\ell_* m_*}^{(m_\varphi)}$, can be obtained by using the normalized binary separation vector\footnote{This expression for $\hat{\textbf{n}}$ is obtained by applying the coordinate transformation (\ref{eqn:Angles}) to its equivalent expression in the usual spherical coordinates, $\hat{\textbf{n}} = (\sin \Theta_* \cos \Phi_*, \sin \Theta_* \sin \Phi_*, \cos \Phi_*)$ .} $\hat{\textbf{n}} = (\cos \iota_* \cos \varphi_*, \sin \varphi_*, \sin \iota_* \cos \varphi_*)$ in the symmetric-trace-free tensor representation of the gravitational perturbation (see e.g.~\cite{Taylor:2008xy, Thorne:1980ru}). The sizes of these reduced moments affect the magnitudes of the the overlap (\ref{eqn:etaDef}), while the oscillatory term $e^{-i m_\varphi \varphi_* }$ are shown explicitly in (\ref{eqn:etaDef}). Interestingly, the true anomaly $\varphi_*$ only appears in the exponents of the oscillatory terms. On the other hand, the inclination angle $\iota_*$ determines the strength of the perturbation and implicitly affects the summation over $m_\varphi$ in (\ref{eqn:mgammaDef}). As we shall see, this means that $\iota_*$ also determines the polarization of the perturbation.

\vskip 4pt

It is instructive to illustrate the effect of $\iota_*$ on the tidal moments through an explicit example. Let us therefore consider the quadrupole $\ell_*=2$, where the tidal moments are
\beq
\begin{aligned}
\mathcal{E}_{2 \mp  2} & = \frac{1}{2}\sqrt{\frac{6 \pi }{5}}    \left( \cos \iota_*  \cos \varphi_* \pm  i \sin \varphi_* \right)^2 \, , \\
\mathcal{E}_{2 \mp 1} & = \pm \sqrt{\frac{6 \pi }{5}}  \sin \iota_* \cos \varphi_* \left( \cos \iota_*  \cos
   \varphi_* \pm  i  \sin \varphi_* \right), \\
   \mathcal{E}_{2 0} & =- \frac{1}{4}\sqrt{\frac{\pi}{5}}  
   \left( 1 - 3 \cos 2 \varphi_*  + 6 \cos^2 \varphi_*  \cos 2 \iota_* \right) \, . \label{eqn:Equadrupole}
\end{aligned}
\eeq
For equatorial orbits, $\iota_*=0$, this become 
\beq
\begin{aligned}
\mathcal{E}_{2 m_*}(\iota_*=0) & = \frac{1}{2} \sqrt{\frac{6\pi}{5}} \left\{ e^{ + 2 i \varphi_*}, \, 0 , \, -\sqrt{2/3} , \, 0 , \, e^{-2 i \varphi_*} \right\}  , \label{eqn:ElectricEquator}
\end{aligned}
\eeq
where the list runs from $m_* \in \{ -2, \cdots \hskip -2pt, +2 \}$. Since the tidal moments vanish for $m_* = \pm 1$, states with $\Delta m_{ab} = \pm 1$ decouple from each other---see the selection rules (S1) and (V1) in~\S\ref{sec:gravlevelmix}. On the other hand, all  of the non-vanishing components in (\ref{eqn:ElectricEquator}) consists of a \textit{single} oscillatory term $\propto e^{- i m_* \varphi_*}$, with $m_\varphi = m_*$. These properties generalize to all values of $\ell_*$. In particular, for even (odd) values of $\ell_*$, the odd (even) $m_*$-th component vanishes, and each of the non-vanishing component is a single oscillatory term with $m_\varphi = m_*$. For instance, for $\ell_*=3$, we find 
\beq
\mathcal{E}_{3 m_*}(\iota_*=0) = \frac{1}{2} \sqrt{\frac{5\pi}{7}} \left\{  e^{ + 3 i \varphi_*}, \, 0 , \, - \sqrt{3/5} e^{+ i \varphi_* }  , \, 0 , \,   \sqrt{3/5} e^{- i \varphi_* }  , \, 0 , \, - e^{- 3 i \varphi_* } \right\} \, , \label{eqn:ElectricEquatorOctupole}
\eeq
where states with $\Delta m_{ab} = \pm 1$ are now connected through the weaker octupolar perturbation. This property can be seen most transparently by setting $\iota_*=0$ in (\ref{eqn:Angles}), where $\Phi_* = \varphi_*$ and the spherical harmonics in (\ref{eqn:QuadrupoleTidalMomentOld0}) obeys $Y^*_{\ell_* m_*} \propto e^{- i m_* \Phi_*} \propto e^{- i m_* \varphi_*}$. Since the tidal moments for equatorial orbits induce perturbations with definite frequency $m_* \varphi_*$, these perturbations are said to be \textit{circularly-polarized}. 

\vskip 4pt

The properties described above no longer hold for general inclined orbits with $\iota_* \neq 0$. For example, for $\iota_*=\pi/3$, (\ref{eqn:Equadrupole}) becomes
\beq
\begin{aligned}
\mathcal{E}_{2 \mp 2}(\iota_*=\pi/3) & = \frac{1}{32}\sqrt{\frac{6\pi}{5}} \left( 9 e^{\pm 2 i \varphi_*} + e^{\mp 2 i \varphi_*} - 6 \right)  , \\
\mathcal{E}_{2 \mp 1}(\iota_*=\pi/3) & = \pm \frac{1}{16} \sqrt{\frac{18\pi}{5}}\left( 3 e^{\pm 2 i \varphi_*} - e^{\mp 2 i \varphi_*} + 2  \right)   , \\
\mathcal{E}_{2 0}(\iota_*=\pi/3) & = \frac{1}{48}\sqrt{\frac{9\pi}{5}} \left( 9 e^{+ 2 i \varphi_*} + 9 e^{- 2 i \varphi_*} + 2 \right)  . \label{eqn:ElectricInclined}
\end{aligned} 
\eeq
Not only are all of the tidal components now non-vanishing, they are given by a superposition of $\propto e^{\pm 2 i \varphi_*}$ and constant terms. 
This feature can also be generalized to all $\ell_*$ in inclined orbits, where the multipole moments consist of superpositions of oscillatory terms $\propto e^{- i m_\varphi \varphi_*}$, with $ m_\varphi$ summed over all permissible odd (even) values for odd (even) values of $\ell_*$. In these cases, the off-diagonal terms in the Schr\"odinger frame Hamiltonian, such as (\ref{eqn:Two-State-Schr-H}), consist of a summation over different frequencies.  
Furthermore, unlike for equatorial orbits, these tidal moments now necessarily contain oscillatory terms with opposite values of $\pm | m_\varphi |$. This is because the motion of an inclined orbit, \textit{projected} onto the equatorial plane, can be represented by a superposition of co-rotating and counter-rotating orbits. Since the amplitudes of these oscillatory terms are different (except for $m_*=0$, which is azimuthal about the spin-axis of the cloud), the projected equatorial orbit still has definite orientation. Inclined orbits therefore generate \textit{elliptically-polarized} gravitational perturbations. 

\vskip 4pt
Finally, in the special limit $\iota_*=\pi/2$, (\ref{eqn:Equadrupole}) turns into
\beq
\mathcal{E}_{2 m_*}(\iota_*=\pi/2)  = \frac{1}{2} \sqrt{\frac{6\pi}{5}} \left\{ - \sin ^2 \varphi_*, \,  i \sin 2\varphi_*, \,  (3 \cos 2 \varphi_* +1) / \sqrt{6}, \,   i \sin 2 \varphi_* ,\, - \sin ^2\varphi_* \right\}  . \label{eqn:ElectricOrthogonal}
\eeq
The amplitudes of both $\pm |m_\varphi|$ oscillatory terms now have the same magnitude. In this case, the orbital motion projected onto the equatorial plane is just given by an oscillating perturbation  along a straight line. By having the same magnitudes, the projected co-rotating and counter-rotating orbits cancel the net motion along the direction orthogonal to this straight line. The gravitational perturbation generated in this case is hence \textit{linearly-polarized}. Our explicit example for the $\ell_*=2$ tidal moment, from (\ref{eqn:ElectricEquator}) to (\ref{eqn:ElectricOrthogonal}), represent the continuous behavior of $\mathcal{E}_{\ell_* m_*}$ of all $\ell_*$ under a rotation about $\iota_*$.

\subsection{Higher-Order Couplings} \label{app:HigherOrder}

In \S\ref{sec:gravlevelmix}, we ignored the $\alpha$-corrections to the gravitational potentials (\ref{eqn:ScalarPert}) and (\ref{eqn:VectorPert}). However, these higher-order couplings can still mediate interesting transitions, especially if they mediate resonances that are forbidden by the selection rules of their leading-order counterparts. 
We will now show the expressions of these additional couplings, leaving a detailed investigation of their effects for future work. 

\vskip 4pt

It is useful to first estimate the order in $\alpha$ at which the $\alpha$-expansion should be truncated. As a useful guide, we require the Landau-Zener parameter (\ref{eqn:LZParam}) to be greater than order unity, such that the nature of the transitions are adiabatic. 
Denoting the higher-order overlaps by $\eta_\beta \equiv \alpha^\beta \eta$, where $\eta$ is the leading-order overlap, cf.~(\ref{eqn:etaDef}), we find that 
\beq
\frac{\eta_\beta^2}{\gamma} \simeq 2.8 \times 10^4 \frac{q}{(1+q)^{5/3}}\left(\frac{R_{ab}}{0.3} \right)^2 \left( \frac{|\Delta m_{ab}|}{2}\right)^{5/3} \left| \frac{n_a^2 n^{2}_b}{n^{ 2}_a - n_b^2  } \right|^{5/3} \left( \frac{0.07}{\alpha}\right)^5 \alpha^{2 \beta}   \, , \label{eqn:HigherOrderLZ}
\eeq
where we assumed the transitions are Bohr-like, and have used (\ref{eqn:gamma}) and (\ref{eqn:Eta}). By demanding adiabaticity in (\ref{eqn:HigherOrderLZ}), i.e.~$\eta_\beta^2/\gamma \gtrsim 1$, we find $\beta \lesssim 2$. This conclusion remains unchanged for fine and hyperfine transitions. In what follows, we will therefore expand the gravitational interactions up to order $\alpha^2$.

\vskip 4pt

For the scalar cloud, the gravitational perturbation up to order $\alpha^2$ is
\beq
\begin{aligned}
V_*  = - \frac{1}{4} \mu \bar{h}^{00}  - i \bar{h}^{0i} \partial_i + \mathcal{O}(\alpha^3) \, , \label{eqn:ScalarPert4}
\end{aligned}
\eeq
where the first term is the same as that in (\ref{eqn:ScalarPert}), and the second term arises from the (trace-reversed) gravitomagnetic perturbation $\bar{h}^{0i}$. The latter metric perturbation scales as $\bar{h}^{0i} \sim v \hskip 1pt \bar{h}^{00}$, where $v$ is the typical velocity of the binary. Since $\partial_i \psi \sim \mu \alpha \psi$ and $v \lesssim \alpha$ (see Footnote~\ref{footnote:alpha-scaling}), this new term is $\alpha^2$-suppressed compared to its leading-order counterpart. On the other hand, for the vector cloud, we find that 
\beq
\begin{aligned}
V_*^{il} = &  -\frac{1}{4} \delta^{il} \mu  \bar{h}^{00}   - \frac{\bar{h}^{00}}{ 2\mu}\partial^i \partial^l   - \frac{\partial_r \bar{h}^{00}}{4\mu} {\epsilon^{i}}_{rp} {\epsilon^{l}}_{pq} \partial_q \\
& + \frac{i}{2} \partial^l \bar{h}^{0i} + \frac{i}{2} \bar{h}^{0l} \partial^i  - i  \delta^{il} \bar{h}^{0p} \partial_p   \, , \label{eqn:VSpatial4}
\end{aligned}
\eeq
where $\epsilon_{ijk}$ is the Levi-Civita symbol, and we replace all of the temporal components $\psi^0$ with the Lorenz condition  $\partial_i \psi_i + i \mu\psi_0 =\mathcal{O}(\alpha^3) $. To obtain both (\ref{eqn:ScalarPert4}) and (\ref{eqn:VSpatial4}), we utilized the fact that the gravitomagnetic perturbation is divergence-less, i.e. $\partial_i \bar{h}^{i0}=0$. We also simplified these results by ignoring higher-order corrections that contain time-derivatives, $\partial_0 \psi \sim \mu \alpha^2 \psi$, as the types of resonances that they mediate are already accessible by the leading-order potentials. This is to be contrasted with terms with spatial gradients, which change the angular structures of the field and therefore  induce resonances that obey different selection rules.

\newpage

\section{Landau-Zener Transition}
\label{app:LZ}

As discussed in detail in Section~\ref{sec:gcollider}, a key feature of the evolution of boson clouds in the inspiral of a binary are resonant Landau-Zener transitions.
In this appendix, we collect details of these transitions that are relevant to, but not necessarily suitable for, the main text. Specifically, in \S\ref{app:twoState}, we first review the exact solution of the two-state Landau-Zener transition, cf.~\cite{Landau,Zener}, which provides intuition for understanding the more complicated multi-state transitions. In \S\ref{app:adFloTheo}, we then describe adiabatic Floquet theory. This is necessary for generalizing the dressed frame presented in Section~\ref{sec:gcollider} to connect the dynamics of a general binary-cloud system to those of a series of LZ transitions.

\subsection{The Two-State Model} 
\label{app:twoState}

  The Schr\"{o}dinger equation for the two-state LZ transition presented in \S\ref{sec:twoState} can be reduced to the Weber equation of a quantum harmonic oscillator, and is thus exactly solvable. In the following, we will review this solution.

  \vskip 4pt
  The dressed frame coefficients $d_1(t)$ and $d_2(t)$ evolve according to the Hamiltonian (\ref{eqn:Two-State-Schr-H}): 
  \begin{equation}
     \begin{aligned}
       i \dot{d}_1 &= +\tilde{\gamma} t d_1/2 + \eta d_2\, , \\
      i \dot{d}_2 &= -\tilde{\gamma} t d_2/2 + \eta d_1\,,
     \end{aligned} \label{eq:weberCoupled}
  \end{equation}
where we defined $\tilde{\gamma} \equiv |\Delta m| \gamma$. These equations can be combined into 
the Weber equations for either coefficient,
  \begin{equation}
    \begin{aligned}
     \ddot{d}_1 + \tfrac{1}{4}\left( (\tilde{\gamma} t)^2 + 2 i \tilde{\gamma} + 4\eta^2 \right) d_1 &= 0\,, \\
\ddot{d}_2 + \tfrac{1}{4} \left((\tilde{\gamma} t)^2 - 2 i \tilde{\gamma} + 4 \eta^2 \right) d_2 &= 0\, ,
    \end{aligned}
  \end{equation}
whose solutions are the parabolic cylinder functions \cite{NIST:DLMF},
   \begin{equation}
   \begin{aligned}
     d_{1}(t) &= C_1\, \lab{D}_{-i z}\left((-1)^{1/4}\sqrt{\tilde{\gamma}} t\right) + C_2\, \lab{D}_{i z - 1}\left((-1)^{3/4}\sqrt{\tilde{\gamma}} t\right) , \\ 
      d_{2}(t) &= \frac{(-1)^{1/4} }{\sqrt{z}} C_2 \,\lab{D}_{i z}\left((-1)^{3/4} \sqrt{\tilde{\gamma}} t\right) + (-1)^{1/4} \sqrt{z}\, C_1\, \lab{D}_{-i z-1}\left((-1)^{1/4} \sqrt{\tilde{\gamma}} t\right) ,
   \end{aligned} \label{eq:generalSolution}
  \end{equation}
   where we have defined $z \equiv \eta^2/\tilde{\gamma}$ and related the two solutions using (\ref{eq:weberCoupled}). We are ultimately interested in finding the total population contained in the second state $|d_2(t)|^2$ in the asymptotic future, assuming that the system fully occupied the first state in the asymptotic past, ${|d_{1}(-\infty)|^2 = 1}$. Imposing this initial condition, the undetermined coefficients in  (\ref{eq:generalSolution}) are (up to an arbitrary choice of phase)
   \begin{equation}
      C_1 = \exp\Big(\frac{3 \pi }{4 } z\Big) \quad \text{and} \quad C_2 = \frac{i \sqrt{2 \pi}}{\Gamma(i z )}\exp\Big(\frac{\pi }{4} z\Big)\,.
   \end{equation}
   We thus find that
   \begin{equation}
   \begin{aligned}
    |d_{1}(\infty)|^2 &= e^{-2 \pi z} \, ,\\
     |d_{2}(\infty)|^2 &= 1 - e^{-2\pi z} \,. \label{eq:twoStateResult1}
     \end{aligned}
  \end{equation}
  We see that for large values of the Landau-Zener parameter, $z= \eta^2/\tilde{\gamma} \gg 1$, 
  the ``probability'' that the system remains in the state $|1 \rangle$ is exponentially small
  and the state $|2\rangle$ becomes almost fully occupied.

  \vskip 4pt
  Likewise, if we impose $|d_{2}(-\infty)|^2 = 1$, then the coefficients in (\ref{eq:generalSolution}) are (up to an arbitrary phase)
  \begin{equation}
    C_1 = 0 \quad \text{and} \quad C_2 = z \exp\Big( \hskip -2pt -\frac{\pi}{4} z \Big)\, ,
  \end{equation}
and the asymptotic occupations   are the converse of (\ref{eq:twoStateResult1}):
  \begin{equation}
  \begin{aligned}
    |d_{1}(\infty)|^2 &= 1 - e^{-2 \pi z}\, , \\
    |d_{2}(\infty)|^2 &= e^{-2 \pi z}\, .
    \end{aligned}
    \label{eq:twoStateResult2}
  \end{equation}
  As discussed in \S\ref{sec:Smatrix}, the information contained in (\ref{eq:twoStateResult1}) and (\ref{eq:twoStateResult2}) can be encoded in an S-matrix, $|\psi(\infty)\rangle = S |\psi(-\infty)\rangle$. We will often only consider the modulus of the S-matrix elements
  \begin{equation}
    |S| = \begin{pmatrix} e^{-\pi z} & \displaystyle \sqrt{1 - e^{-2 \pi z}} \\ \displaystyle \sqrt{1 - e^{-2 \pi z}}  & e^{-\pi z}  \end{pmatrix},
  \end{equation}
which have well-defined asymptotic limits. 

  \subsection{Adiabatic Floquet Theory} 
  \label{app:adFloTheo}
  
    Throughout Section~\ref{sec:gcollider}, it was convenient to work in a dressed frame that rotated along with the orbital motion of the gravitational perturbation, making it clear that the system could evolve adiabatically. However, states are generically connected by perturbations that oscillate at multiple frequencies, especially when we consider more general inspiral configurations with eccentric and inclined orbits, cf.~\S\ref{app:TidalMoments}. In that case, there are multiple dressed frames and one would have to awkwardly switch between them depending on which resonance one is interested in. Fortunately, the Floquet Hamiltonian provides a natural generalization of the dressed frame to arbitrary inspiral configurations, and we now review its construction.

    \vskip 4pt
    Floquet theory relies on the existence of a compact variable or phase to trivialize some part of the dynamics. Throughout, we will concentrate on the Hamiltonian studied in the main text,
    \begin{equation}
      \mathcal{H}_{ab}(t, \varphi_*(t)) = E_a \delta_{ab} + \sum_{m_\varphi \in \mathbbm{Z}} \eta_{ab}^{(m_\varphi)}(t) e^{-i m_\varphi \varphi_*(t)}\,, \label{eq:appMainHam}
    \end{equation}
    where we now emphasize its dependence on the angular variable $\varphi_*(t)$. This Hamiltonian describes generic orbits, with inclination and eccentricity. Intuitively, it would be useful to expand in the Fourier modes $e^{i k \varphi_*(t)}$, as this will trivialize the oscillatory part of the dynamics. This is how Floquet theory is typically presented---if a Hamiltonian is time-periodic, $\mathcal{H}(t) = \mathcal{H}(t + 2\pi)$, we may expand the wavefunction as $\psi(x, t) = \sum_{k \in \mathbbm{Z}} \psi_k(x, t) e^{i k t}$ to derive a ``Floquet Hamiltonian'' for the modes $\psi_k(x, t)$ that does not depend on time. However, it is not clear what the Fourier modes are for the phase $\varphi_*(t)$, which can be an arbitrary non-monotonic function of time. It will thus be useful to present an alternative construction of the Floquet counterpart of (\ref{eq:appMainHam}), which is more thoroughly reviewed in \cite{Guerin:2003}.

    \vskip 4pt
    It will be useful to extend (\ref{eq:appMainHam}) to a family of Hamiltonians $\mathcal{H}_{ab}(t, \varphi_*(t) + \theta)$, parameterized by a new phase $\theta \in [0, 2 \pi)$, each with its own time evolution operator
    \begin{equation}
      i \es \partial_t U(t, t_0; \theta) = \mathcal{H}(t, \varphi_*(t) + \theta) U(t, t_0; \theta)\,.
    \end{equation}
    This is useful because we can then extend our original Hilbert space $\mathscr{H}$ to $\mathscr{K} = \mathscr{H} \otimes \mathscr{L}$, where $\mathscr{L} = L_2(S_1, \ud \theta/2 \pi)$ is the space of square-integrable $2 \pi$-periodic functions. This space is generated by $e^{i k \theta}$, with $k \in \mathbbm{Z}$. That is, we now think of $\theta$ as an additional variable in our problem. The dressed-frame coefficients then become functions of this variable, $c_a (t) \to c_a(t, \theta)$, and inner products between two states are defined by
    \begin{equation}
      \langle \psi_1 | \psi_2 \rangle = \sum_a \int_{0}^{2 \pi}\!\frac{\ud \theta}{2 \pi} c_{1, a}^*(t, \theta) c_{2, a}(t, \theta)\,,
    \end{equation}
    where $|\psi_i\rangle \equiv \sum_a c_{i, a}(t, \theta) |a \rangle$.
    Time evolution is again generated by $U(t, t_0; \theta)$, such that 
    \begin{equation}
      c_a(t, \theta) = U_{ab}(t, t_0; \theta) c_{b}(t_0, \theta)\,,
    \end{equation}
    where the time-evolution operator has been extended to the new Hilbert space $\mathscr{K}$ by treating its dependence on $\theta$ as multiplication.

    \vskip 4pt
    The reason this is useful is that we may now define a phase translation operator
    \begin{equation}
      \mathcal{T}_{\phi} = \exp(\phi\, \partial_\theta)\,,
    \end{equation}    
    which acts on a state $\psi(t, \theta)$ like $\mathcal{T}_\phi \psi(t, \theta) = \psi(t, \theta + \phi)$. Crucially, this operator allows us to remove the dependence of $\varphi_*(t)$ of the Hamiltonian (\ref{eq:appMainHam}):
    \begin{equation}
      \mathcal{H}(t, \varphi_*(t) + \theta) = \mathcal{T}_{\varphi_{*}(t)} \mathcal{H}(t, \theta) \mathcal{T}_{-\varphi_{*}(t)} \, . \label{eq:dressedFrameExt}
    \end{equation}
    This allows us to rewrite the full time evolution operator in terms of the \emph{Floquet time evolution operator}
    \begin{equation}
      U(t, t_0; \theta) = \mathcal{T}_{\varphi_*(t)} U_{\mathcal{K}}(t, t_0) \mathcal{T}_{-\varphi_*(t_0)}\,,
    \end{equation}
    which evolves $i \,\partial_t U_{\mathcal{K}}(t, t_0) = \mathcal{K}(t, \theta) U_{\mathcal{K}}(t, t_0)$ according to the \emph{Floquet Hamiltonian} 
   \begin{equation}
      \mathcal{K}_{ab}(t, \theta) =  \delta_{ab} \big(E_a -i \dot{\varphi}_*(t) \partial_\theta\big) + \sum_{m_\varphi \in \mathbbm{Z}} \eta_{ab}^{(m_\varphi)}(t)\, e^{-i m_\varphi \theta}\,. \label{eq:floquetHam}
    \end{equation}
    The transformation (\ref{eq:dressedFrameExt}) and the Hamiltonian (\ref{eq:floquetHam}) generalize the dressed frame transformation (\ref{eq:multiDressedUnitary}) and the Hamiltonian (\ref{eq:multiStateDressedHam}), respectively. All of the ``fast'' motion due to $\varphi_*(t)$ has now been removed from (\ref{eq:floquetHam}), which evolves slowly in time. 

    \vskip 4pt
    Given a solution of the Schr\"{o}dinger equation 
    \begin{equation}
      i \,\partial_t c_a(t, \theta) = \mathcal{K}_{ab}(t, \theta) c_{b}(t, \theta)\,,
    \end{equation}
    we can then generate a solution for the original Hamiltonian by simple substitution, $c_{a}(t) = c_{a}(t, \varphi_*(t))$. If we expand $c_{a}(t, \theta)$ into the Floquet eigenbasis $e^{i k \theta}$,
    \begin{equation}
      c_{a}(t, \theta) = \sum_{k \in \mathbbm{Z}} c_{a}^{(k)}(t) e^{i k \theta}\,,
    \end{equation}
    then it is clear that this substitution is simply the Fourier expansion in $\varphi_*(t)$ we originally wanted.

   \vskip 4pt
   It is particularly useful to analyze the dynamics in this Floquet basis, in which the Floquet Hamiltonian is
   \begin{equation}
      \mathcal{K}_{ab}^{f_1, f_2}(t) =  \int_{0}^{2 \pi}\!\frac{\ud \theta}{2\pi} \,e^{i f_1 \theta} \mathcal{K}_{ab}(t, \theta) e^{-i f_2 \theta} = 
      \delta_{ab} \delta_{f_1, f_2} \left(E_a - f_2 \dot{\varphi}_*(t)\right) + \eta_{ab}^{(f_1 - f_2)}(t)\,.
   \end{equation}
   Clearly, this is of the same form as the dressed Hamiltonian we considered in Section~\ref{sec:gcollider}, and the same decoupling arguments apply here. Indeed, we may consider the instantaneous (quasi)-energy eigenstates and describe the cloud's evolution throughout the entire inspiral as a series of isolated scattering events. Crucially, if the system is in a state with definite Floquet number, the translation in $\theta$ does not affect populations, $|c_a(t, \varphi_*(t))|^2 = |c_{a}(t, \theta)|^2$. We can avoid working with the Schr\"{o}dinger-frame coefficients entirely, if we are only interested in the population contained within a state.

  \section{Angular Momentum Transfer} 
  \label{app:AMTransfer}  

Our analysis of the backreaction in Section~\ref{sec:Backreaction} relied crucially on the balance of angular momentum between the orbit and the cloud. In this appendix, we provide a more detailed analysis of this balance. We first derive the cloud-orbit coupling through conservation of angular momentum, and then analyze the orbit's behavior during an adiabatic transition. We find that angular momentum conservation during the transition can cause the orbit to either float or kick.

\subsection{Orbital Dynamics}

  The motion of the binary companion satisfies (see e.g.~\cite{Blanchet:2013haa, Galley:2016zee})
   \begin{align}
      \ddot{R}_* - R_* \Omega^2 &= -\frac{(1+q)M}{R_*^2} + \frac{64 q (1+q) M^3}{15} \frac{\dot{R}_*}{R_*^4} + \frac{16 q M^2 }{5} \frac{\dot{R}^3_*}{R_*^3} + \frac{16 q M^2 }{5 R_*} \dot{R}_* \Omega^2\,, \label{eq:eomNewtonLaw}\\
      R_*^2 \dot{\Omega} + 2 R_* \dot{R}_* \Omega &= -\frac{24 q (1+q) M^3}{5 R_*^2} \Omega - \frac{8 q M^2}{5 R_*} \dot{R}_*^2 \Omega - \frac{8 q M^2}{5} R_* \Omega^3\,, \label{eq:eomAngMomCons}
    \end{align}
    where we have included the leading Newtonian and radiation reaction forces. Both the radial separation, $R_* = R_*(t)$, and instantaneous frequency, $\Omega = \Omega(t)$, are functions of time and, since the orbital plane does not precess, they fully describe the orbital motion. We should interpret (\ref{eq:eomAngMomCons}) as a statement of angular momentum conservation: the orbital angular momentum of the binary on the left-hand side decreases due to gravitational wave emission on the right-hand side.

    \vskip 4pt
    Fortunately, most of the terms in (\ref{eq:eomNewtonLaw}) and (\ref{eq:eomAngMomCons}) can be dropped for the large, quasi-circular orbits we consider. Let us consider such an orbit of characteristic size $R_{r}$ and frequency $\Omega_{r}$. We take the frequency to be near a Bohr resonance, $\Omega_{r} \sim \mu \alpha^2$. Because the orbit is quasi-circular, we have $\Omega_{r}^2 R_{r}^3 = (1+q)M$, and so the dimensionless quantity 
    \begin{equation}
      \varpi \equiv \frac{2 q^{1/5}  R_{r} \Omega_{r}}{(1+q)^{2/5}} \propto \frac{q^{1/5} \alpha}{(1+q)^{1/15}}\,, \label{eq:alphaSuppApp}
    \end{equation}
    is suppressed $\alpha \ll 1$.  
    Because the orbit changes on the timescale of order $\Omega_r/\gamma$, where $\gamma$ was defined in~(\ref{eqn:circleRate}), it will also be useful to define the dimensionless time
    \begin{equation}
      \tau \equiv \frac{\gamma t}{\Omega_r}\,.
    \end{equation}
 Finally, we define the dimensionless radius and frequency 
    \begin{equation}
      R_*(t) = R_{r} \es r_*(\tau) \quad \text{and} \quad \Omega_*(t) = \Omega_{r} \es \omega_*(\tau)\, ,
    \end{equation}
    so that the equations of motion become 
    \begin{align}
      0 &= 1 - \omega_*^2 r_*^3 + \frac{\varpi^{10}}{50} \left( 18 r^2 r_*'' - \frac{(4 + 3 \omega_*^2 r_*^3)r_*'}{r_*^2}\right)   - \frac{27 \varpi^{20}}{1250} \frac{r_*'^{3}}{r_*} \, , \\
      0 &= \omega_*' + \frac{2 \omega_* r_*'}{r^*} + \frac{\omega_*(3 + \omega_*^2 r_*^3)}{12 r_*^4} + \frac{3 \varpi^{10}}{100}\frac{\omega_* r_*'^{2}}{r_*^3}\,,
    \end{align}
    where primes denote derivatives with respect to the time $\tau$.
    Clearly, the terms dressed by (high) powers of $\varpi$  are suppressed\footnote{This suppression is lost for extremely large mass ratios. However, for $\alpha \lesssim 0.2$, such mass ratios~$q \gtrsim 10^{5}$ are far beyond those that we consider in this paper, and so we can safely ignore these terms.} for $\alpha \ll 1$, and we may ignore them. In that case, (\ref{eq:eomNewtonLaw}) reduces to
    \begin{equation}
      \Omega^2 R_*^3 = (1+q)M\,, \label{eq:eomNewtonLaw2}
    \end{equation}
    while (\ref{eq:eomAngMomCons}) becomes\hskip 1pt\footnote{Strictly, this is only true in the adiabatic limit, where we also ignore angular momentum lost due to gravitational wave emission of the cloud. Fortunately, these effects are negligible on the timescales we are interested in.}
    \begin{align}
      R_*^2 \dot{\Omega} + 2 R_* \dot{R}_* \Omega &= -\frac{24 q (1+q) M^3}{5 R_*^2} \Omega  - \frac{8 q M^2}{5} R_* \Omega^3\,. \label{eq:eomAngMomCons2}
    \end{align}
Including the angular momentum of the cloud (\ref{eqn:Scloud}), the last equation receives an extra contribution 
     \begin{align}
      R_*^2 \dot{\Omega} + 2 R_* \dot{R}_* \Omega \pm \frac{(1+q)\dot{S}_\lab{c}(t)}{q M} &= -\frac{24 q (1+q) M^3}{5 R_*^2} \Omega  - \frac{8 q M^2}{5} R_* \Omega^3\,, \label{eq:eomAngMomCons2X}
    \end{align}
    where the upper (lower) sign denotes co-rotating (counter-rotating) orbits. Using (\ref{eq:eomNewtonLaw2}), the orbital angular momentum $L = q M R_r^2 \Omega_r/(1+q)$, and the cloud's angular momentum (\ref{eqn:Scloud}),  this equation can then be rewritten as
    \begin{equation}
     \frac{\ud \Omega}{\ud t} = \gamma \left(\frac{\Omega}{\Omega_r}\right)^{11/3}\!\! \pm  3 R_J \es \Omega_r \left(\frac{\Omega}{\Omega_r}\right)^{4/3} \frac{\ud}{\ud t}\left[m_1 |c_1|^2 + m_2 |c_2|^2 + \dots + m_N |c_N|^2\right]\,, \label{eq:angMomCons2}
    \end{equation}
    where we have introduced $R_J$, the ratio of the cloud and orbital angular momenta (\ref{eq:ratioAngMom}).

    \vskip 4pt
Assuming $m_2 = \dots = m_N$ and $|c_1|^2 +  \dots + |c_N|^2 = 1$, we may write (\ref{eq:angMomCons2}) as
    \begin{equation}
  \frac{\ud \Omega}{\ud t} = \gamma \left(\frac{\Omega}{\Omega_r}\right)^{11/3} \!\! \pm 3 R_J \Delta m \Omega_r \left(\frac{\Omega}{\Omega_r}\right)^{4/3}\! \left(-\frac{\ud |c_1|^2}{\ud t}\right), \label{eq:dotOmegaInterp}
    \end{equation}
    where $\Delta m =m_2 - m_1$.  To get an intuition for the physics behind this equation, let us assume that the orbit is co-rotating, so that $+$ sign is appropriate above.  The population $|c_{1}|^2$ depletes during the transition, so its time derivative will be negative. If $\Delta m < 0 $, the cloud \emph{loses} angular momentum, and we see that the frequency will increase more slowly---the angular momentum lost by the cloud resupplies the orbital angular momentum, which depletes due to gravitational wave emission. Conversely, if $\Delta m > 0$, the cloud \emph{gains} angular momentum during the transition, causing the orbit to speed up and shrink faster.

\subsection{Backreaction Time}

We would like to understand how long the transition takes when we include the cloud's backreaction. Assuming $|c_1|^2$ to be a function only of $\Omega$, we may write (\ref{eq:dotOmegaInterp}) as
    \begin{equation}
\frac{\ud \Omega}{\ud t}= \gamma \left(\frac{\Omega}{\Omega_r}\right)^{11/3} \left[1 \pm 3 R_J \Delta m \Omega_r \left(\frac{\Omega}{\Omega_r}\right)^{4/3} \frac{\ud |c_1|^2}{\ud \Omega}\right]^{-1}\,. \label{eq:omegaDotEquation}
    \end{equation}
    It will be convenient to define the effective coupling $\eta_\lab{eff}$ by
    \begin{equation}
      \left.\frac{\ud |c_1|^2}{\ud \Omega}\right|_{\Omega_r} \!\!\equiv -\frac{|\Delta m|}{4 \eta_\lab{eff}}\,, \label{eq:etaEffDef}
    \end{equation}
     such that $\eta_\lab{eff} = |\eta|$ for the two-state system (\ref{eq:dressedFrameHam}). An exact form for $\eta_\lab{eff}$ for multi-state systems is difficult to find. However, numerical experiments show that $\eta_\lab{eff}^2 \approx  \tilde{\eta}_{1 2}^2 + \tilde{\eta}_{1 3}^2 + \cdots + \tilde{\eta}_{1 N}^2$ is a good approximation, where the parameters $\tilde{\eta}_{1 a}$ are the diagonalized couplings between the initial state and the degenerate subspace, cf.~(\ref{eq:diagDressFrameHam}).

     \vskip 4pt
   Near $\Omega = \Omega_r$, the equation of motion~(\ref{eq:omegaDotEquation}) simplifies to
    \begin{equation}
  \frac{\ud \Omega}{\ud t} = \frac{\gamma}{1 \mp 3 R_J \Delta m |\Delta m| \, \Omega_r/(4 \eta_{\lab{eff}})} +  \mathcal{O}(\Omega - \Omega_r)\,,
    \end{equation}
which integrates to
    \begin{equation}
      \Omega(t) \approx \Omega_r + \frac{\gamma t}{1 \mp 3 R_J |\Delta m| \Delta m \,\Omega_r/(4 \eta_\lab{eff})}\, + \mathcal{O}(t^2)\,. \label{equ:C16}
    \end{equation}
    As we explained before, if the orbit is co-rotating ($-$ sign) and $\Delta m < 0$, the cloud will \emph{lose} angular momentum to the  orbit, and the instantaneous frequency will increase more slowly. If $\Delta m >0$ however, the cloud will \emph{receive} angular momentum from the orbit, and the instantaneous frequency will grow more quickly.
The state of the cloud changes rapidly when the instantaneous frequency is in the resonance band $|\Omega(t) - \Omega_r| \lesssim \eta_\lab{eff}$. The time  it takes to cross this band decomposes into two pieces, $\Delta t_\lab{tot} = \Delta t \pm \Delta t_c$, where we have defined both the time it takes the binary to cross the resonance band without including backreaction, $\Delta t$, and the time added or subtracted by the cloud, $\Delta t_c$. In terms of the parameters in (\ref{equ:C16}), we get
   \begin{equation}
		\Delta t = \frac{2 \eta_\lab{eff}}{\gamma}  \quad {\rm and} \quad 
		\Delta t_c = \frac{3 R_J |\Delta m^2 \Omega_r|}{2 \gamma}\,.
	\end{equation}
	We discuss the observable consequences of this backreaction in Sections~\ref{sec:Backreaction} and~\ref{sec:unravel}.

\newpage
\phantomsection
\addcontentsline{toc}{section}{References}
\bibliographystyle{utphys}
\bibliography{collider}

\providecommand{\href}[2]{#2}\begingroup\raggedright\begin{thebibliography}{10}

\bibitem{Aad:2012tfa}
{G.~Aad {\it et al.}~[ATLAS Collaboration]}, ``{Observation of a New Particle
  in the Search for the Standard Model Higgs Boson with the ATLAS Detector at
  the LHC},'' \href{http://dx.doi.org/10.1016/j.physletb.2012.08.020}{{\em
  Phys. Lett.} {\bfseries B716} (2012) 1--29},
\href{http://arxiv.org/abs/1207.7214}{{\ttfamily arXiv:1207.7214 [hep-ex]}}.

\bibitem{Chatrchyan:2012xdj}
{S.~Chatrchyan {\it et al.}~[CMS Collaboration]}, ``{Observation of a New Boson
  at a Mass of 125 GeV with the CMS Experiment at the LHC},''
  \href{http://dx.doi.org/10.1016/j.physletb.2012.08.021}{{\em Phys. Lett.}
  {\bfseries B716} (2012) 30--61},
\href{http://arxiv.org/abs/1207.7235}{{\ttfamily arXiv:1207.7235 [hep-ex]}}.

\bibitem{Essig:2013lka}
R.~Essig {\em et~al.}, ``{Working Group Report: New Light Weakly Coupled
  Particles},'' in {\em {Community Summer Study 2013: Snowmass on the
  Mississippi (CSS2013) Minneapolis, MN, USA, July 29-August 6, 2013}}.
\newblock 2013.
\newblock
\href{http://arxiv.org/abs/1311.0029}{{\ttfamily arXiv:1311.0029 [hep-ph]}}.
\newblock

\bibitem{Abbott:2016blz}
{B.~Abbott {\it et al.}~(LIGO/Virgo Collaboration)}, ``{Observation of
  Gravitational Waves from a Binary Black Hole Merger},''
  \href{http://dx.doi.org/10.1103/PhysRevLett.116.061102}{{\em Phys. Rev.
  Lett.} {\bfseries 116} (2016) 061102},
\href{http://arxiv.org/abs/1602.03837}{{\ttfamily arXiv:1602.03837 [gr-qc]}}.

\bibitem{Abbott:2016nmj}
{B.~Abbott {\it et al.}~(LIGO/Virgo Collaboration)}, ``{GW151226: Observation
  of Gravitational Waves from a 22-Solar-Mass Binary Black Hole Coalescence},''
  \href{http://dx.doi.org/10.1103/PhysRevLett.116.241103}{{\em Phys. Rev.
  Lett.} {\bfseries 116} (2016) 241103},
\href{http://arxiv.org/abs/1606.04855}{{\ttfamily arXiv:1606.04855 [gr-qc]}}.

\bibitem{Abbott:2017vtc}
{B.~Abbott {\it et al.}~(LIGO/Virgo Collaboration)}, ``{GW170104: Observation
  of a 50-Solar-Mass Binary Black Hole Coalescence at Redshift 0.2},''
  \href{http://dx.doi.org/10.1103/PhysRevLett.118.221101}{{\em Phys. Rev.
  Lett.} {\bfseries 118} (2017) 221101},
\href{http://arxiv.org/abs/1706.01812}{{\ttfamily arXiv:1706.01812 [gr-qc]}}.

\bibitem{Abbott:2017oio}
{B.~Abbott {\it et al.}~(LIGO/Virgo Collaboration)}, ``{GW170814: A
  Three-Detector Observation of Gravitational Waves from a Binary Black Hole
  Coalescence},'' \href{http://dx.doi.org/10.1103/PhysRevLett.119.141101}{{\em
  Phys. Rev. Lett.} {\bfseries 119} (2017) 141101},
\href{http://arxiv.org/abs/1709.09660}{{\ttfamily arXiv:1709.09660 [gr-qc]}}.

\bibitem{TheLIGOScientific:2017qsa}
{B.~Abbott {\it et al.}~(LIGO/Virgo Collaboration)}, ``{GW170817: Observation
  of Gravitational Waves from a Binary Neutron Star Inspiral},''
  \href{http://dx.doi.org/10.1103/PhysRevLett.119.161101}{{\em Phys. Rev.
  Lett.} {\bfseries 119} (2017) 161101},
\href{http://arxiv.org/abs/1710.05832}{{\ttfamily arXiv:1710.05832 [gr-qc]}}.

\bibitem{GBM:2017lvd}
{B.~Abbott {\it et al.}}, ``{Multi-Messenger Observations of a Binary Neutron
  Star Merger},'' \href{http://dx.doi.org/10.3847/2041-8213/aa91c9}{{\em
  Astrophys. J.} {\bfseries 848} (2017) L12},
\href{http://arxiv.org/abs/1710.05833}{{\ttfamily arXiv:1710.05833
  [astro-ph.HE]}}.

\bibitem{Sathyaprakash:2019rom}
B.~Sathyaprakash {\em et~al.}, ``{Multimessenger Universe with Gravitational
  Waves from Binaries},''
\href{http://arxiv.org/abs/1903.09277}{{\ttfamily arXiv:1903.09277
  [astro-ph.HE]}}.

\bibitem{review}
R.~A. Porto, ``{The Effective Field Theorist's Approach to Gravitational
  Dynamics},'' \href{http://dx.doi.org/10.1016/j.physrep.2016.04.003}{{\em
  Phys. Rept.} {\bfseries 633} (2016) 1},
\href{http://arxiv.org/abs/1601.04914}{{\ttfamily arXiv:1601.04914 [hep-th]}}.

\bibitem{Porto:2016zng}
R.~A. Porto, ``{The Tune of Love and the Nature(ness) of Spacetime},''
  \href{http://dx.doi.org/10.1002/prop.201600064}{{\em Fortsch. Phys.}
  {\bfseries 64} no.~10, (2016) 723--729},
\href{http://arxiv.org/abs/1606.08895}{{\ttfamily arXiv:1606.08895 [gr-qc]}}.

\bibitem{Porto:2017lrn}
R.~A. Porto, ``{The Music of the Spheres: The Dawn of Gravitational Wave
  Science},''
\href{http://arxiv.org/abs/1703.06440}{{\ttfamily arXiv:1703.06440
  [physics.pop-ph]}}.

\bibitem{Barack:2018yly}
L.~Barack {\em et~al.}, ``{Black Holes, Gravitational Waves and Fundamental
  Physics: A Roadmap},'' \href{http://dx.doi.org/10.1088/1361-6382/ab0587}{{\em
  Class. Quant. Grav.} {\bfseries 36} no.~14, (2019) 143001},
\href{http://arxiv.org/abs/1806.05195}{{\ttfamily arXiv:1806.05195 [gr-qc]}}.

\bibitem{Sathyaprakash:2019yqt}
B.~Sathyaprakash {\em et~al.}, ``{Extreme Gravity and Fundamental Physics},''
\href{http://arxiv.org/abs/1903.09221}{{\ttfamily arXiv:1903.09221
  [astro-ph.HE]}}.

\bibitem{Bertone:2019irm}
G.~Bertone {\em et~al.}, ``{Gravitational Wave Probes of Dark Matter:
  Challenges and Opportunities},''
\href{http://arxiv.org/abs/1907.10610}{{\ttfamily arXiv:1907.10610
  [astro-ph.CO]}}.

\bibitem{Zeldovich:1971a}
Y.~{Zel'Dovich}, ``{Generation of Waves by a Rotating Body},'' {\em JETP
  Letters} {\bfseries 14} (1971) 180.

\bibitem{Zeldovich:1972spj}
Y.~{Zel'Dovich}, ``{Amplification of Cylindrical Electromagnetic Waves
  Reflected from a Rotating Body},'' {\em Sov. Phys. JETP} {\bfseries 35}
  (1972) 1085.

\bibitem{Brito:2015oca}
R.~Brito, V.~Cardoso, and P.~Pani, ``{Superradiance},''
  \href{http://dx.doi.org/10.1007/978-3-319-19000-6}{{\em Lect. Notes Phys.}
  {\bfseries 906} (2015) 1},
\href{http://arxiv.org/abs/1501.06570}{{\ttfamily arXiv:1501.06570 [gr-qc]}}.

\bibitem{Arvanitaki:2009fg}
A.~Arvanitaki, S.~Dimopoulos, S.~Dubovsky, N.~Kaloper, and J.~March-Russell,
  ``{String Axiverse},''
  \href{http://dx.doi.org/10.1103/PhysRevD.81.123530}{{\em Phys. Rev. D}
  {\bfseries 81} (2010) 123530},
\href{http://arxiv.org/abs/0905.4720}{{\ttfamily arXiv:0905.4720 [hep-th]}}.

\bibitem{Arvanitaki:2010sy}
A.~Arvanitaki and S.~Dubovsky, ``{Exploring the String Axiverse with Precision
  Black Hole Physics},''
  \href{http://dx.doi.org/10.1103/PhysRevD.83.044026}{{\em Phys. Rev. D}
  {\bfseries 83} (2011) 044026},
\href{http://arxiv.org/abs/1004.3558}{{\ttfamily arXiv:1004.3558 [hep-th]}}.

\bibitem{Yoshino:2014}
H.~Yoshino and H.~Kodama, ``{Gravitational Radiation from an Axion Cloud around
  a Black Hole: Superradiant Phase},''
  \href{http://dx.doi.org/10.1093/ptep/ptu029}{{\em PTEP} {\bfseries 2014}
  (2014) 043E02},
\href{http://arxiv.org/abs/1312.2326}{{\ttfamily arXiv:1312.2326 [gr-qc]}}.

\bibitem{Baumann:2018vus}
D.~Baumann, H.~S. Chia, and R.~A. Porto, ``{Probing Ultralight Bosons with
  Binary Black Holes},''
  \href{http://dx.doi.org/10.1103/PhysRevD.99.044001}{{\em Phys. Rev.}
  {\bfseries D99} (2019) 044001},
\href{http://arxiv.org/abs/1804.03208}{{\ttfamily arXiv:1804.03208 [gr-qc]}}.

\bibitem{Chen:2009zp}
X.~Chen and Y.~Wang, ``{Quasi-Single-Field Inflation and Non-Gaussianities},''
  \href{http://dx.doi.org/10.1088/1475-7516/2010/04/027}{{\em JCAP} {\bfseries
  1004} (2010) 027},
\href{http://arxiv.org/abs/0911.3380}{{\ttfamily arXiv:0911.3380 [hep-th]}}.

\bibitem{Baumann:2011nk}
D.~Baumann and D.~Green, ``{Signatures of Supersymmetry from the Early
  Universe},'' \href{http://dx.doi.org/10.1103/PhysRevD.85.103520}{{\em Phys.
  Rev.} {\bfseries D85} (2012) 103520},
\href{http://arxiv.org/abs/1109.0292}{{\ttfamily arXiv:1109.0292 [hep-th]}}.

\bibitem{Noumi:2012vr}
T.~Noumi, M.~Yamaguchi, and D.~Yokoyama, ``{Effective Field Theory Approach to
  Quasi-Single-Field Inflation and Effects of Heavy Fields},''
  \href{http://dx.doi.org/10.1007/JHEP06(2013)051}{{\em JHEP} {\bfseries 06}
  (2013) 051},
\href{http://arxiv.org/abs/1211.1624}{{\ttfamily arXiv:1211.1624 [hep-th]}}.

\bibitem{Flauger:2013hra}
R.~Flauger, D.~Green, and R.~A. Porto, ``{On Squeezed Limits in Single-Field
  Inflation. Part I},'' \href{http://dx.doi.org/10.1088/1475-7516/2013/08/032,
  10.1088/1475-7516/2013/08/032/}{{\em JCAP} {\bfseries 1308} (2013) 032},
\href{http://arxiv.org/abs/1303.1430}{{\ttfamily arXiv:1303.1430 [hep-th]}}.

\bibitem{Arkani-Hamed:2015bza}
N.~Arkani-Hamed and J.~Maldacena, ``{Cosmological Collider Physics},''
\href{http://arxiv.org/abs/1503.08043}{{\ttfamily arXiv:1503.08043 [hep-th]}}.

\bibitem{Arkani-Hamed:2018kmz}
N.~Arkani-Hamed, D.~Baumann, H.~Lee, and G.~L. Pimentel, ``{The Cosmological
  Bootstrap: Inflationary Correlators from Symmetries and Singularities},''
\href{http://arxiv.org/abs/1811.00024}{{\ttfamily arXiv:1811.00024 [hep-th]}}.

\bibitem{Baumann:2019oyu}
D.~Baumann, C.~Duaso~Pueyo, A.~Joyce, H.~Lee, and G.~L. Pimentel, ``{The
  Cosmological Bootstrap: Weight-Shifting Operators and Scalar Seeds},''
\href{http://arxiv.org/abs/1910.14051}{{\ttfamily arXiv:1910.14051 [hep-th]}}.

\bibitem{Landau}
L.~Landau, ``{Zur Theorie der Energie\"ubertragung II},'' {\em Z. Sowjetunion}
  {\bfseries 2} (1932) 46--51. \url{https://ci.nii.ac.jp/naid/10011873546/en/}.

\bibitem{Zener}
C.~{Zener}, ``{Non-Adiabatic Crossing of Energy Levels},''
  \href{http://dx.doi.org/10.1098/rspa.1932.0165}{{\em Proceedings of the Royal
  Society of London Series A} {\bfseries 137} no.~833, (Sep, 1932) 696--702}.

\bibitem{Demirtas:2018akl}
M.~Demirtas, C.~Long, L.~McAllister, and M.~Stillman, ``{The Kreuzer-Skarke
  Axiverse},''
\href{http://arxiv.org/abs/1808.01282}{{\ttfamily arXiv:1808.01282 [hep-th]}}.

\bibitem{Buchbinder:2008jf}
I.~Buchbinder, E.~Kirillova, and N.~Pletnev, ``{Quantum Equivalence of Massive
  Antisymmetric Tensor Field Models in Curved Space},''
  \href{http://dx.doi.org/10.1103/PhysRevD.78.084024}{{\em Phys. Rev.}
  {\bfseries D78} (2008) 084024},
\href{http://arxiv.org/abs/0806.3505}{{\ttfamily arXiv:0806.3505 [hep-th]}}.

\bibitem{Baumann:2019eav}
D.~Baumann, H.~S. Chia, J.~Stout, and L.~ter Haar, ``{The Spectra of
  Gravitational Atoms},''
  \href{http://dx.doi.org/10.1088/1475-7516/2019/12/006}{{\em JCAP} no.~12,
  (2019) 006},
\href{http://arxiv.org/abs/1908.10370}{{\ttfamily arXiv:1908.10370 [gr-qc]}}.

\bibitem{Detweiler:1980uk}
S.~Detweiler, ``{Klein-Gordon Equation and Rotating Black Holes},''
\href{http://dx.doi.org/10.1103/PhysRevD.22.2323}{{\em Phys. Rev.} {\bfseries
  D22} (1980) 2323--2326}.

\bibitem{Dolan:2007mj}
S.~Dolan, ``{Instability of the Massive Klein-Gordon Field on the Kerr
  Spacetime},'' \href{http://dx.doi.org/10.1103/PhysRevD.76.084001}{{\em Phys.
  Rev.} {\bfseries D76} (2007) 084001},
\href{http://arxiv.org/abs/0705.2880}{{\ttfamily arXiv:0705.2880 [gr-qc]}}.

\bibitem{Yoshino:2012kn}
H.~Yoshino and H.~Kodama, ``{Bosenova Collapse of Axion Cloud Around a Rotating
  Black Hole},'' \href{http://dx.doi.org/10.1143/PTP.128.153}{{\em Prog. Theor.
  Phys.} {\bfseries 128} (2012) 153--190},
\href{http://arxiv.org/abs/1203.5070}{{\ttfamily arXiv:1203.5070 [gr-qc]}}.

\bibitem{Dolan:2012yt}
S.~Dolan, ``{Superradiant Instabilities of Rotating Black Holes in the Time
  Domain},'' \href{http://dx.doi.org/10.1103/PhysRevD.87.124026}{{\em Phys.
  Rev.} {\bfseries D87} no.~12, (2013) 124026},
\href{http://arxiv.org/abs/1212.1477}{{\ttfamily arXiv:1212.1477 [gr-qc]}}.

\bibitem{Okawa:2014nda}
H.~Okawa, H.~Witek, and V.~Cardoso, ``{Black Holes and Fundamental Fields in
  Numerical Relativity: Initial Data Construction and Evolution of Bound
  States},'' \href{http://dx.doi.org/10.1103/PhysRevD.89.104032}{{\em Phys.
  Rev.} {\bfseries D89} no.~10, (2014) 104032},
\href{http://arxiv.org/abs/1401.1548}{{\ttfamily arXiv:1401.1548 [gr-qc]}}.

\bibitem{Brito:2014wla}
R.~Brito, V.~Cardoso, and P.~Pani, ``{Black Holes as Particle Detectors:
  Evolution of Superradiant Instabilities},''
  \href{http://dx.doi.org/10.1088/0264-9381/32/13/134001}{{\em Class. Quant.
  Grav.} {\bfseries 32} (2015) 134001},
\href{http://arxiv.org/abs/1411.0686}{{\ttfamily arXiv:1411.0686 [gr-qc]}}.

\bibitem{Arvanitaki:2014wva}
A.~Arvanitaki, M.~Baryakhtar, and X.~Huang, ``{Discovering the QCD Axion with
  Black Holes and Gravitational Waves},''
  \href{http://dx.doi.org/10.1103/PhysRevD.91.084011}{{\em Phys. Rev. D}
  {\bfseries 91} (2015) 084011},
\href{http://arxiv.org/abs/1411.2263}{{\ttfamily arXiv:1411.2263 [hep-ph]}}.

\bibitem{Arvanitaki:2016qwi}
A.~Arvanitaki, M.~Baryakhtar, S.~Dimopoulos, S.~Dubovsky, and R.~Lasenby,
  ``{Black Hole Mergers and the QCD Axion at Advanced LIGO},''
  \href{http://dx.doi.org/10.1103/PhysRevD.95.043001}{{\em Phys. Rev. D}
  {\bfseries 95} (2017) 043001},
\href{http://arxiv.org/abs/1604.03958}{{\ttfamily arXiv:1604.03958 [hep-ph]}}.

\bibitem{Brito:2017wnc}
R.~Brito {\em et~al.}, ``{Stochastic and Resolvable Gravitational Waves from
  Ultralight Bosons},''
  \href{http://dx.doi.org/10.1103/PhysRevLett.119.131101}{{\em Phys. Rev.
  Lett.} {\bfseries 119} (2017) 131101},
\href{http://arxiv.org/abs/1706.05097}{{\ttfamily arXiv:1706.05097 [gr-qc]}}.

\bibitem{Brito:2017zvb}
R.~Brito, S.~Ghosh, E.~Barausse, E.~Berti, V.~Cardoso, I.~Dvorkin, A.~Klein,
  and P.~Pani, ``{Gravitational Wave Searches for Ultralight Bosons with LIGO
  and LISA},'' \href{http://dx.doi.org/10.1103/PhysRevD.96.064050}{{\em Phys.
  Rev.} {\bfseries D96} no.~6, (2017) 064050},
\href{http://arxiv.org/abs/1706.06311}{{\ttfamily arXiv:1706.06311 [gr-qc]}}.

\bibitem{Witek:2012tr}
H.~Witek, V.~Cardoso, A.~Ishibashi, and U.~Sperhake, ``{Superradiant
  Instabilities in Astrophysical Systems},''
  \href{http://dx.doi.org/10.1103/PhysRevD.87.043513}{{\em Phys. Rev.}
  {\bfseries D87} no.~4, (2013) 043513},
\href{http://arxiv.org/abs/1212.0551}{{\ttfamily arXiv:1212.0551 [gr-qc]}}.

\bibitem{Pani:2012vp}
P.~Pani, V.~Cardoso, L.~Gualtieri, E.~Berti, and A.~Ishibashi, ``{Black Hole
  Bombs and Photon Mass Bounds},''
  \href{http://dx.doi.org/10.1103/PhysRevLett.109.131102}{{\em Phys. Rev.
  Lett.} {\bfseries 109} (2012) 131102},
\href{http://arxiv.org/abs/1209.0465}{{\ttfamily arXiv:1209.0465 [gr-qc]}}.

\bibitem{Pani:2012bp}
P.~Pani, V.~Cardoso, L.~Gualtieri, E.~Berti, and A.~Ishibashi, ``{Perturbations
  of Slowly Rotating Black Holes: Massive Vector Fields in the Kerr Metric},''
  \href{http://dx.doi.org/10.1103/PhysRevD.86.104017}{{\em Phys. Rev.}
  {\bfseries D86} (2012) 104017},
\href{http://arxiv.org/abs/1209.0773}{{\ttfamily arXiv:1209.0773 [gr-qc]}}.

\bibitem{Endlich:2016jgc}
S.~Endlich and R.~Penco, ``{A Modern Approach to Superradiance},''
  \href{http://dx.doi.org/10.1007/JHEP05(2017)052}{{\em JHEP} {\bfseries 05}
  (2017) 052},
\href{http://arxiv.org/abs/1609.06723}{{\ttfamily arXiv:1609.06723 [hep-th]}}.

\bibitem{East:2017ovw}
W.~East and F.~Pretorius, ``{Superradiant Instability and Backreaction of
  Massive Vector Fields around Kerr Black Holes},''
  \href{http://dx.doi.org/10.1103/PhysRevLett.119.041101}{{\em Phys. Rev.
  Lett.} {\bfseries 119} (2017) 041101},
\href{http://arxiv.org/abs/1704.04791}{{\ttfamily arXiv:1704.04791 [gr-qc]}}.

\bibitem{East:2017mrj}
W.~East, ``{Superradiant Instability of Massive Vector Fields Around Spinning
  Black Holes in the Relativistic Regime},''
  \href{http://dx.doi.org/10.1103/PhysRevD.96.024004}{{\em Phys. Rev.}
  {\bfseries D96} no.~2, (2017) 024004},
\href{http://arxiv.org/abs/1705.01544}{{\ttfamily arXiv:1705.01544 [gr-qc]}}.

\bibitem{East:2018glu}
W.~East, ``{Massive Boson Superradiant Instability of Black Holes: Nonlinear
  Growth, Saturation, and Gravitational Radiation},''
  \href{http://dx.doi.org/10.1103/PhysRevLett.121.131104}{{\em Phys. Rev.
  Lett.} {\bfseries 121} no.~13, (2018) 131104},
\href{http://arxiv.org/abs/1807.00043}{{\ttfamily arXiv:1807.00043 [gr-qc]}}.

\bibitem{Baryakhtar:2017ngi}
M.~Baryakhtar, R.~Lasenby, and M.~Teo, ``{Black Hole Superradiance Signatures
  of Ultralight Vectors},''
  \href{http://dx.doi.org/10.1103/PhysRevD.96.035019}{{\em Phys. Rev.}
  {\bfseries D96} no.~3, (2017) 035019},
\href{http://arxiv.org/abs/1704.05081}{{\ttfamily arXiv:1704.05081 [hep-ph]}}.

\bibitem{Cardoso:2018tly}
V.~Cardoso, {\'O}.~Dias, G.~Hartnett, M.~Middleton, P.~Pani, and J.~Santos,
  ``{Constraining the Mass of Dark Photons and Axion-Like Particles Through
  Black Hole Superradiance},''
  \href{http://dx.doi.org/10.1088/1475-7516/2018/03/043}{{\em JCAP} {\bfseries
  1803} no.~03, (2018) 043},
\href{http://arxiv.org/abs/1801.01420}{{\ttfamily arXiv:1801.01420 [gr-qc]}}.

\bibitem{Dolan:2018dqv}
S.~Dolan, ``{Instability of the Proca Field on Kerr Spacetime},''
  \href{http://dx.doi.org/10.1103/PhysRevD.98.104006}{{\em Phys. Rev.}
  {\bfseries D98} no.~10, (2018) 104006},
\href{http://arxiv.org/abs/1806.01604}{{\ttfamily arXiv:1806.01604 [gr-qc]}}.

\bibitem{Siemonsen:2019cr}
N.~Siemonsen and W.~East, ``{Gravitational Wave Signatures of Ultralight Vector
  Bosons from Black Hole Superradiance},''
\href{http://arxiv.org/abs/1910.09476}{{\ttfamily arXiv:1910.09476 [gr-qc]}}.

\bibitem{Thorne:1980ru}
K.~Thorne, ``{Multipole Expansions of Gravitational Radiation},''
\href{http://dx.doi.org/10.1103/RevModPhys.52.299}{{\em Rev. Mod. Phys.}
  {\bfseries 52} (1980) 299--339}.

\bibitem{PhysRev.131.435}
P.~Peters and J.~Mathews, ``{Gravitational Radiation from Point Masses in a
  Keplerian Orbit},'' \href{http://dx.doi.org/10.1103/PhysRev.131.435}{{\em
  Phys. Rev.} {\bfseries 131} (Jul, 1963) 435--440}.
  \url{https://link.aps.org/doi/10.1103/PhysRev.131.435}.

\bibitem{landau2013quantum}
L.~Landau and E.~Lifshitz, {\em Quantum Mechanics: Non-Relativistic Theory},
  vol.~3.
\newblock Elsevier, 2013.

\bibitem{Brundobler:1993smat}
S.~Brundobler and V.~Elser, ``{S-matrix for Generalized Landau-Zener
  Problem},'' \href{http://dx.doi.org/10.1088/0305-4470/26/5/037}{{\em Journal
  of Physics A: Mathematical and General} {\bfseries 26} no.~5, (Mar, 1993)
  1211--1227}. \url{https://doi.org/10.1088/0305-4470/26/5/037}.

\bibitem{Press:1972zz}
W.~Press and S.~Teukolsky, ``{Floating Orbits, Superradiant Scattering and the
  Black Hole Bomb},''
\href{http://dx.doi.org/10.1038/238211a0}{{\em Nature} {\bfseries 238} (1972)
  211--212}.

\bibitem{Zhang:2018kib}
J.~Zhang and H.~Yang, ``{Gravitational Floating Orbits Around Hairy Black
  Holes},'' \href{http://dx.doi.org/10.1103/PhysRevD.99.064018}{{\em Phys.
  Rev.} {\bfseries D99} no.~6, (2019) 064018},
\href{http://arxiv.org/abs/1808.02905}{{\ttfamily arXiv:1808.02905 [gr-qc]}}.

\bibitem{Cardoso:2011xi}
V.~Cardoso, S.~Chakrabarti, P.~Pani, E.~Berti, and L.~Gualtieri, ``{Floating
  and Sinking: The Imprint of Massive Scalars around Rotating Black Holes},''
  \href{http://dx.doi.org/10.1103/PhysRevLett.107.241101}{{\em Phys. Rev.
  Lett.} {\bfseries 107} (2011) 241101},
\href{http://arxiv.org/abs/1109.6021}{{\ttfamily arXiv:1109.6021 [gr-qc]}}.

\bibitem{Poisson:1997ha}
E.~Poisson, ``{Gravitational Waves from Inspiraling Compact Binaries: The
  Quadrupole Moment Term},''
  \href{http://dx.doi.org/10.1103/PhysRevD.57.5287}{{\em Phys. Rev. D}
  {\bfseries 57} (1998) 5287},
\href{http://arxiv.org/abs/gr-qc/9709032}{{\ttfamily arXiv:gr-qc/9709032
  [gr-qc]}}.

\bibitem{Flanagan:2007ix}
E.~Flanagan and T.~Hinderer, ``{Constraining Neutron Star Tidal Love Numbers
  with Gravitational Wave Detectors},''
  \href{http://dx.doi.org/10.1103/PhysRevD.77.021502}{{\em Phys. Rev. D}
  {\bfseries 77} (2008) 021502},
\href{http://arxiv.org/abs/0709.1915}{{\ttfamily arXiv:0709.1915 [astro-ph]}}.

\bibitem{Hinderer:2007mb}
T.~Hinderer, ``{Tidal Love Numbers of Neutron Stars},''
  \href{http://dx.doi.org/10.1086/533487}{{\em Astrophys. J.} {\bfseries 677}
  (2008) 1216},
\href{http://arxiv.org/abs/0711.2420}{{\ttfamily arXiv:0711.2420 [astro-ph]}}.

\bibitem{nrgr}
W.~Goldberger and I.~Rothstein, ``{An Effective Field Theory of Gravity for
  Extended Objects},'' \href{http://dx.doi.org/10.1103/PhysRevD.73.104029}{{\em
  Phys. Rev. D} {\bfseries 73} (2006) 104029},
\href{http://arxiv.org/abs/hep-th/0409156}{{\ttfamily arXiv:hep-th/0409156
  [hep-th]}}.

\bibitem{nrgrs}
R.~A. Porto, ``{Post-Newtonian Corrections to the Motion of Spinning Bodies in
  NRGR},'' \href{http://dx.doi.org/10.1103/PhysRevD.73.104031}{{\em Phys. Rev.
  D} {\bfseries 73} (2006) 104031},
\href{http://arxiv.org/abs/gr-qc/0511061}{{\ttfamily arXiv:gr-qc/0511061
  [gr-qc]}}.

\bibitem{nrgrs2}
R.~A. Porto and I.~Rothstein, ``{Next-to-Leading Order Spin(1)Spin(1) Effects
  in the Motion of Inspiralling Compact Binaries},''
  \href{http://dx.doi.org/10.1103/PhysRevD.81.029905,
  10.1103/PhysRevD.78.044013}{{\em Phys. Rev. D} {\bfseries 78} (2008) 044013},
  \href{http://arxiv.org/abs/0804.0260}{{\ttfamily arXiv:0804.0260 [gr-qc]}}.
[Erratum: Phys. Rev. D {\bf 81} (2010) 029905].

\bibitem{Berti:2019wnn}
E.~Berti, R.~Brito, C.~Macedo, G.~Raposo, and J.~Rosa, ``{Ultralight Boson
  Cloud Depletion in Binary Systems},''
  \href{http://dx.doi.org/10.1103/PhysRevD.99.104039}{{\em Phys. Rev.}
  {\bfseries D99} no.~10, (2019) 104039},
\href{http://arxiv.org/abs/1904.03131}{{\ttfamily arXiv:1904.03131 [gr-qc]}}.

\bibitem{Bekenstein:1973mi}
J.~Bekenstein, ``{Extraction of Energy and Charge from a Black Hole},''
\href{http://dx.doi.org/10.1103/PhysRevD.7.949}{{\em Phys. Rev.} {\bfseries D7}
  (1973) 949--953}.

\bibitem{Sennett:2017etc}
N.~Sennett, T.~Hinderer, J.~Steinhoff, A.~Buonanno, and S.~Ossokine,
  ``{Distinguishing Boson Stars from Black Holes and Neutron Stars from Tidal
  Interactions in Inspiraling Binary Systems},''
  \href{http://dx.doi.org/10.1103/PhysRevD.96.024002}{{\em Phys. Rev. D}
  {\bfseries 96} (2017) 024002},
\href{http://arxiv.org/abs/1704.08651}{{\ttfamily arXiv:1704.08651 [gr-qc]}}.

\bibitem{Zhang:2019eid}
J.~Zhang and H.~Yang, ``{Dynamic Signatures of Black Hole Binaries with
  Superradiant Clouds},''
\href{http://arxiv.org/abs/1907.13582}{{\ttfamily arXiv:1907.13582 [gr-qc]}}.

\bibitem{Ficarra:2018rfu}
G.~Ficarra, P.~Pani, and H.~Witek, ``{Impact of Multiple Modes on the
  Black-Hole Superradiant Instability},''
  \href{http://dx.doi.org/10.1103/PhysRevD.99.104019}{{\em Phys. Rev.}
  {\bfseries D99} no.~10, (2019) 104019},
\href{http://arxiv.org/abs/1812.02758}{{\ttfamily arXiv:1812.02758 [gr-qc]}}.

\bibitem{Taylor:2008xy}
S.~Taylor and E.~Poisson, ``{Nonrotating Black Hole in a Post-Newtonian Tidal
  Environment},'' \href{http://dx.doi.org/10.1103/PhysRevD.78.084016}{{\em
  Phys. Rev. D} {\bfseries 78} (2008) 084016},
\href{http://arxiv.org/abs/0806.3052}{{\ttfamily arXiv:0806.3052 [gr-qc]}}.

\bibitem{NIST:DLMF}
{\em {\it NIST Digital Library of Mathematical Functions}}.
\newblock \url{http://dlmf.nist.gov/}.

\bibitem{Guerin:2003}
S.~Gu\'{e}rin and H.~Jauslin, {\em Control of Quantum Dynamics by Laser Pulses:
  Adiabatic Floquet Theory},
  \href{http://dx.doi.org/10.1002/0471428027.ch3}{ch.~3, pp.~147--267}.
\newblock John Wiley \& Sons, Ltd, 2003.

\bibitem{Blanchet:2013haa}
L.~Blanchet, ``{Gravitational Radiation from Post-Newtonian Sources and
  Inspiralling Compact Binaries},''
  \href{http://dx.doi.org/10.12942/lrr-2014-2}{{\em Living Rev. Rel.}
  {\bfseries 17} (2014) 2},
\href{http://arxiv.org/abs/1310.1528}{{\ttfamily arXiv:1310.1528 [gr-qc]}}.

\bibitem{Galley:2016zee}
C.~Galley and I.~Rothstein, ``{Deriving Analytic Solutions for Compact Binary
  Inspirals without Recourse to Adiabatic Approximations},''
  \href{http://dx.doi.org/10.1103/PhysRevD.95.104054}{{\em Phys. Rev.}
  {\bfseries D95} no.~10, (2017) 104054},
\href{http://arxiv.org/abs/1609.08268}{{\ttfamily arXiv:1609.08268 [gr-qc]}}.

\end{thebibliography}\endgroup
\end{document}